\documentclass[12pt]{article}

\usepackage[utf8]{inputenc}
\usepackage[T1]{fontenc}

\usepackage{amsmath,amssymb,amsfonts}
\usepackage{bm,bbm}
\usepackage{booktabs}
\usepackage{color}
\usepackage[dvips,letterpaper]{geometry}
\usepackage{epsfig}
\usepackage{fullpage}
\usepackage{graphicx,psfrag,epsf}
\usepackage{indentfirst}
\usepackage[authoryear]{natbib}
\usepackage{setspace}
\usepackage{subfigure}
\usepackage{times}
\usepackage{url}
\usepackage{verbatim}
\usepackage[title,titletoc,toc]{appendix}
\usepackage{titlesec}
\usepackage{abstract}
\usepackage{graphics}
\graphicspath{{./figures/}}
\usepackage{algorithm,algcompatible,algpseudocode}

\usepackage{adjustbox}
\usepackage{caption}
\usepackage{multicol,multirow}
\usepackage{float}
\usepackage{mathtools}

\renewcommand{\sinh}{\text{senh}}

\input cyracc.def

\def \build#1#2#3{\mathrel{\mathop{#1}\limits^{#2}_{#3}}}

\def \p {\partial}

\def \build#1#2#3{\mathrel{\mathop{#1}\limits^{#2}_{#3}}}

\newtheorem{The}{Theorem}[section]
\newtheorem{Proof}{Proof}[section]

\title{\bf Parametric quantile autoregressive moving average models with exogenous terms applied to Walmart sales data}

\author{
\normalsize
\textbf{Alan Dasilva}$^{1}$, \textbf{Helton Saulo}$^{2}$, \textbf{Roberto Vila}$^{2}$, \textbf{Jose A. Fiorucci}$^{2}$ and \textbf{Suvra Pal}$^{3}$\\[-0.05cm]
{\small $^{1}$Institute of Mathematics and Statistics, Universidade de S\~{a}o Paulo, S\~{a}o Paulo, Brazil}\\[-0.05cm]
{\small $^{2}$Department of Statistics, Universidade de Bras\'{i}lia, Bras\'{i}lia, Brazil}\\[-0.05cm]
{\small $^{2}$Department of Mathematics, University of Texas at Arlington, Arlington, TX, USA}\\[-0.05cm]
}

\date{}

\begin{document}

\maketitle

\maketitle
\begin{abstract}
Parametric autoregressive moving average models with exogenous terms (ARMAX) have been widely used in the literature. Usually, these models consider a conditional mean or median dynamics, which limits the analysis. In this paper, we introduce a class of quantile ARMAX models based on log-symmetric distributions. This class is indexed by quantile and dispersion parameters. It not only accommodates the possibility to model bimodal and/or light/heavy-tailed distributed data but also accommodates heteroscedasticity. We estimate the model parameters by using the conditional maximum likelihood method. Furthermore, we carry out an extensive Monte Carlo simulation study to evaluate the performance of the proposed models and the estimation method in retrieving the true parameter values. Finally, the proposed class of models and the estimation method are applied to a dataset on the competition ``M5 Forecasting - Accuracy'' that corresponds to the daily sales history of several Walmart products. The results indicate that the proposed log-symmetric quantile ARMAX models have good performance in terms of model fitting and forecasting.

\end{abstract}

\section{Introduction}\label{sec:01}
Autoregressive moving average models with exogenous terms (ARMAX) have been widely used in several fields of study. These models can be seen as a generalization of regression models by incorporating a temporal dependency structure; see \cite{sauloetlal:20}. In the past two decades, the number of non-Gaussian ARMAX models in the literature has increased considerably. Much of this increase was due to the work of \cite{brs:03}, where the authors proposed a class of generalized ARMAX models by assuming the dependent variable to follow a conditional exponential family of distributions given the past history of the process. \cite{maiorcysneiros:18} considered conditional symmetric distributions for the dependent variable, whereas \cite{gmct:18} and \cite{cordeiroandrade:09} studied transformed symmetric ARMAX models. \cite{rochacribari:09}, \cite{bayeretal:17} and \cite{zmka:19} discussed ARMAX models based on the beta distributions, Kumaraswamy and skew normal distributions, respectively. \cite{rb:18} and \cite{leivaelal:21} introduced ARMAX models based on the Birnbaum-Saunders distribution. Finally, \cite{sauloetlal:20} proposed a class of ARMAX models based on log-symmetric distributions. This class contains many conditional distributions as special cases, such as the log-normal, log-Student-$t$, log-power-exponential, log-hyperbolic, log-slash, log-contaminated-normal, extended Birnbaum-Saunders and extended Birnbaum-Saunders-$t$ distributions, among others.

Although there are several works on ARMAX models in the literature, they are usually specified in terms of a conditional mean or median, which limits the analysis. In cases where a broader analysis is desired, that is, along the spectrum of the dependent variable, it is necessary to use a quantile approach;
interested readers may refer to the books by \cite{koenker:05}, \cite{hn:07} and \cite{dfv:14} for elaborate details on quantile regression models.

In this work, we propose a new class of ARMAX models by modeling the conditional quantile rather than the traditionally employed conditional median or mean. The proposed model is based on the class of quantile-based log-symmetric distributions introduced by \cite{saulo:21}. This class is indexed by quantile and dispersion parameters. It accommodates the possibility to model bimodal and/or light/heavy-tailed distributed data, in addition to accommodating heteroscedasticity; see \cite{vanegaspaula:17} and \cite{cunhaetal:21}. We illustrate the usefulness of the proposed log-symmetric quantile ARMAX models by using a dataset on the competition "M5 Forecasting - Accuracy" \citep{makridakis2020m5} that corresponds to the daily sales history of several Walmart products. Note that in the literature there are at least three possible techniques to tackle quantile modeling, namely, the distribution-free technique, the pseudo-likelihood technique via an asymmetric Laplace distribution, and the parametric technique with the maximum likelihood framework; see \cite{cunhaetal:21}. The proposed methodology falls into the last category and it is a generalization of the work of \cite{sauloetlal:20} to a quantile environment.

The rest of this paper is organized as follows. In Section \ref{sec:2}, we briefly describe the quantile-based log-symmetric distributions, and then introduce the log-symmetric quantile ARMAX models. In this section, we also discuss stationary conditions, inference, prediction and residual analysis. In Section \ref{sec:3}, we carry out Monte Carlo simulation studies to assess the accuracy and precision of the conditional maximum likelihood (CML) estimators as well as to assess the empirical distribution of the residuals. In Section \ref{sec:4}, we apply our proposed models to analyze Walmart sales data. Finally, in Section \ref{sec:5}, we make some concluding remarks and discuss future research in this direction.

\section{Log-symmetric quantile ARMAX models}\label{sec:2}

\subsection{Quantile-based log-symmetric distributions}

A random variable $Y$ follows a quantile-based log-symmetric {\color{black}(QLS)} distribution, with quantile parameter $Q>0$ and power parameter $\kappa>0$, if it's probability density function (PDF) is given by
\begin{equation}\label{QLS-distribution}
	f_{Y}(y;Q,\kappa)
	=
	\frac{\xi_{nc}}{\sqrt{\kappa} y} \,
	g\left( \frac{1}{\kappa} \left[ \log \left( \frac{y}{Q} \right) + \sqrt{\kappa}z_{\tau} \right]^2  \right), \quad y > 0,
\end{equation}
where $g(u)>0,u>0$ is a kernel density generator function usually associated with an additional parameter $\vartheta$ (or extra parameter vector $\boldsymbol{\vartheta}$) such that $\xi_{nc}$ is a normalization constant
and $G(\omega) = \xi_{nc} \int_{-\infty}^{\omega} g(z^2) \mathrm{d}z$ with  $\omega \in \mathbb{R}$, and $z_{\tau}=G^{-1}(\tau)$. Let us denote $Y \sim \mathrm{QLS}(Q,\kappa,g)$. The parameter $Q$ is the $100\tau$-th quantile of $Y$ and the parameter $\kappa$ represents the skewness (or the relative dispersion). Some special cases of the quantile-based log-symmetric family of distributions can be obtained by assuming some particular forms for $g(\cdot)$, which are given below

\begin{itemize}
\item Log-normal($\lambda,\kappa$), $g(u)=\exp\left( -\frac{1}{2}u\right)$;
\item Log-Student-$t$($\lambda,\kappa,\vartheta$), $g(u)=\left(1+\frac{u}{\vartheta} \right)^{-\frac{\vartheta+1}{2}}$, $\vartheta>0$;
\item Log-power-exponential($\lambda,\kappa,\vartheta$), $g(u)=\exp\left( -\frac{1}{2}u^{\frac{1}{1+\vartheta}}\right)$, $-1<{\vartheta}\leq{1}$;
\item Log-hyperbolic($\lambda,\kappa,\vartheta$), $g(u)=\exp(-\vartheta\sqrt{1+u})$, $\vartheta>0$;
\item Log-slash($\lambda,\kappa,\vartheta$), $g(u)=\textrm{IGF}\left(\vartheta+\frac{1}{2} ,\frac{u}{2}\right)$, $\vartheta>0$;
\item Log-contaminated-normal($\lambda,\kappa,{\bm\vartheta}=(\vartheta_1,\vartheta_2)^\top$), $g(u)=\sqrt{\vartheta_2}\exp\left(-\frac{1}{2}\vartheta_2 u\right)+\frac{(1-\vartheta_1)}{\vartheta_1},
\exp\left(-\frac{1}{2} u\right)$, $0<\vartheta_1,\vartheta_2<1$;
\item Extended Birnbaum-Saunders($\lambda,\kappa,\vartheta$), $g(u)=\cosh(u^{1/2})\exp\left(-\frac{2}{\vartheta^2}\sinh^2(u^{1/2}) \right)$, $\vartheta>0$;
\item Extended Birnbaum-Saunders-$t$($\lambda,\kappa,{\bm\vartheta}=(\vartheta_{1},\vartheta_{2})^{\top}$), $g(u)=\cosh(u^{1/2})\left(\vartheta_{2}\vartheta_{1}^2+4\sinh^2(u^{1/2})\right)^{-\frac{\vartheta_{2}+1}{2}} $, $\vartheta_{1},\vartheta_{2}>0$.
\end{itemize}

{\color{black}
The following properties follow immediately from the definition of the QLS distribution:
\begin{itemize}
	\item[(P1)] The normalization constant $\xi_{nc}$ is expressed as
	$\xi_{nc}=1/\int_{-\infty}^\infty g(z^2){\rm d}z=1/\int_{0}^\infty u^{-1/2}g(u){\rm d}u$;
	\item[(P2)]
		There is a real-valued function $r(\cdot)$ so that all modes of the {QLS} distribution satisfy the following identity:
	\begin{align*}
	r(x^2)x=\dfrac{\sqrt{\kappa}}{2}
	\quad	\text{with }
	x={\log(y/Q) + \sqrt{\kappa}z_{\tau} \over\sqrt{\kappa}}.
	\end{align*}
	For example, in the cases of Log-normal,
	Log-Student-$t$,
	Log-power-exponential,
	Log-hyperbolic and
	Extended Birnbaum-Saunders distributions, we have
	$r(x)=-1/2$,
	$r(x)=-(\nu+1)((x/\nu)+1)^{-1}/(2\nu)$,
	$r(x)=x^{-\nu/(\nu+1)}/(2(\nu+1))$,
	$r(x)=-\nu(x+1)^{-1/2}/2$ and
	$r(x)=\sinh(\sqrt{x})\cosh(\sqrt{x})$ $\times\big((\nu^2/\cosh^2(\sqrt{x}))-4\big)/(2\nu^2\sqrt{x})$,
	respectively;
\end{itemize}
and, analogously to \cite{vanegas2016},
\begin{itemize}
	\item[(P3)]
	The CDF of $Y \sim \mathrm{QLS}(Q,\kappa,g)$, denoted by $F_Y(y;Q,\kappa)$,  is expressed as $F_Y(y;Q,\kappa)=
	G\big((\log(y/{Q})+ \sqrt{\kappa}z_{\tau})/\sqrt{\kappa}\big)
	=\mathrm{P}(Z\leqslant\log({y}/{Q}) + \sqrt{\kappa}z_{\tau})$ with $Z=(Y-\log({Q})+ \sqrt{\kappa}z_{\tau})/\sqrt{\kappa}$;
	\item[(P4)]
	The random variable $(T/Q)^{1/\sqrt{\kappa}}$ follows the standard QLS distribution. That is, $(T/Q)^{1/\sqrt{\kappa}}\sim {\rm QLS}(1,1,g)$;
	\item[(P5)]
	For all $c>0$, $cT\sim {\rm QLS}(cQ,\kappa,g)$;
	\item[(P6)]
	For all $c\neq 0$, $T^c\sim {\rm QLS}(Q^c,c^2\kappa,g)$;
	\item[(P7)]
	The random variables $T \exp(\sqrt{\kappa} z_\tau)/Q$ and $Q/\big(T \exp(\sqrt{\kappa} z_\tau)\big)$ are identically distributed.
\end{itemize}
}

\subsection{Log-symmetric quantile ARMAX models}

Let $\left\lbrace Y_t \right\rbrace_{t=1}^n $ be a sequence of random variables defined in the probability space $\{ \Omega,\mathcal{A},\mathbb{P}\}$, and $\mathcal{A}_{t}=\sigma(Y_1,\ldots,Y_t)$ be a $\sigma$-algebra generated by information observed up to time $t$. Moreover, we define $\mathcal{A}_{0}=\{\Omega,\emptyset\}$. Then, by assuming that the {\color{black} conditional random variable} $Y_t$ given $\mathcal{A}_{t-1}=\{Y_{t-1},\ldots,Y_1,Q_{t-1},\ldots,Q_1,\kappa_{t-1},\ldots,\kappa_1\}$
{\color{black} follows the QLS distribution in \eqref{QLS-distribution}},
denoted by $Y_t|\mathcal{A}_{t-1} \sim \mathrm{QLS}(Q_t,\kappa_t,g)$, we have the following PDF:
\begin{equation}
	f_{Y_t|\mathcal{A}_{t-1}}(y_t;Q_t,\kappa_t|\mathcal{A}_{t-1}) = \frac{\xi_{nc}}{\sqrt{\kappa_t} y_t} g\left( \frac{1}{\kappa_t} \left[ \log \left( \frac{y_t}{Q_t} \right) + \sqrt{\kappa_t}z_p \right]^2  \right),\, y_t > 0, \nonumber
\end{equation}
where $Q_t = Q_{Y}(q|\mathcal{A}_{t-1})$ and $\kappa_t$ represents the conditional quantile and the skewness (or the relative dispersion), respectively. Then, we define the QLS-ARMA as follows:
\begin{eqnarray}\label{eq:eta}
	\eta_t &=& h(Q_t) = \boldsymbol{x_t^\top \beta} + \varrho_t, \; t = m+1, \ldots, n,\\ \nonumber
	\gamma_t &=& d(\kappa_t) = \boldsymbol{w_t^\top \tau},
	\end{eqnarray}
where $\boldsymbol{\beta} = (\beta_0, \beta_1, \ldots,\beta_k)^\top$ and $\boldsymbol{\tau} = (\tau_0,\tau_1,\ldots,\tau_l)^\top$ are vectors of unknown parameters, $\boldsymbol{x_t^\top} = (1,x_{t1},x_{t2},\ldots,x_{tk})^\top$ and $\boldsymbol{w_t^\top} = (1,w_{t1},w_{t2},\ldots,w_{tl})^\top$ are vectors containing the values of $k< n$ and $l<n$ covariates. In addition, $h: \mathbb{R} \rightarrow \mathbb{R}^{+}$ and $d: \mathbb{R} \rightarrow \mathbb{R}^{+}$ denotes two continuously twice differentiable monotone link functions with inverses given by $h^{-1}: \mathbb{R}^{+}  \rightarrow \mathbb{R}$ and $d^{-1}: \mathbb{R}^{+}  \rightarrow \mathbb{R}$, respectively, which are twice continuously differentiable as well. In \eqref{eq:eta}, $\varrho_t$ follows a dynamic ARMA structure as
\begin{equation}\label{eq:varrho}
	\varrho_t = \sum_{i=1}^{p} \phi_i \left[ h(Y_{t-i}) - \boldsymbol{x_{t-i}^\top \beta} \right] + \sum_{j = 1}^{q} \theta_j r_{t-j},
\end{equation}
where $\boldsymbol{\phi} = (\phi_1,\ldots,\phi_p)^\top \in \mathbb{R}^p$ and $\boldsymbol{\theta} = (\theta_1,\ldots,\theta_q)^\top \in \mathbb{R}^q$ are ARMA parameters with $p$ and $q$ denoting their respective orders, $r_t$ is a martingale difference sequence (MDS), that is, ${\rm E}|r_t|<\infty$, and ${\rm E}[r_t|\mathcal{A}_{t-1}]=0$, a.s., for all $t$. Then, ${\rm E}[r_t]=0$ for all $t$, and $\mathrm{Cov}[r_s,r_t]=0$ (uncorrelatedness of the sequence) for all $t\neq s$. Therefore, from Equations \eqref{eq:eta} and \eqref{eq:varrho}, we have
\begin{equation}\label{eq:qlsarma}
	\eta_t =   \boldsymbol{x_t^\top \beta} + \sum_{i=1}^{p} \phi_i \left[ h(Y_{t-i}) - \boldsymbol{x_{t-i}^\top \beta} \right] + \sum_{j = 1}^{q} \theta_j r_{t-j},
\end{equation}
which leads to the notation QLS-ARMA($p, q$).

\subsection{Stationarity conditions}

\begin{The}\label{teo1}
	The marginal mean of $Y_t$ in the QLS-ARMA($p,q$) model is given by
	\begin{equation}
		\mathrm{E}[Y_t] = h^{-1}(\boldsymbol{x_t^\top \beta}), \nonumber
	\end{equation}
	where $\Phi(B):\mathbb{R}\to\mathbb{R}$ is an invertible operator (the autoregressive polynomial) defined by $\Phi(B) = -\sum_{i=0}^{p}\phi_i B^i$ with $\phi_0=-1$, and $B^i$ is the lag operator such that $B^i y_t = y_{t-i}$.
\end{The}

\begin{The}\label{teo2}
	The marginal variance of $Y_t$ in the QLS-ARMA($p,q$) model is given by
	\begin{equation}
		\mathrm{Var}[Y_t] = \sum_{i = 0}^{\infty} \psi_i^2 \mathrm{E}[\mathrm{Var}[Y_{t-i}|\mathcal{B}_{t-i-1}]], \nonumber
	\end{equation}
	where $\boldsymbol{\Psi}(B) = \boldsymbol{\Theta}(B) \boldsymbol{\Phi}(B)^{-1} = \sum_{i = 0}^{\infty} \psi_i B^{i}$, $\boldsymbol{\Phi}(B)$ is invertible and $\mathcal{B}_t = \sigma(Y_t,Y_{t-1},\ldots,)$ is the $\sigma$-algebra generated by the information up to time $t$.
\end{The}

\begin{The}\label{teo3}
	The covariance and correlation of $Y_t$ and $Y_{t-k}$ in the QLS-ARMA($p,q$) model are given by
	\begin{eqnarray}
		\mathrm{Cov}[Y_t,Y_{t-k}] &=& \sum_{i = 0}^{\infty} \psi_i \psi_{i-k} \mathrm{E}[\mathrm{Var}[Y_t|\mathcal{B}_{t-1}]], \; k > 0, \ \ \text{and} \nonumber \\
		\mathrm{Corr}[Y_t,Y_{t-k}] &=& \frac{\sum_{i=0}^{\infty} \psi_i \psi_{i+k} \mathrm{E}[\mathrm{Var}[Y_{t-i}|\mathcal{B}_{t-i-1}]]}{\prod_{j \in \left\lbrace 0,k \right\rbrace } \sqrt{ \sum_{i = 0}^{\infty} \psi_{i+j}^2 \mathrm{E}[\mathrm{Var}[Y_{t-j-i}|\mathcal{B}_{t-j-i-1}]]}}, \nonumber
	\end{eqnarray}
	respectively, where $\mathcal{B}_t = \sigma(Y_t,Y_{t-1},\ldots,)$ is a $\sigma$-algebra generated by the information up to time $t$.
\end{The}

Theorems \ref{teo1}-\ref{teo3}, as well as their proofs, are discussed in more detail in Appendix \ref{sec:apendxa}.

\subsection{Estimation and inference}

The estimates of the parameters of the QLS-ARMA($p,q$) model can be obtained by using the CML method based on the first $m = \max \left\lbrace p,q \right\rbrace $ ($m<n$) observations. Let $\boldsymbol{\zeta} = \boldsymbol{(\beta^\top,\tau^\top,\phi^\top,\theta^\top)^\top}$ denote the parameter vector. Then, the conditional likelihood function is given by
\begin{equation}
	\bm{L}(\boldsymbol{\zeta})_{m,n} = \prod_{t = m+1}^{n} f_{Y_t|\mathcal{A}_{t-1}}(y_t;Q_t,\kappa_t|\mathcal{A}_{t-1}),\quad y_t > 0, Y_t|\mathcal{A}_{t-1} \sim \mathrm{QLS}(Q_t,\kappa_t,g),
\end{equation}
which implies that the conditional log-likelihood function (without the constant) can be expressed as
\begin{equation}\label{eq:loglik}
		\bm{\ell}(\boldsymbol{\zeta})_{m,n} = \sum_{t = m+1}^{n} \log(g(z_t^2)) - \frac{1}{2} \sum_{t = m+1}^{n} \log(\kappa_t),
\end{equation}
where $z_t = \left[ \log(y_t/Q_t) + \sqrt{\kappa_t}z_{\tau} \right]/\sqrt{\kappa_t} $, $t = m+1, \ldots,n$, $Q_t$ and $\kappa_t$ are as given in \eqref{eq:qlsarma}.

An estimate of $\boldsymbol{\zeta} = \boldsymbol{(\beta^\top,\tau^\top,\phi^\top,\theta^\top)^\top}$ can be obtained by equating the score vector containing the first-order partial derivatives of $\bm{\ell}(\boldsymbol{\zeta})_{m,n}$, denoted by $\dot{\bm{\ell}}_{\zeta} = (\dot{\bm \ell}_{\beta_s}^\top,\dot{\bm \ell}_{\tau_s}^\top,\dot{\bm \ell}_{\phi_s}^\top,\dot{\bm \ell}_{\theta_s}^\top)^\top$, to zero {\color{black} vector}, leading to the likelihood equations. In the given context, they need to be solved by an iterative procedure for non-linear optimization, such as the Broyden-Fletcher-Goldfarb-Shanno (BFGS) algorithm. In this regard, starting values are required to initiate the iterative procedure. These can be obtained from the least squares estimates (see Subsection~\ref{asymleastsquares}) or the \texttt{R} functions \texttt{arima} and \texttt{ssym.l}, the latter being associated with the \texttt{ssym} package; see \cite{r:2021}.
{\color{black}
The first-order partial derivatives with respect to each parameter
}
are given by
\begin{eqnarray}
	\dot{\bm \ell}_{\beta_s} &=& \sum_{t = m+1}^{n} \frac{v(z_t)}{\sqrt{\kappa_t}} \frac{z_t}{Q_t} \frac{\p Q_t}{\p \eta_t} \left( x_{t,s} -\sum_{i=1}^{p} \phi_i  x_{t-i,s} - \sum_{j=1}^{q} \theta_j \frac{\p Q_{t-j}}{\p \eta_{t-j}} \frac{\p \eta_{t-j}}{\p \beta_s} \right), \; s = 0, \ldots,k, \label{eq:beta_score}
	\\
	\dot{\bm \ell}_{\tau_s} &=& \frac{1}{2} \sum_{t = m+1}^{n} \frac{1}{\kappa_t} \frac{\p \kappa_t}{\p \gamma_t}
	\big( v(z_t)z_t(z_t - z_p) - 1 \big) w_{t,s}, \; s = 0,1,\ldots,l, \label{eq:tau_score} \\
	\dot{\bm \ell}_{\phi_s} &=& \sum_{t = m + 1}^{n} \frac{v(z_t)}{\sqrt{\kappa_t}} \frac{z_t}{Q_t} \frac{\p Q_t}{\p \eta_t} \left( \big(h(y_{t-s}) - \boldsymbol{x_{t-s}^\top \beta}\big) - \sum_{j=1}^{q} \theta_j \frac{\p Q_{t-j}}{\p \eta_{t-j}} \frac{\p \eta_{t-j}}{\p \phi_s} \right) , \; s = 1,2, \ldots, p, \label{eq:phi_score} \\
	\dot{\bm \ell}_{\theta_s} &=& \sum_{t = m + 1}^{n} \frac{v(z_t)}{\sqrt{\kappa_t}} \frac{z_t}{Q_t} \frac{\p Q_t}{\p \eta_t} \left( r_{t-s} + \theta_s \frac{\p r_{t-s}}{\p \theta_s} \right),   \; s = 1,2,\ldots,q, \label{eq:theta_score}
\end{eqnarray}
where
 $v(u) = -2g'{(u^2)}/g(u^2)$, with $g'(u) = \mathrm{d} g(u)/\mathrm{d}u$, are weights associated with the $g(\cdot)$ function.

Inference on $\boldsymbol{\zeta}$ corresponding to the QLS-ARMA model can be based on the asymptotic distribution of the CML estimator $\boldsymbol{\widehat{\zeta}}$. Considering the regularity conditions and sufficiently large $n$, the CML estimator $\boldsymbol{\widehat{\zeta}}$ converges in distribution to a multivariate normal distribution, that is,
\begin{equation}
	\sqrt{n}[\boldsymbol{\widehat{\zeta} - \zeta}] \build{\to}{\cal D}{} \mathrm{N}_{2+k+l+p+q}(\boldsymbol{0}, {\cal I}(\boldsymbol{\zeta})^{-1}) \nonumber
\end{equation}
with $n \to \infty$, where $\build{\to}{\cal D}{}$ denotes ``convergence in distribution'' and ${\cal I}(\boldsymbol{\zeta})$ denotes the expected Fisher information matrix. In practice, one may approximate the expected Fisher information matrix by its observed version obtained from the Hessian matrix ${\cal J}(\boldsymbol{\zeta})$; see more details in Appendix \ref{sec:apendxb}.

{\color{black}
\subsection{Asymptotic properties of least squares estimators}\label{asymleastsquares}
The QLS-ARMA model satisfies the following general regression model structure:
\begin{align*}
Y_t=f_t(\boldsymbol{\zeta} )+\varepsilon_t, \quad t=m+1,\dots,n,
\end{align*}
where $f_t(\boldsymbol{\zeta} )
=h^{-1}({\bm x}_t^{\top}\bm\beta  + \varrho_t)=Q_t$ or $f_t(\boldsymbol{\zeta} )
=d^{-1}( \boldsymbol{w_t^\top \tau})=\kappa_t$,
with additional dynamic ARMA component $\varrho_t $ given by \eqref{eq:varrho}. Here, $f_t$ is a random $\mathcal{A}_ {t-1}$-measurable function of an unknown parameter vector
$\boldsymbol{\zeta} = \boldsymbol{(\beta^\top,\tau^\top,\phi^\top,\theta^\top)^\top}$.

The statistic $\widehat{\boldsymbol{\zeta} }_t$  that minimizes the quantity
\begin{align*}
S_t(\boldsymbol{\zeta} )
=
\sum_{t=1}^{n}[Y_t-f_t(\boldsymbol{\zeta} )]^2
\end{align*}
is known as the least squares estimator of $\boldsymbol{\zeta} $.
Assuming $f_t$ to be differentiable, $\widehat{\boldsymbol{\zeta}}_t$ is computed by iterative solution of the following equation
\begin{align*}
\nabla S_t(\boldsymbol{\zeta} )
=
-2\sum_{t=1}^{n} [Y_t-f_t(\boldsymbol{\zeta} )]\nabla f_t(\boldsymbol{\zeta} )=0.
\end{align*}

By assuming $f_t$ to be sufficiently smooth, as in  \citet[Theorems 1 and 2]{lai:30} in which the random disturbances $\varepsilon_t$ form a martingale difference sequence with respect to $\sigma$-fields $\{\mathcal{A}_ {t}\}$, by Theorem 1  of \cite{lai:30}, we have
\begin{align*}
\sup_t\mathbb{E}(\varepsilon_t^2\vert \mathcal{A}_ {t-1})<\infty , \quad
\mathbb{E}(\varepsilon^2_t\vert \mathcal{A}_ {t-1})\stackrel{\rm P}{\longrightarrow} \sigma^2,
\end{align*}
that is, the least squares estimator $\widehat{\boldsymbol{\zeta} }_t$ converges  almost surely to $\boldsymbol{\zeta} $,
whenever $\boldsymbol{\zeta} $ belongs to a compact subset $\Theta$ of a specific Euclidean space.
Moreover, by Theorem 2 of \cite{lai:30}, we can say that
$$
\left\{\sum_{t=1}^{n} [\nabla f_t(\boldsymbol{\zeta} )][\nabla f_t(\boldsymbol{\zeta} )]^{\top}\right\}^{1/2}(\widehat{\boldsymbol{\zeta} }_t-\boldsymbol{\zeta} )
$$
converges in distribution, as $n\to \infty$, to a multivariate normal distribution with zero-mean vector and covariance matrix $\sigma^2 \bm I$, whenever $\boldsymbol{\zeta} $ belongs to the interior of $\Theta$.

}

\subsection{Prediction}

After computing the CML estimates of the parameters of the QLS-ARMAX($p,q$) model, we next discuss the predictions of the response from time $t$ to time $t+h$, which we denote by $\widehat{y}_{t+h}$, where
\begin{equation}
	\widehat{y}_{t+h} =
	\begin{cases}
		\widehat{y}_t(h), & \text{if} \; h > 0; \\
		y_{t+h}, & \text{if}\; h \leq 0;
	\end{cases} \nonumber
	\qquad \text{with} \quad
	\widehat{r}_{t+h} =
	\begin{cases}
		0, & \text{if}\; h > 0; \\
		\widehat{r}_{t+h}, &  \text{if}\; h \leq 0.
	\end{cases}
\end{equation}
From the CML estimates of the model parameters, we can easily obtain the estimates of the quantile $Q_t$, $t = m+1,\ldots,n$, as
\begin{equation}\label{eq:Qest}
	\widehat{Q}_t =  h^{-1} \left(  \boldsymbol{x_t^\top \widehat{\beta}} + \sum_{i=1}^{p} \widehat{\phi}_i \left[ h(y_{t-i}) - \boldsymbol{x_{t-i}^\top \widehat{\beta}} \right] + \sum_{j = 1}^{q} \widehat{\theta}_j \widehat{r}_{t-j} \right).
\end{equation}
From \eqref{eq:Qest}, we get the estimates of $ \widehat{r}_t = h(y_t) - \widehat{\eta}_t$, for $t = m+1,\ldots, n$, that justifies the fact that $\widehat{r}_{t+h} = 0$ if $h>0$. From the estimated $\widehat{Q}_t$ and $\widehat{r}_t$, we can obtain the prediction of $y_{n+1}$ as
\begin{equation}
	\widehat{y}_{n+1} =  h^{-1} \left(  \boldsymbol{x_{n+1}^\top \widehat{\beta}} + \sum_{i=1}^{p} \widehat{\phi}_i \left[ h(y_{n+1-i}) - \boldsymbol{x_{n+1-i}^\top \widehat{\beta}} \right] + \sum_{j = 1}^{q} \widehat{\theta}_j \widehat{r}_{n+1-j} \right). \nonumber
\end{equation}
Once the prediction for the time $n+1$ is obtained, it follows, in an analogous way, that for $n+2$ we have
\begin{equation}
	\widehat{y}_{n+2} =  h^{-1} \left(  \boldsymbol{x_{n+2}^\top \widehat{\beta}} + \sum_{i=1}^{p} \widehat{\phi}_i \left[ h(\widehat{y}_{n+2-i}) - \boldsymbol{x_{n+2-i}^\top \widehat{\beta}} \right] + \sum_{j = 1}^{q} \widehat{\theta}_j \widehat{r}_{n+2 -j} \right), \nonumber
\end{equation}
and so on for times greater than $n+2$.

\subsection{Residual analysis}

Residual analysis is an important tool to assess goodness of fit of a model. In this paper, we consider two types of residuals. The first type of residual that we consider is the randomized quantile (RQ) residuals,
which are often used for generalized additive models; see \cite{dunn:1996}. These residuals are defined as
\begin{equation}
	r_t^{RQ} = \Phi^{-1}(\widehat{S}(y_t|\mathcal{B}_{t-1})), \quad t = m + 1, \ldots, n,
 \nonumber
\end{equation}
where $\Phi^{-1}(\cdot)$ is the inverse of the cumulative distribution function of the normal standard distribution and $\widehat{S}$ is the estimate of the survival function. If a model is correctly specified, the randomized quantile residuals are normally distributed; see \cite{dunn:1996}. The second type of residual that we consider is the generalized Cox-Snell (GCS) residuals, which are defined as
\begin{equation}
	r_t^{GCS} = -\log(\widehat{S}(y_t|\mathcal{B}_{t-1})), \quad t = m+1,\ldots n. \nonumber
\end{equation}
If a model is correctly specified, the GCS residuals have an exponential distribution with unit mean; see \cite{saulo:21}.

\section{Monte Carlo simulation}\label{sec:3}

In this section, we carry out two Monte Carlo simulation studies. All results are based on $R = 5\,000$ Monte Carlo runs. For both studies, we consider the QLS-ARMAX$(1,1)$ model. Hence, the model is given by
\begin{eqnarray*}
	\eta_t &=& \log(Q_t) = \beta_0 + \beta_1 x_t + \phi [\log(y_{t-1}) - (\beta_0 + \beta_1 x_{t-1})] + \theta r_{t-1}, \quad t = 2,\ldots, n, \nonumber \\
	\gamma_t &=& \log(\kappa_t) = \tau_0 + \tau_1 w_t, \quad t = 2,\ldots, n. \nonumber
\end{eqnarray*}
We consider the following true choices of the parameters: $\beta_0 = 1$, $\beta_1 = 0.7$, $\tau_0 = 0.5$, $\tau_1 = 1.5$, $\phi = 0.6$ and $\theta = 0.3$. We also consider different sample sizes $n$ and different values for the quantile $q$:
$n \in \left\lbrace 50, 100, 200 \right\rbrace$ and $q \in \left\lbrace 0.25, 0.5, 0.75 \right\rbrace$. The covariates are generated from a uniform distribution in the interval (0,1).
Furthermore, we consider the following special cases of the log-symmetric distributions:
log-normal (Log-NO), log-student-$t$ (Log-$t$, $\vartheta = 4$), log-power-exponential (Log-PE, $\vartheta = 0.5$), log-contamined-normal (Log-NC, $\vartheta_{1} = 0.3, \vartheta_{2} = 0.5$), log-hyperbolic (Log-HP, $\vartheta = 1$), log-slash (log-SL, $\vartheta = 2$), log-sinh-normal (log-SN, $\vartheta = 0.1$), and log-sinh-$t$ (log-ST, $\vartheta_1 = 0.1$, $\vartheta_2 = 4$).

\begin{figure}[H]
	\centering
	\subfigure[Log-NO, $q = 0.25$]{\includegraphics[scale = 0.65]{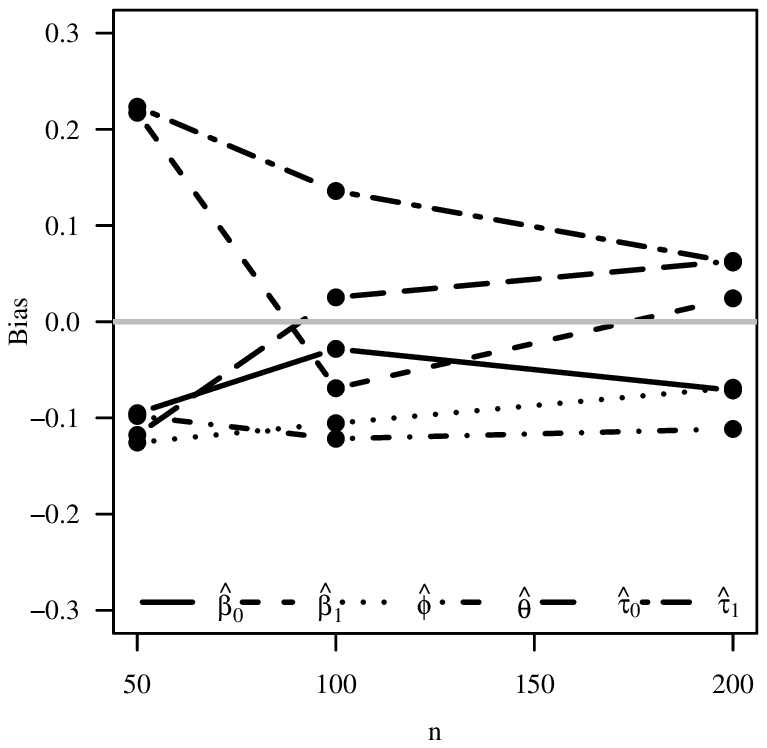}}
	\subfigure[Log-NO, $q = 0.5$]{\includegraphics[scale = 0.65]{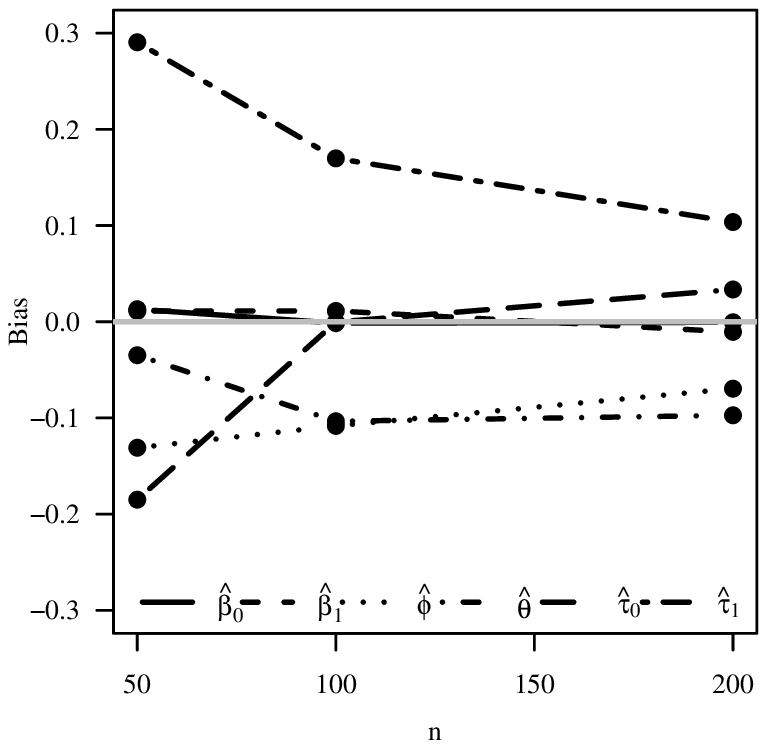}}
	\subfigure[Log-NO, $q = 0.75$]{\includegraphics[scale = 0.65]{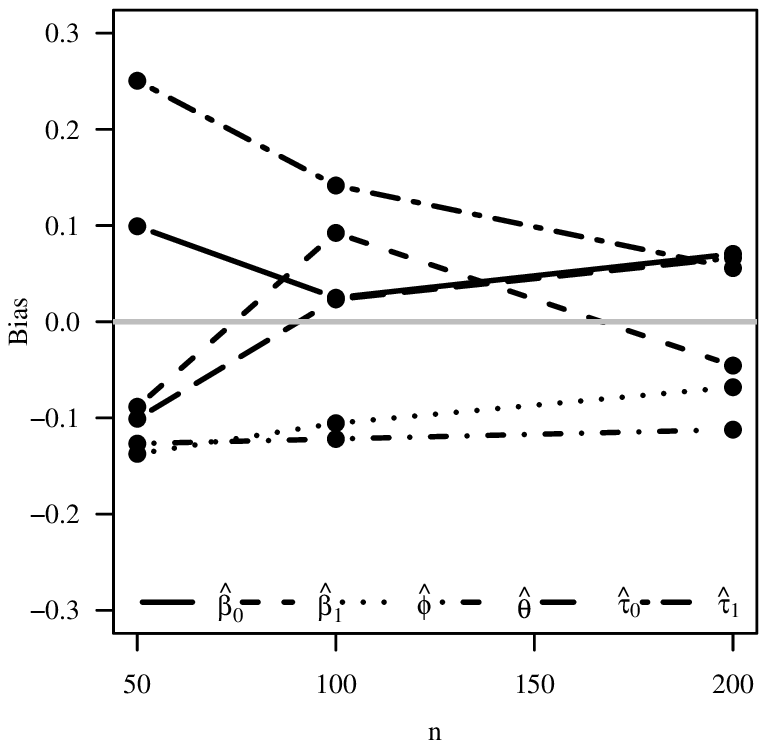}}
	\subfigure[Log-$t$, $q = 0.25$]{\includegraphics[scale = 0.65]{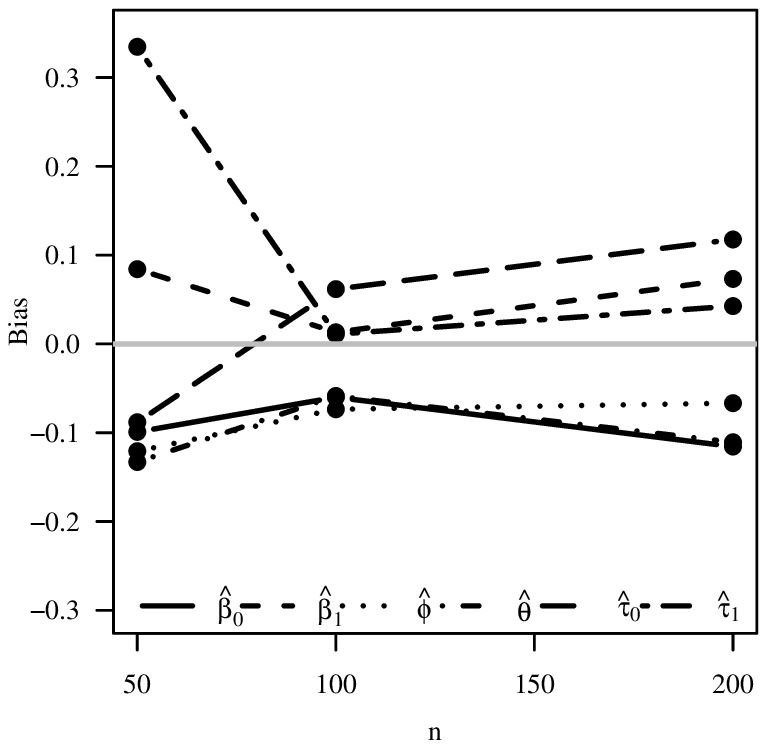}}
	\subfigure[Log-$t$, $q = 0.5$]{\includegraphics[scale = 0.65]{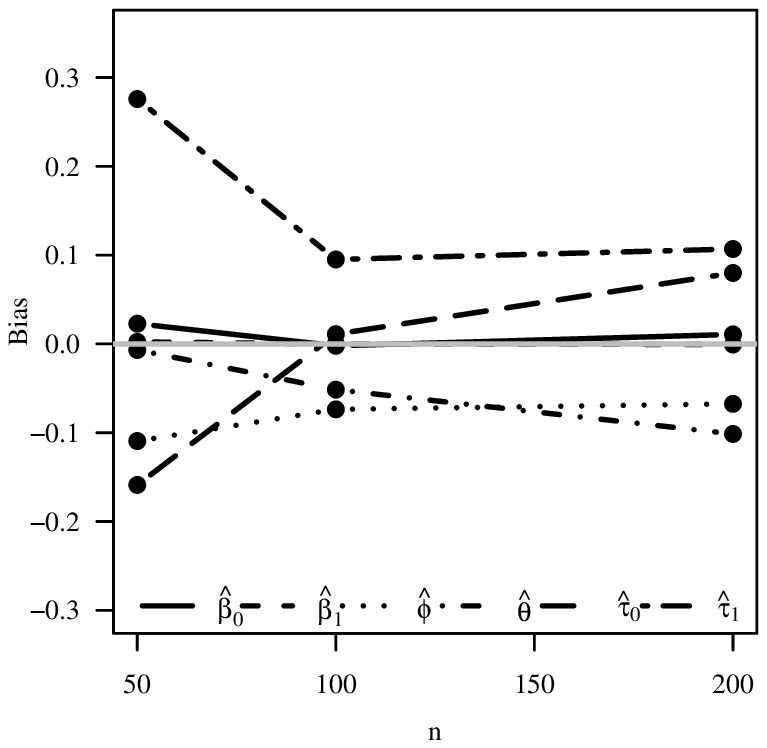}}
	\subfigure[Log-$t$, $q = 0.75$]{\includegraphics[scale = 0.65]{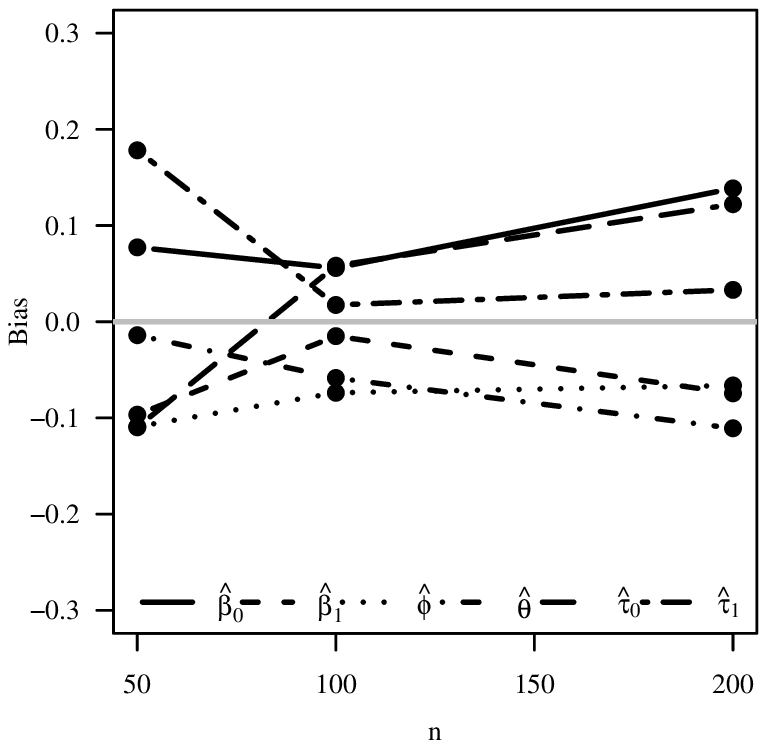}}
	\subfigure[Log-PE, $q = 0.25$]{\includegraphics[scale = 0.65]{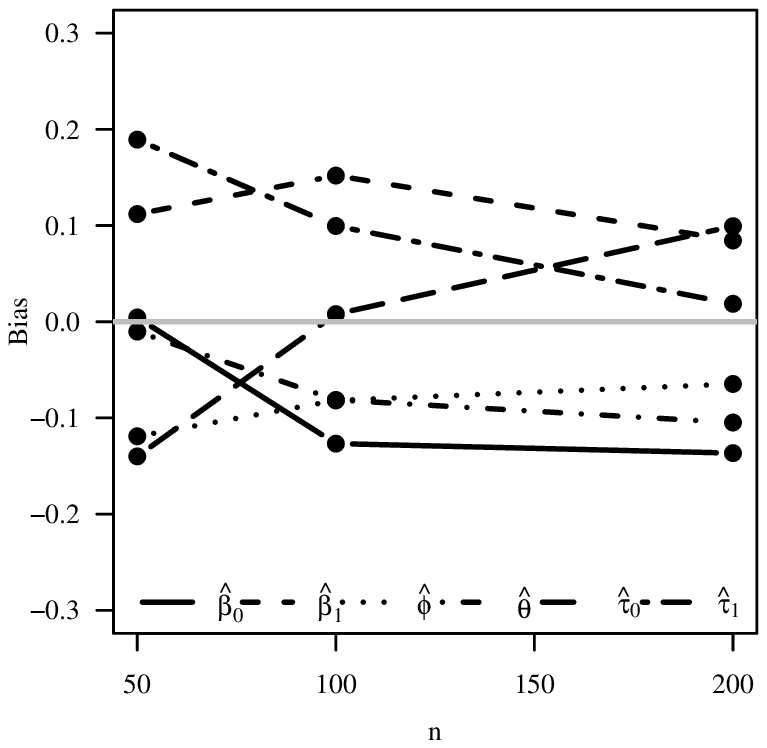}}
	\subfigure[Log-PE, $q = 0.5$]{\includegraphics[scale = 0.65]{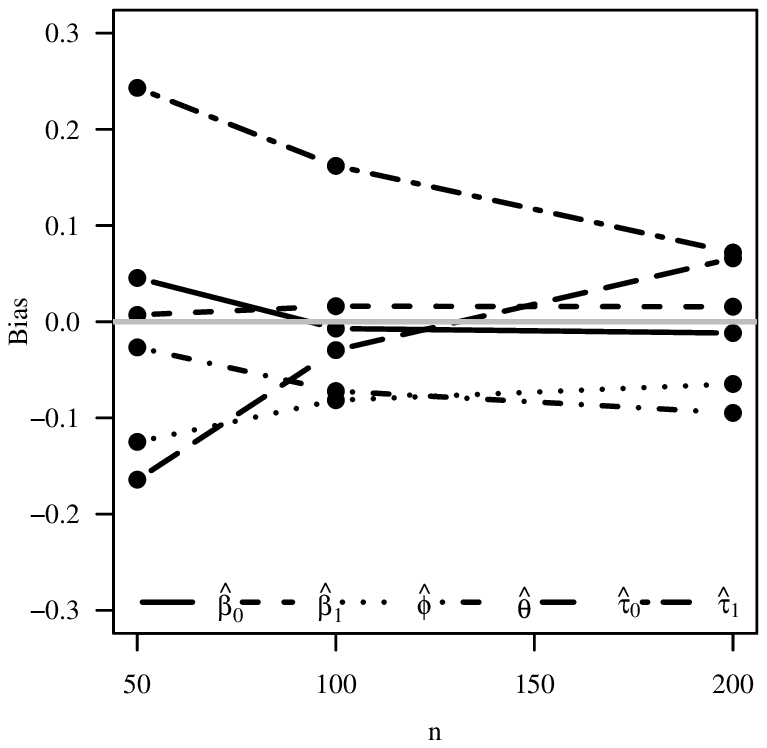}}
	\subfigure[Log-PE, $q = 0.75$]{\includegraphics[scale = 0.65]{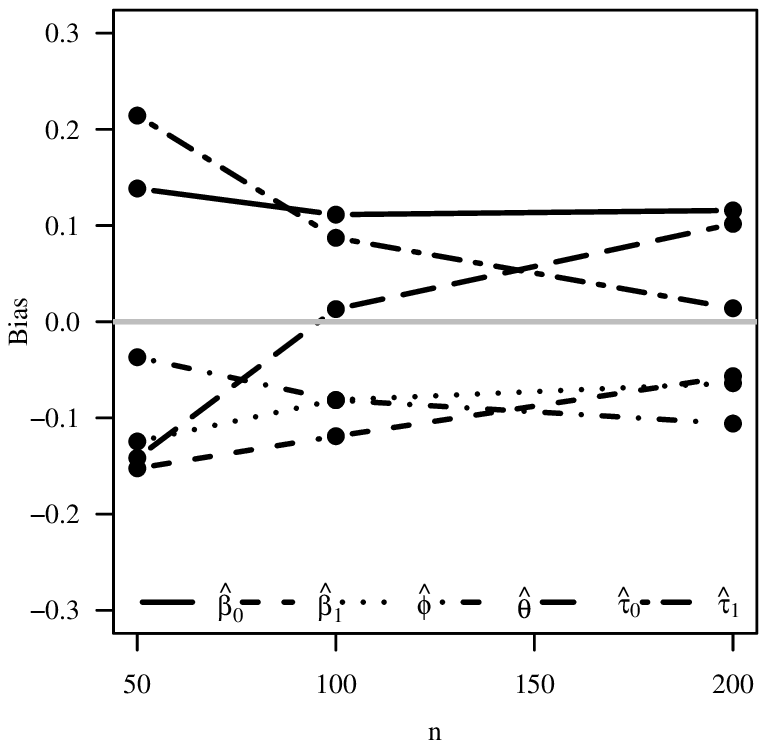}}
	\caption{Biases of the CML estimators for the model QLS-ARMAX(1,1).}
	\label{fig:01}
\end{figure}
\begin{figure}[H]
	\centering
	\subfigure[Log-HP, $q = 0.25$]{\includegraphics[scale = 0.65]{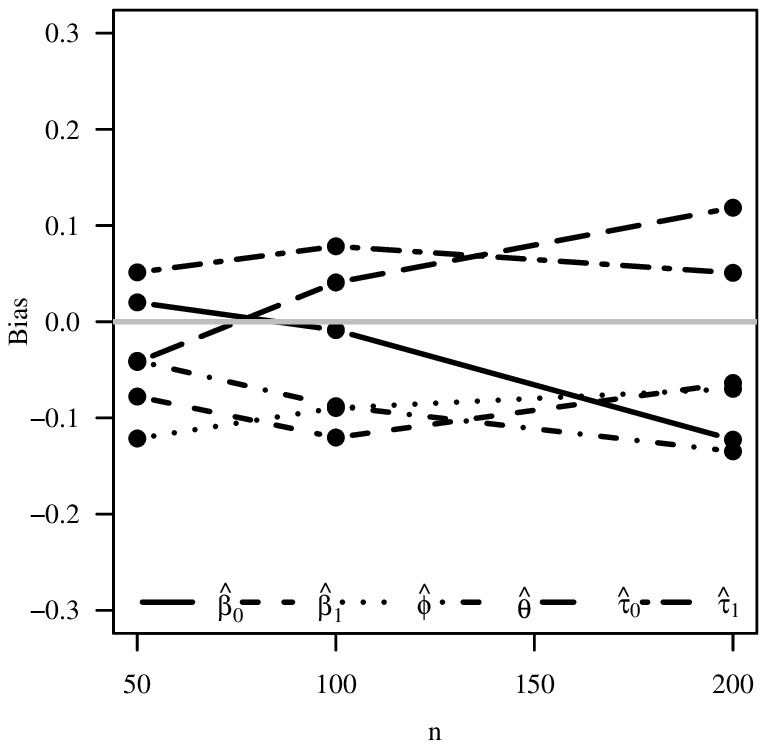}}
	\subfigure[Log-HP, $q = 0.5$]{\includegraphics[scale = 0.65]{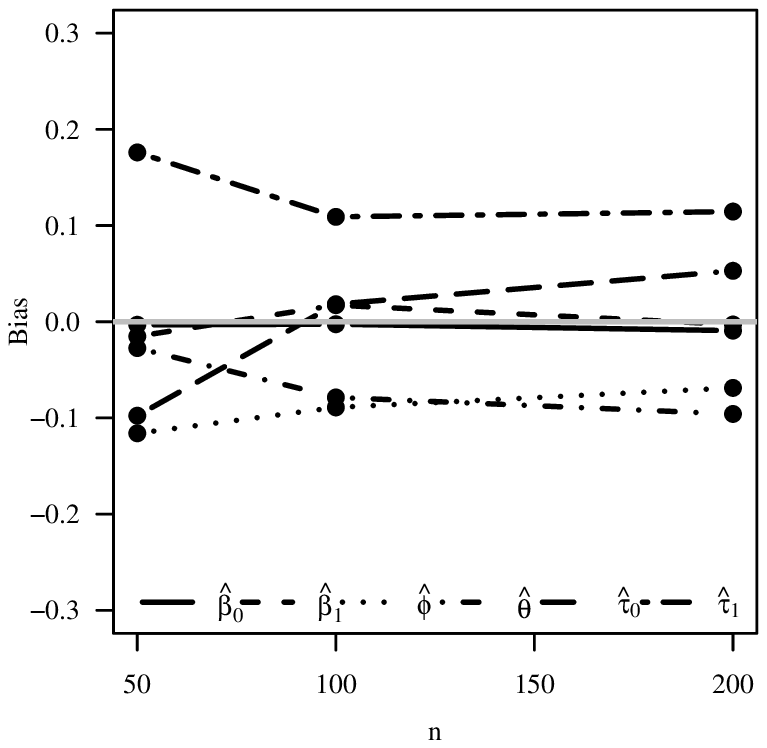}}
	\subfigure[Log-HP, $q = 0.75$]{\includegraphics[scale = 0.65]{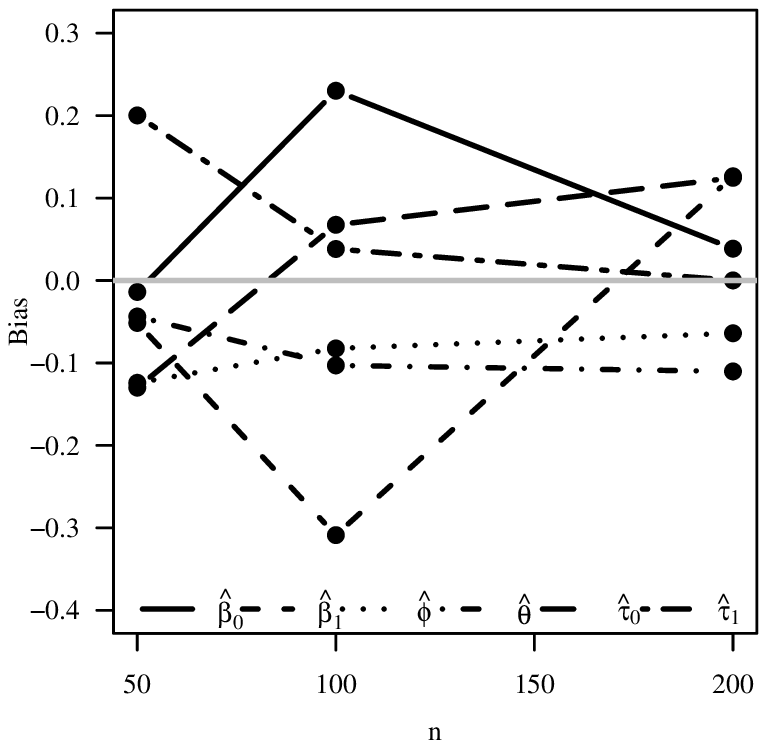}}
	\subfigure[Log-CN, $q = 0.25$]{\includegraphics[scale = 0.65]{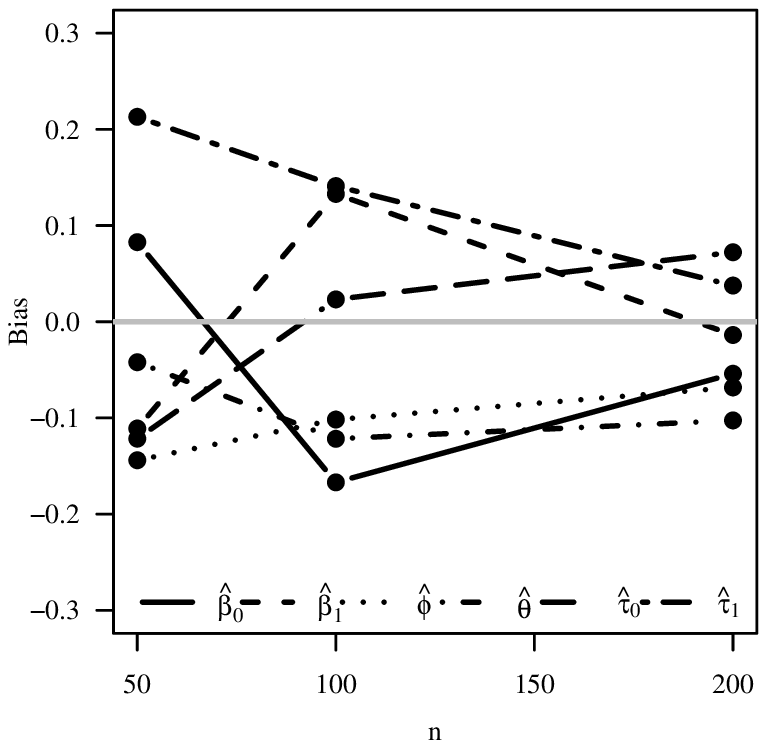}}
	\subfigure[Log-CN, $q = 0.5$]{\includegraphics[scale = 0.65]{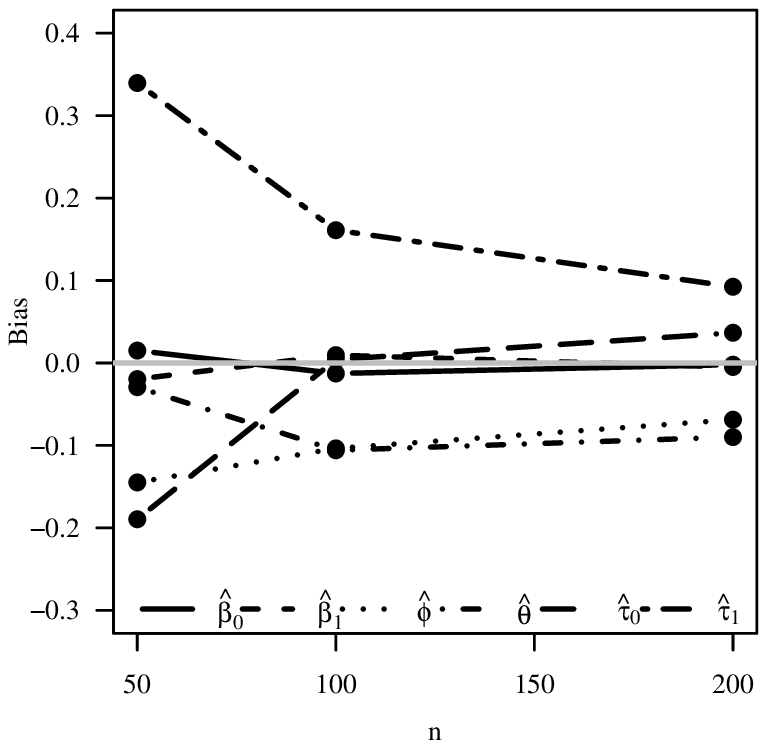}}
	\subfigure[Log-CN, $q = 0.75$]{\includegraphics[scale = 0.65]{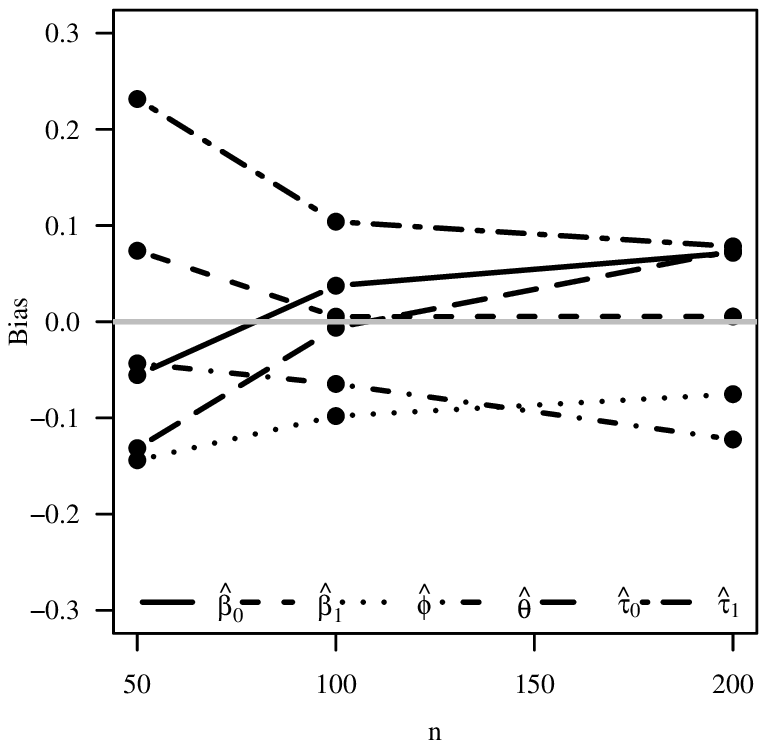}}
	\subfigure[Log-SL, $q = 0.25$]{\includegraphics[scale = 0.65]{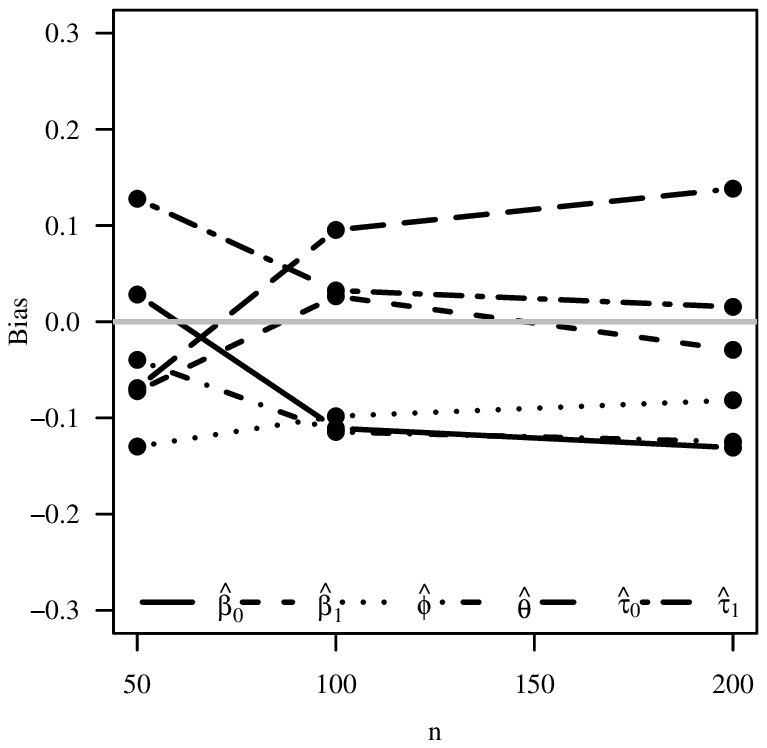}}
	\subfigure[Log-SL, $q = 0.5$]{\includegraphics[scale = 0.65]{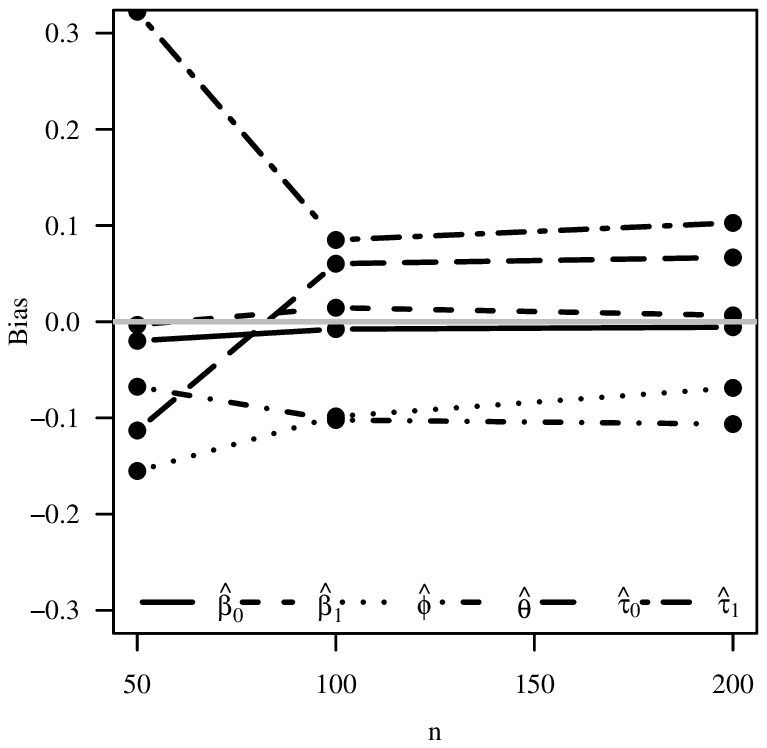}}
	\subfigure[Log-SL, $q = 0.75$]{\includegraphics[scale = 0.65]{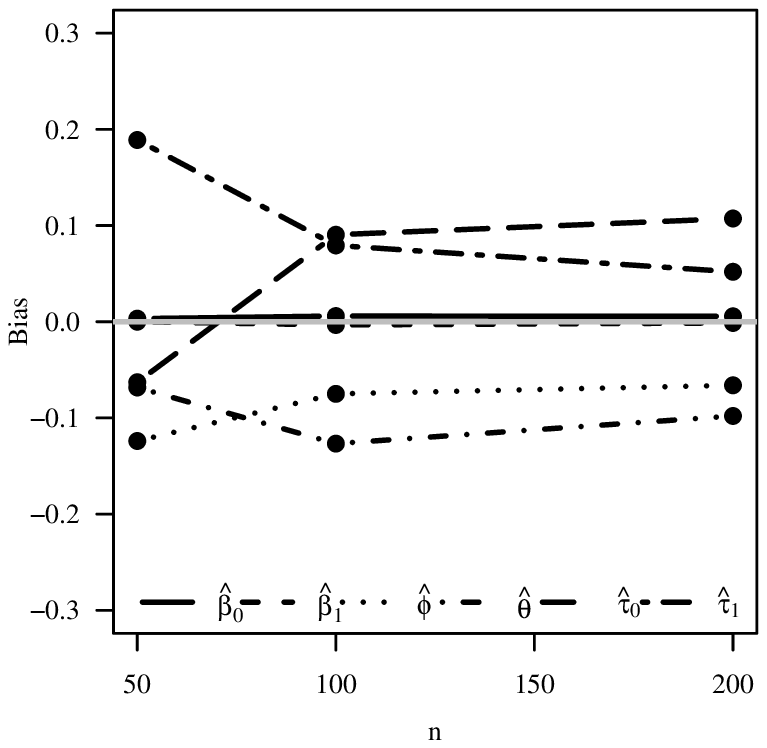}}
	\caption{Biases of the CML estimators for the model QLS-ARMAX(1,1). (continuation)}
	\label{fig:02}
\end{figure}
\begin{figure}[H]
	\centering
	\subfigure[Log-SN, $q = 0.25$]{\includegraphics[scale = 0.65]{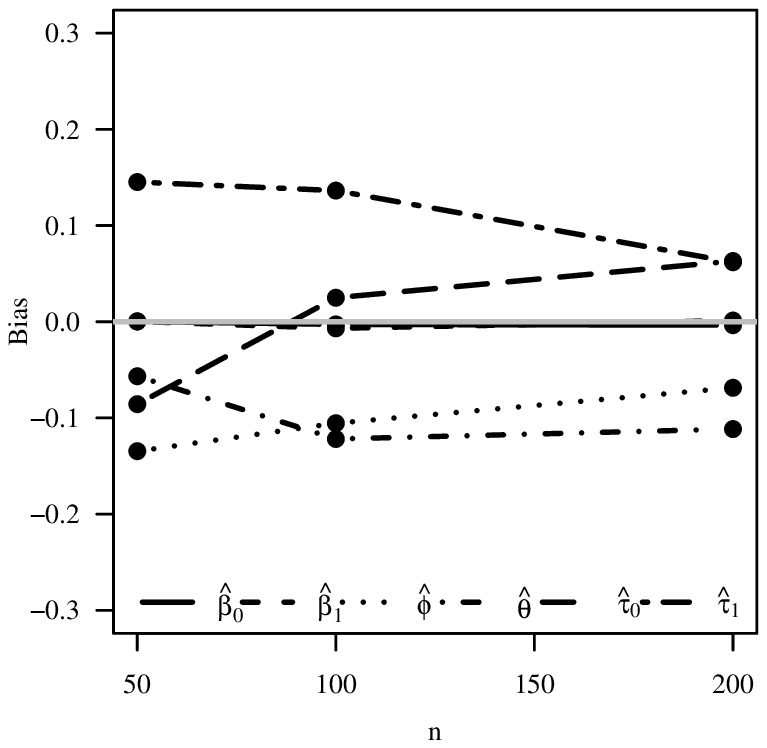}}
	\subfigure[Log-SN, $q = 0.5$]{\includegraphics[scale = 0.65]{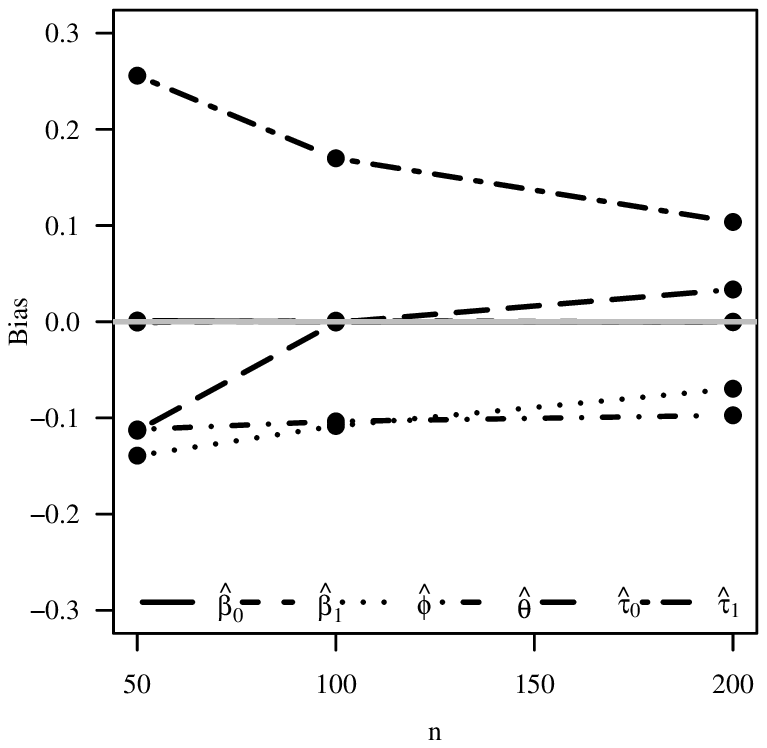}}
	\subfigure[Log-SN, $q = 0.75$]{\includegraphics[scale = 0.65]{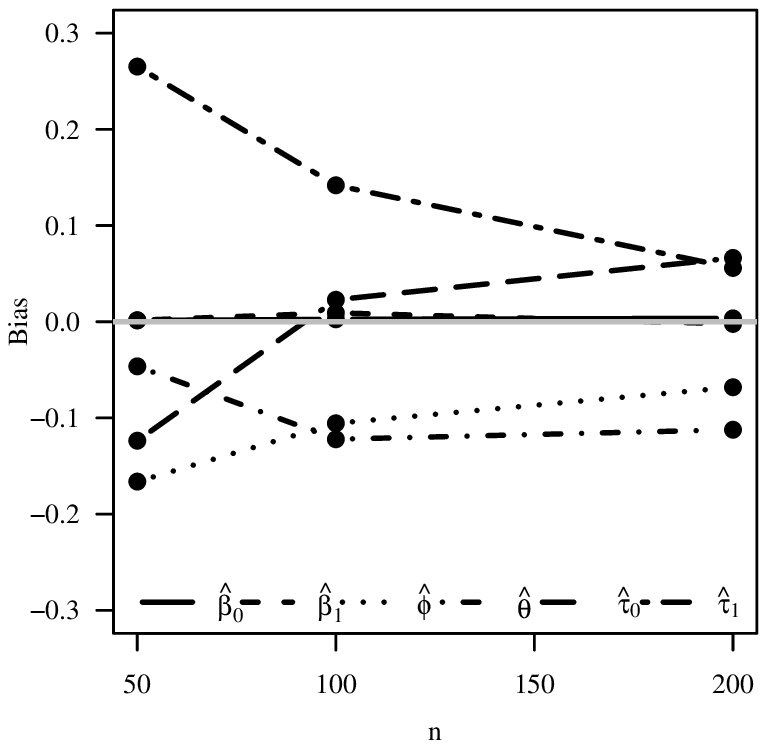}}
	\subfigure[Log-ST, $q = 0.25$]{\includegraphics[scale = 0.65]{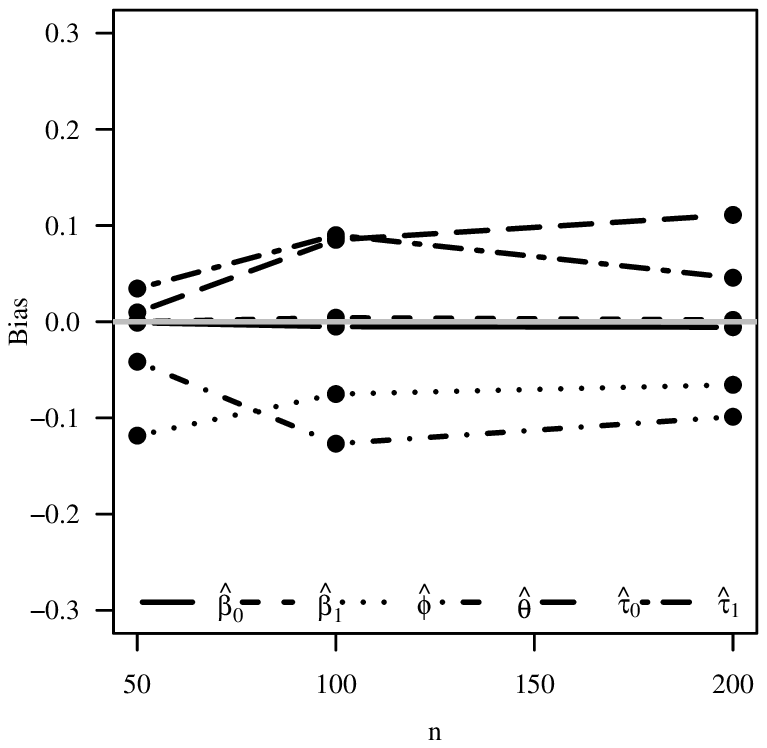}}
	\subfigure[Log-ST, $q = 0.5$]{\includegraphics[scale = 0.65]{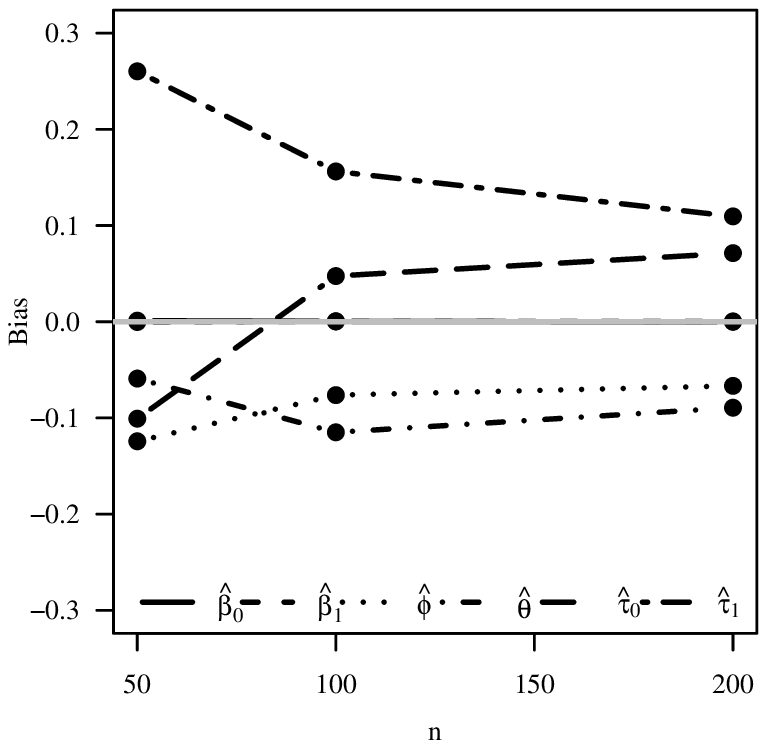}}
	\subfigure[Log-ST, $q = 0.75$]{\includegraphics[scale = 0.65]{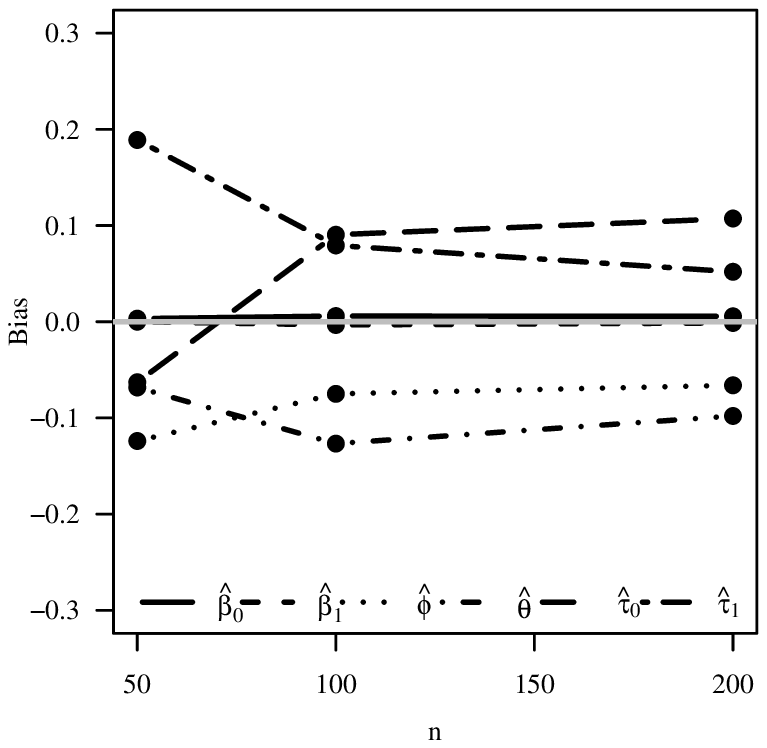}}
	\caption{Biases of the CML estimators for the model QLS-ARMAX(1,1). (continuation)}
	\label{fig:03}
\end{figure}
\begin{figure}[H]
	\centering
	\subfigure[Log-NO, $q = 0.25$]{\includegraphics[scale = 0.65]{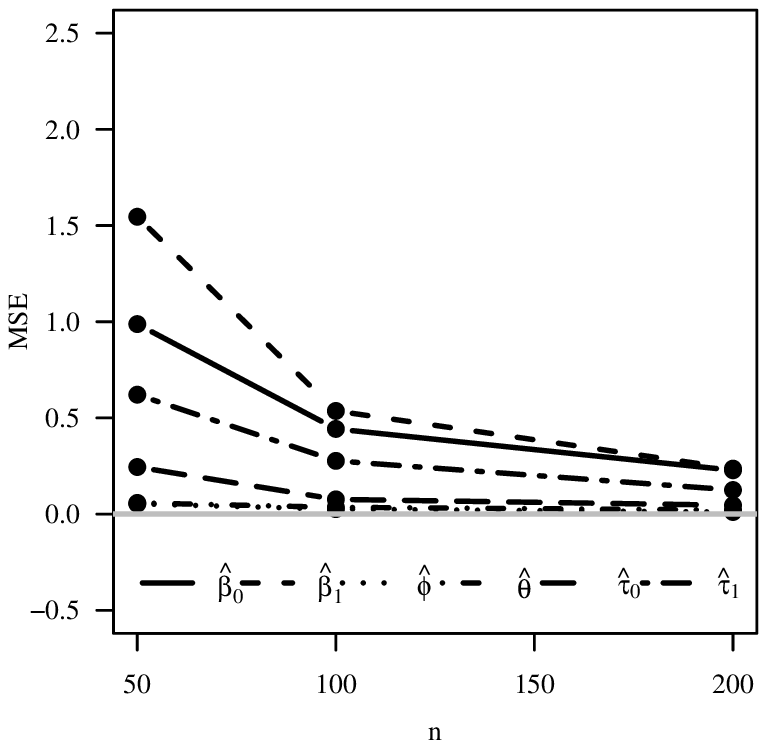}}
	\subfigure[Log-NO, $q = 0.5$]{\includegraphics[scale = 0.65]{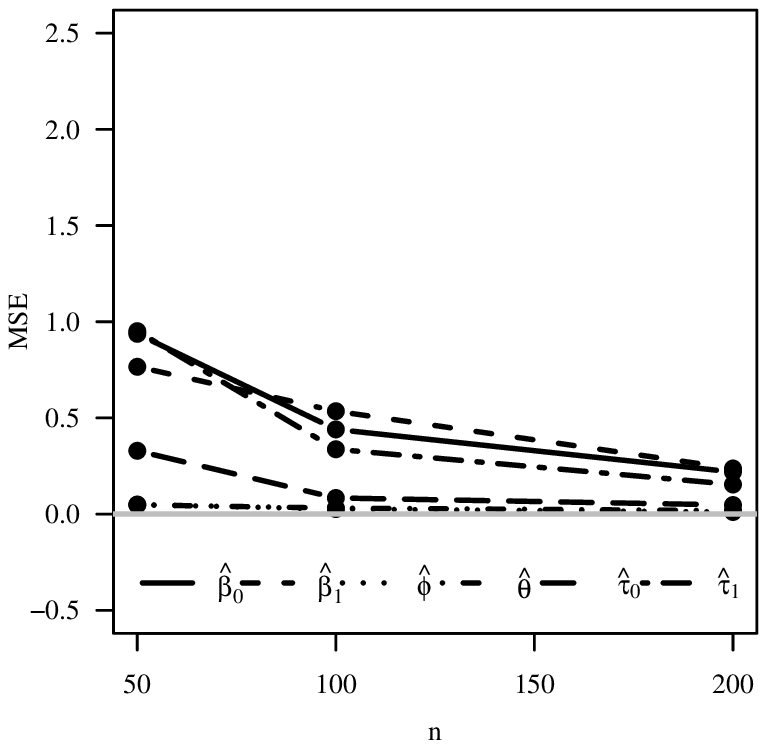}}
	\subfigure[Log-NO, $q = 0.75$]{\includegraphics[scale = 0.65]{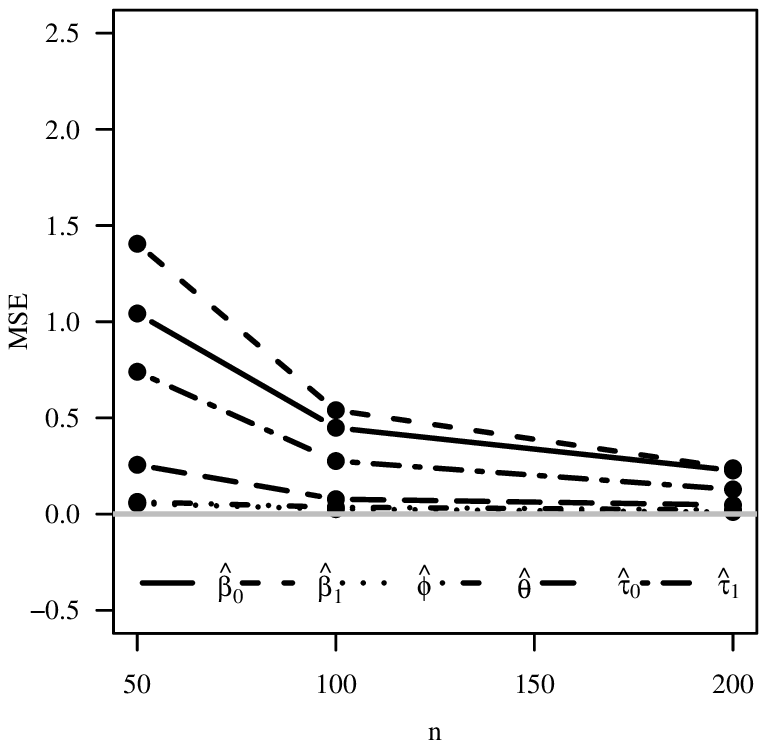}}
	\subfigure[Log-$t$, $q = 0.25$]{\includegraphics[scale = 0.65]{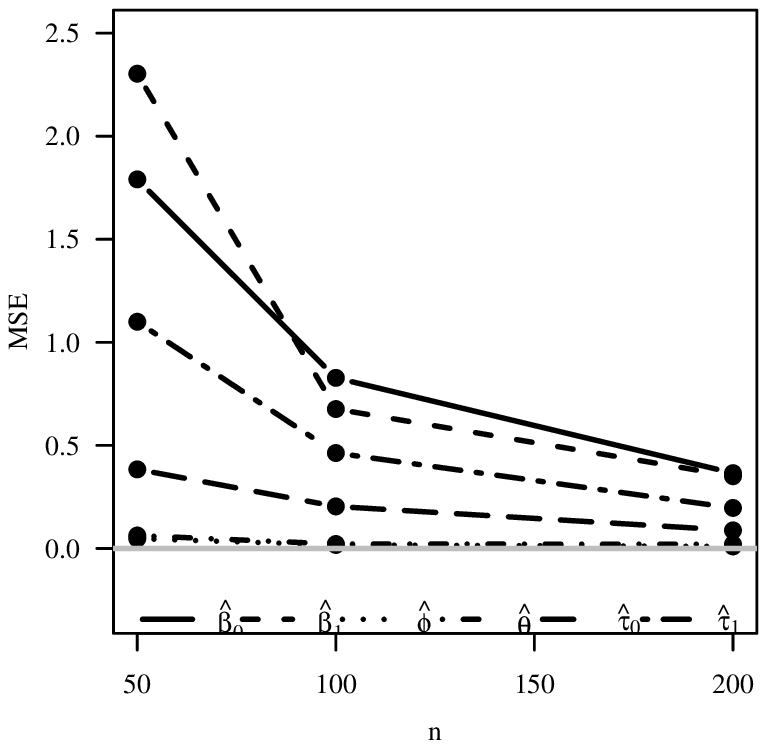}}
	\subfigure[Log-$t$, $q = 0.5$]{\includegraphics[scale = 0.65]{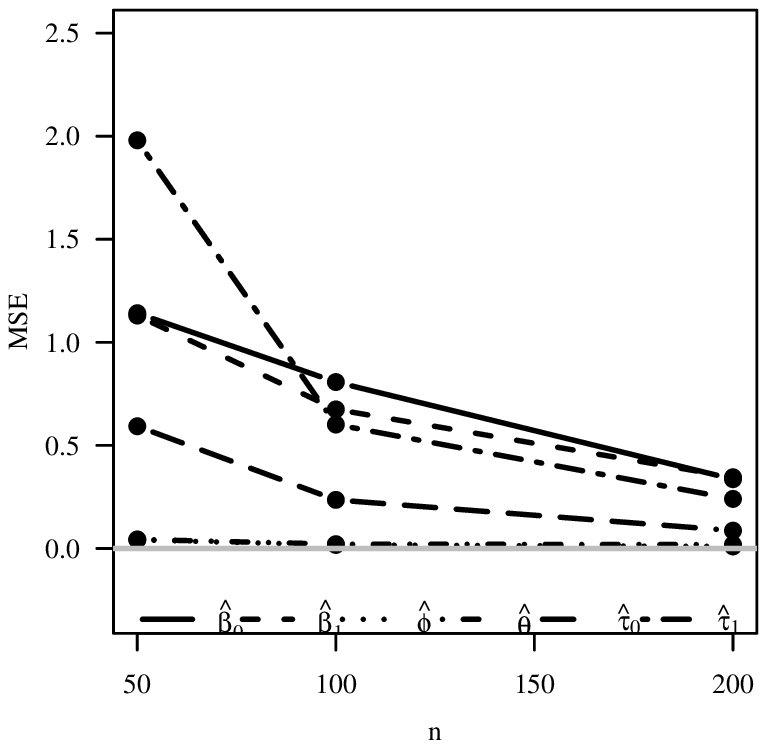}}
	\subfigure[Log-$t$, $q = 0.75$]{\includegraphics[scale = 0.65]{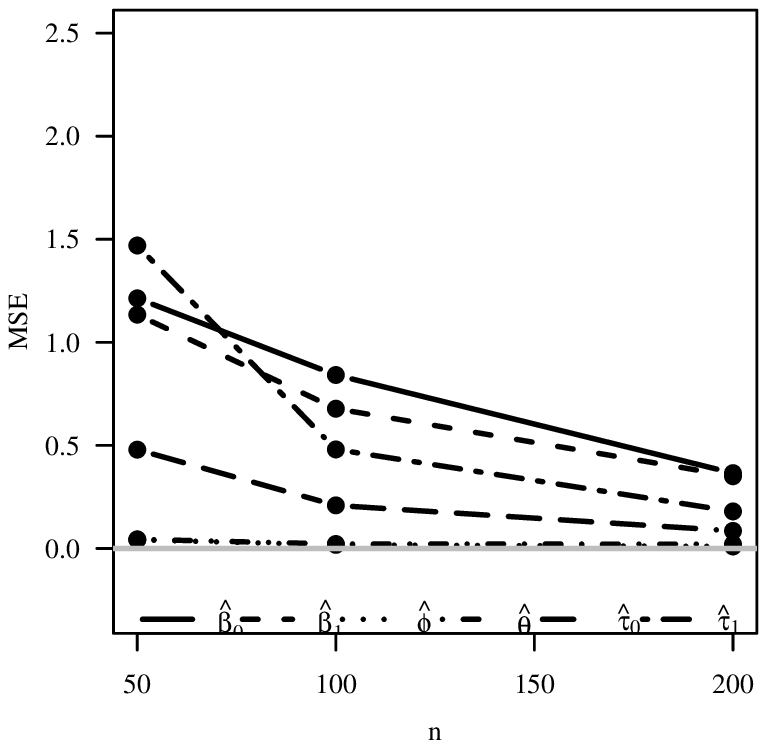}}
	\subfigure[Log-PE, $q = 0.25$]{\includegraphics[scale = 0.65]{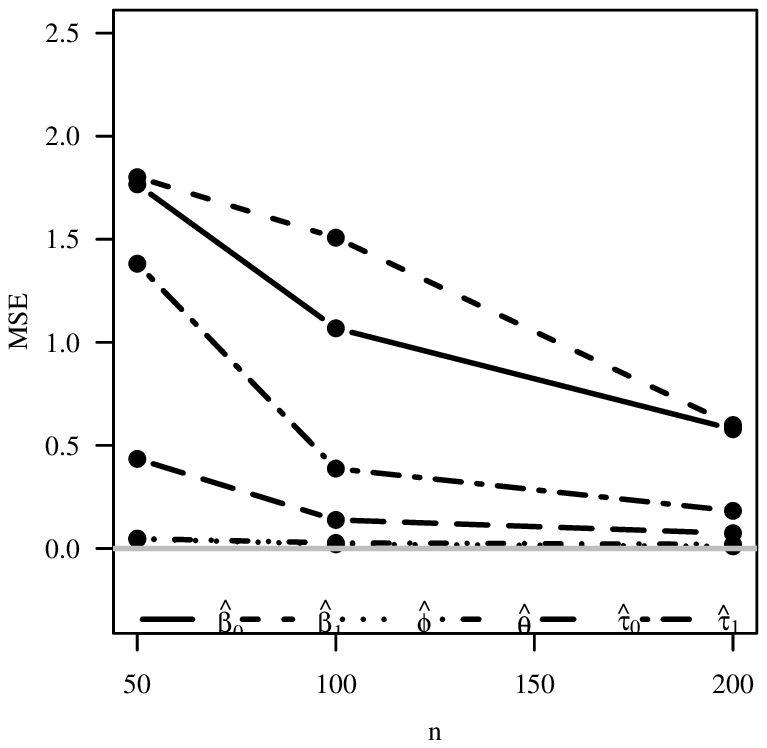}}
	\subfigure[Log-PE, $q = 0.5$]{\includegraphics[scale = 0.65]{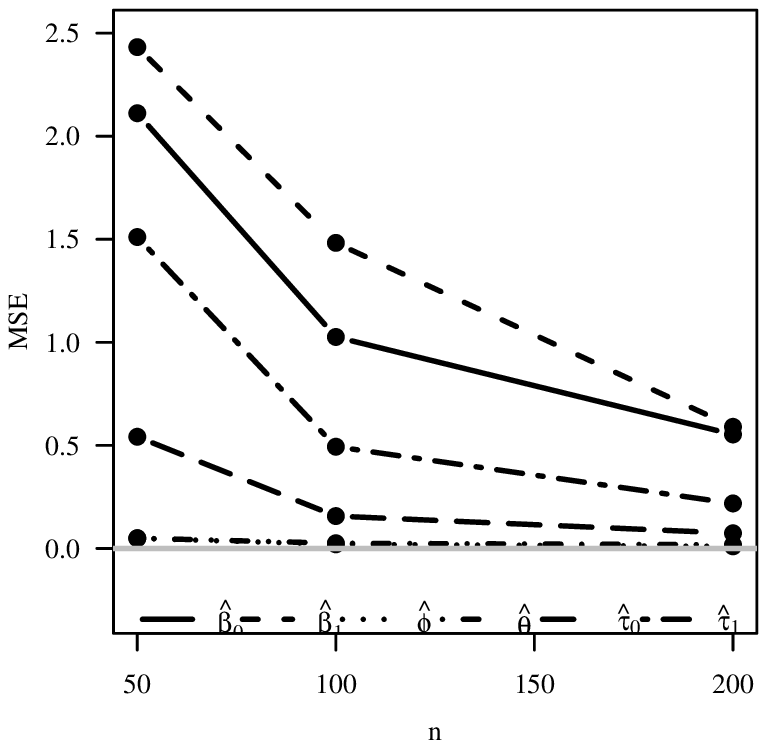}}
	\subfigure[Log-PE, $q = 0.75$]{\includegraphics[scale = 0.65]{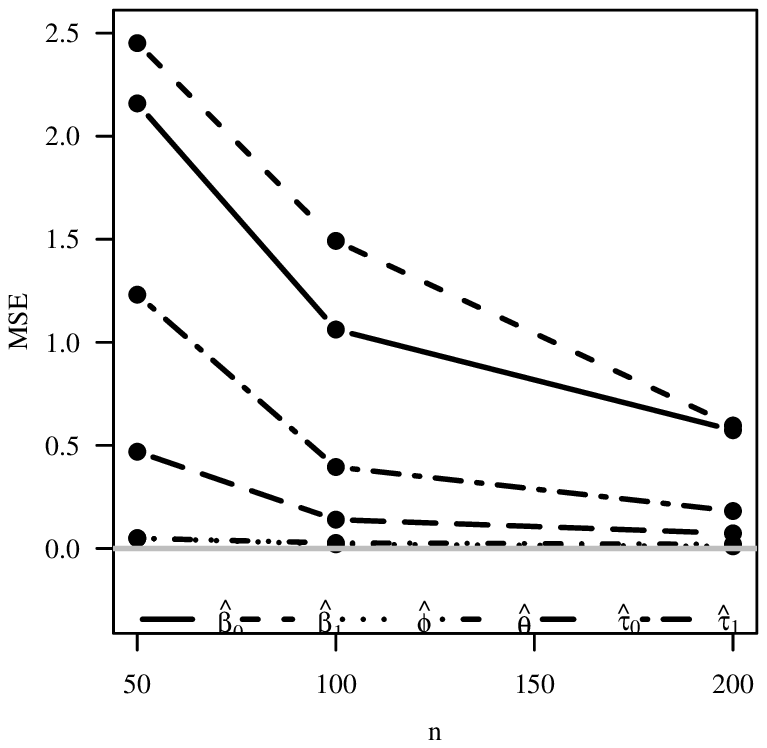}}
	\caption{MSEs of the CML estimators for the model QLS-ARMAX(1,1).}
	\label{fig:04}
\end{figure}
\begin{figure}[H]
	\centering
	\subfigure[Log-HP, $q = 0.25$]{\includegraphics[scale = 0.65]{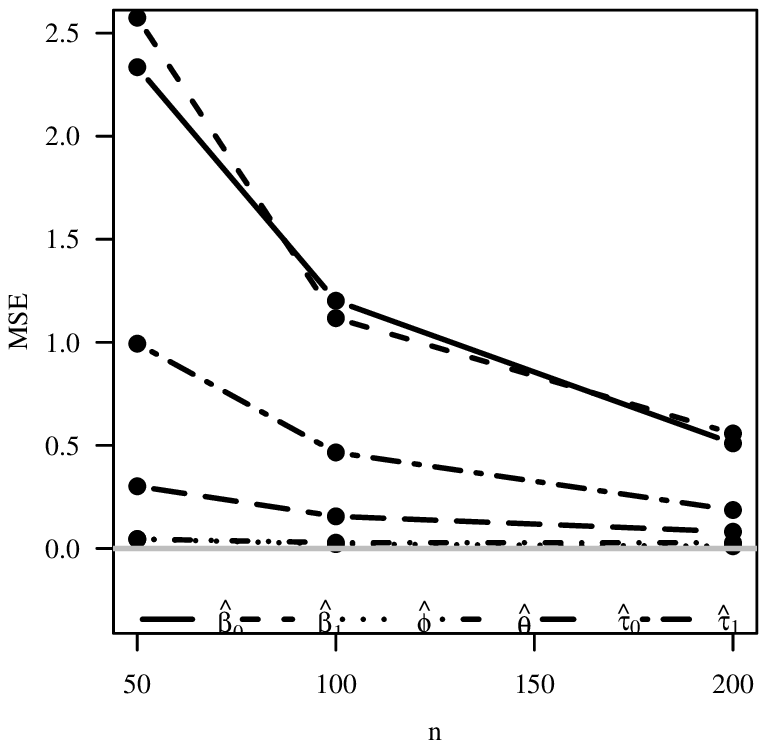}}
	\subfigure[Log-HP, $q = 0.5$]{\includegraphics[scale = 0.65]{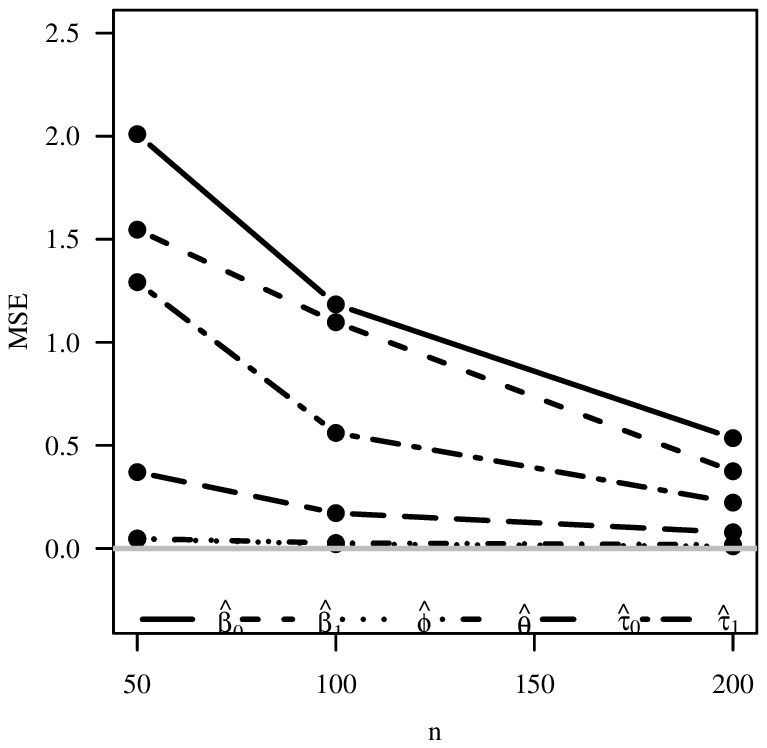}}
	\subfigure[Log-HP, $q = 0.75$]{\includegraphics[scale = 0.65]{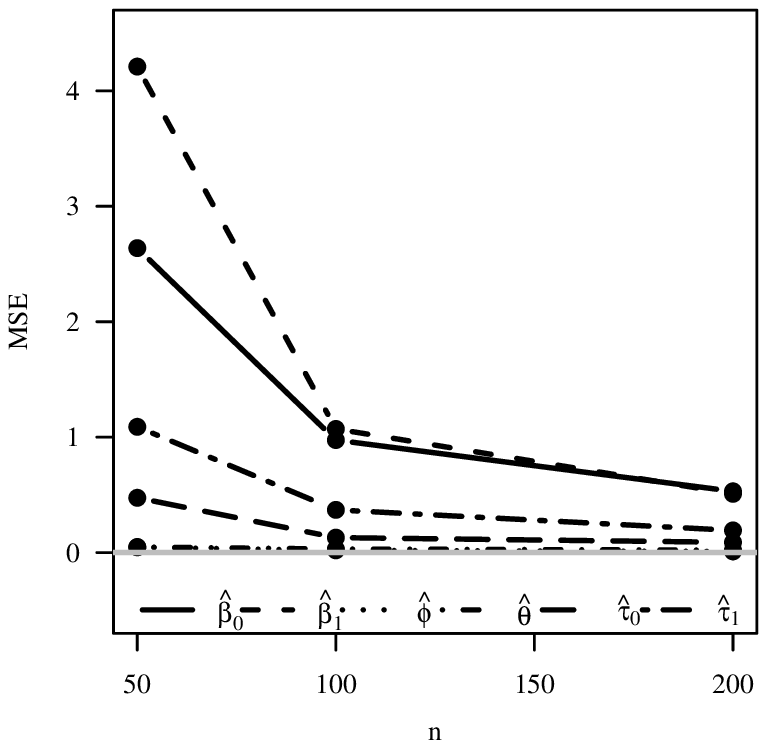}}
	\subfigure[Log-CN, $q = 0.25$]{\includegraphics[scale = 0.65]{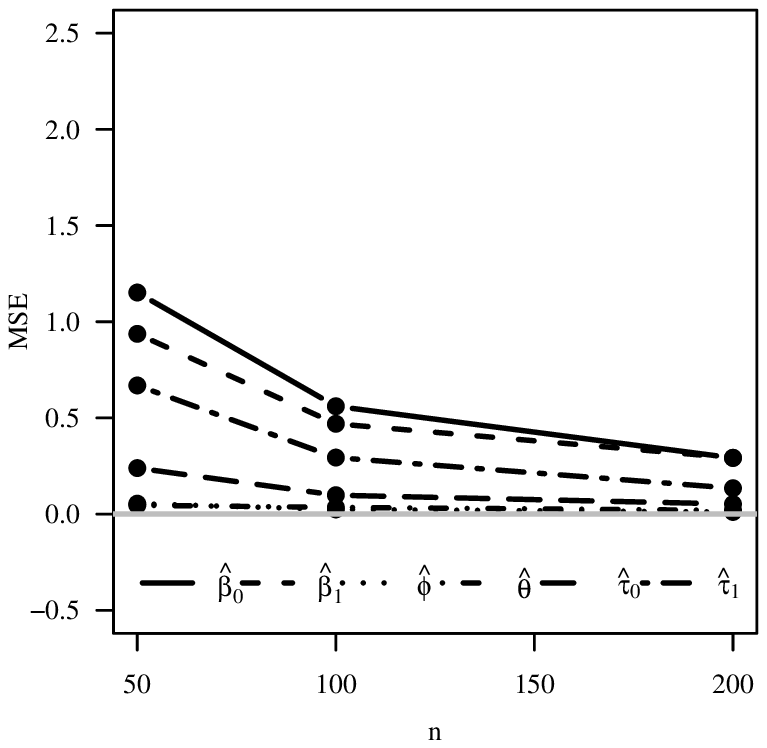}}
	\subfigure[Log-CN, $q = 0.5$]{\includegraphics[scale = 0.65]{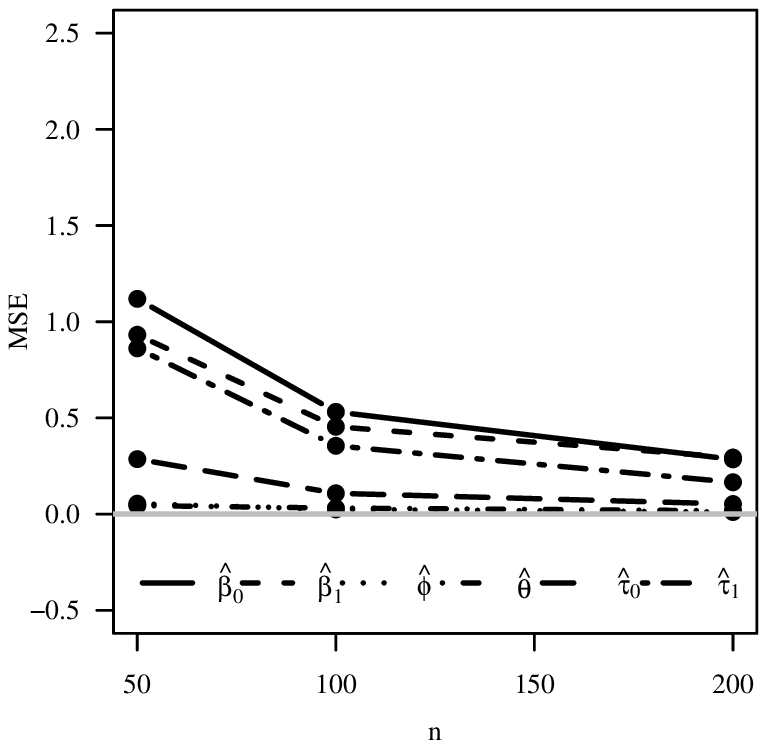}}
	\subfigure[Log-CN, $q = 0.75$]{\includegraphics[scale = 0.65]{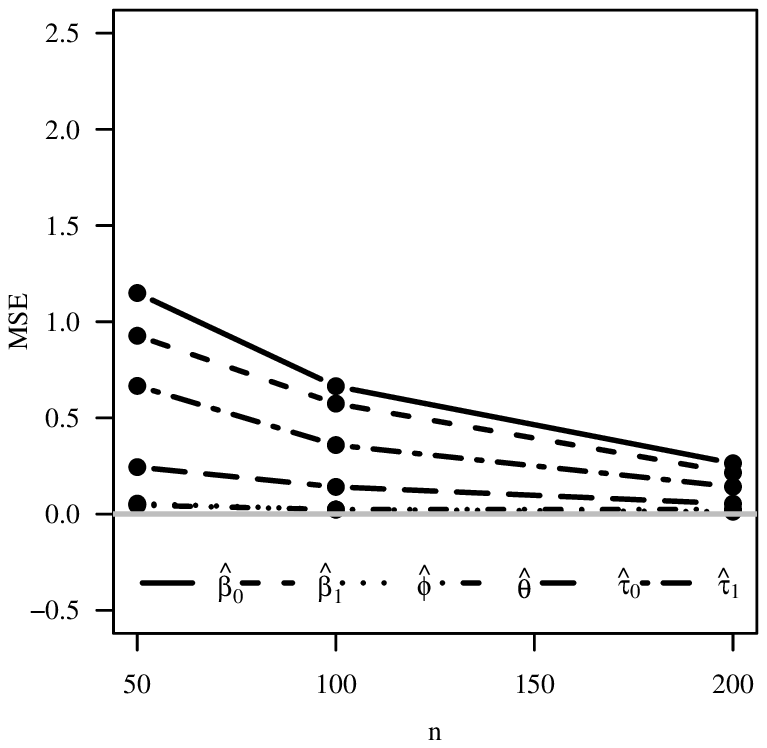}}
	\subfigure[Log-SL, $q = 0.25$]{\includegraphics[scale = 0.65]{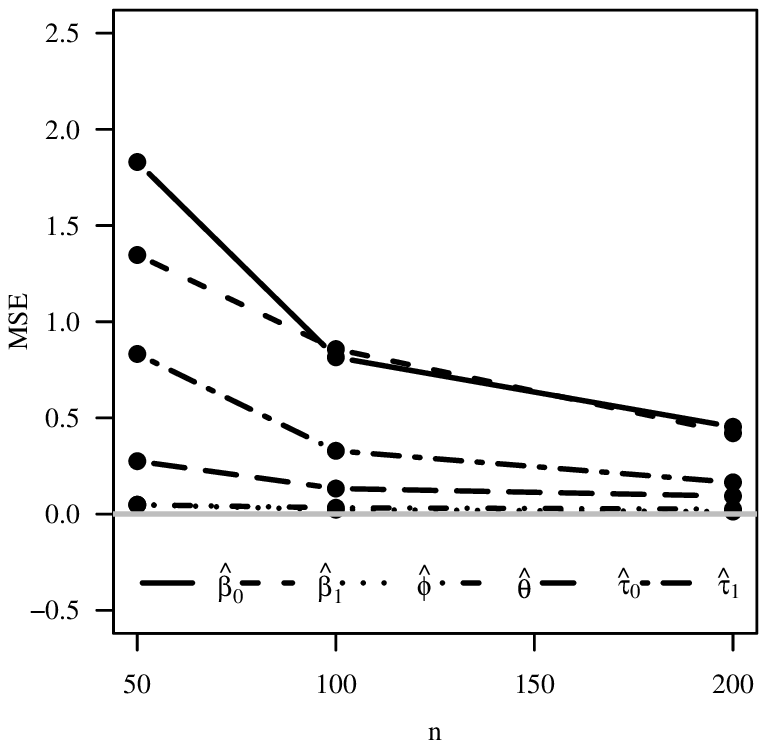}}
	\subfigure[Log-SL, $q = 0.5$]{\includegraphics[scale = 0.65]{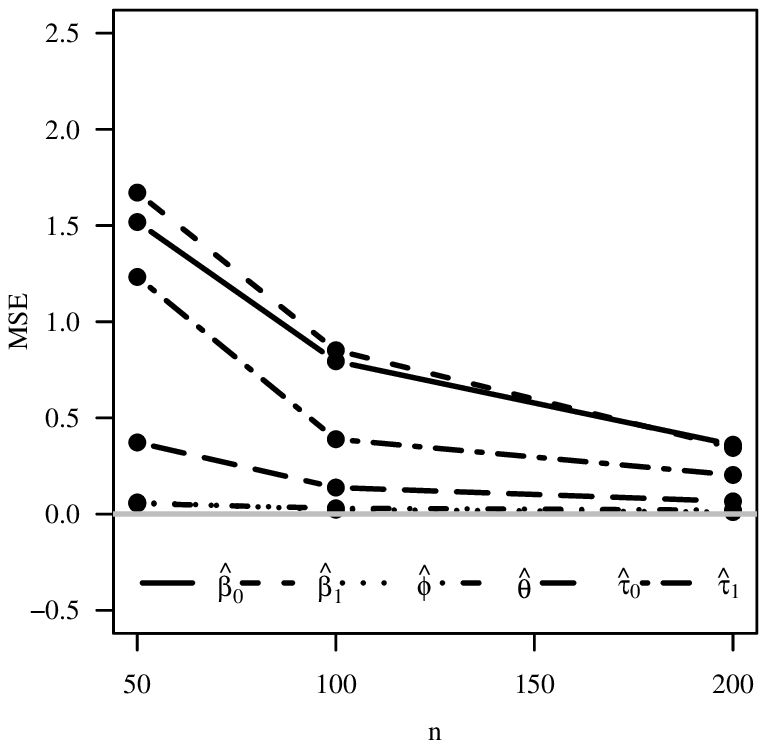}}
	\subfigure[Log-SL, $q = 0.75$]{\includegraphics[scale = 0.65]{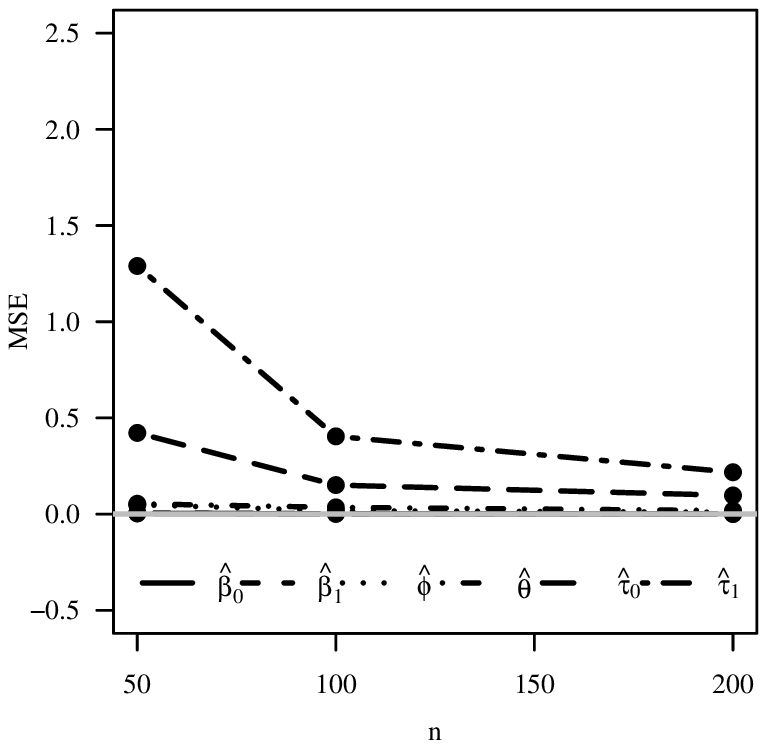}}
	\caption{MSEs of the CML estimators for the model QLS-ARMAX(1,1). (continuation).}
	\label{fig:05}
\end{figure}
\begin{figure}[H]
	\centering
	\subfigure[Log-SN, $q = 0.25$]{\includegraphics[scale = 0.65]{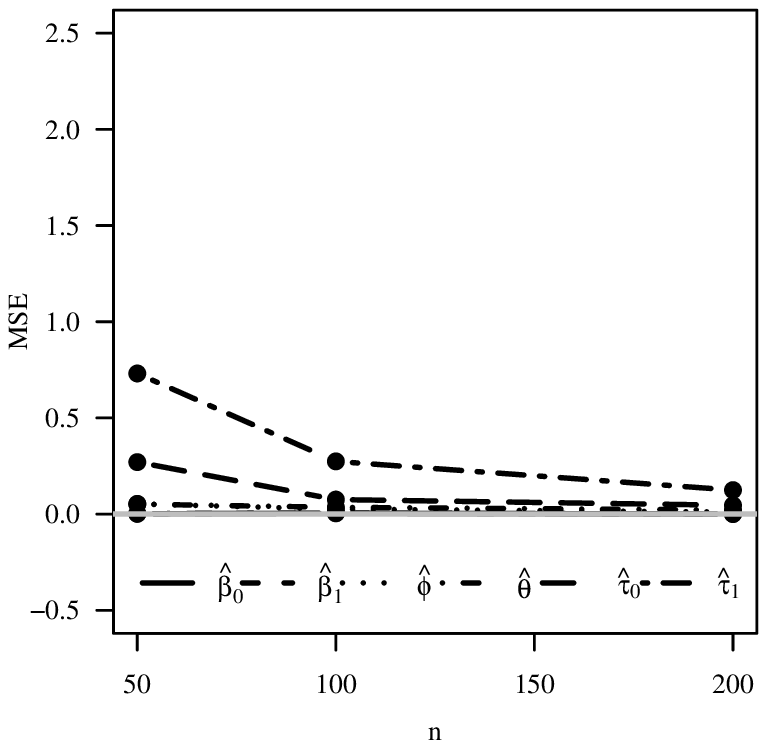}}
	\subfigure[Log-SN, $q = 0.5$]{\includegraphics[scale = 0.65]{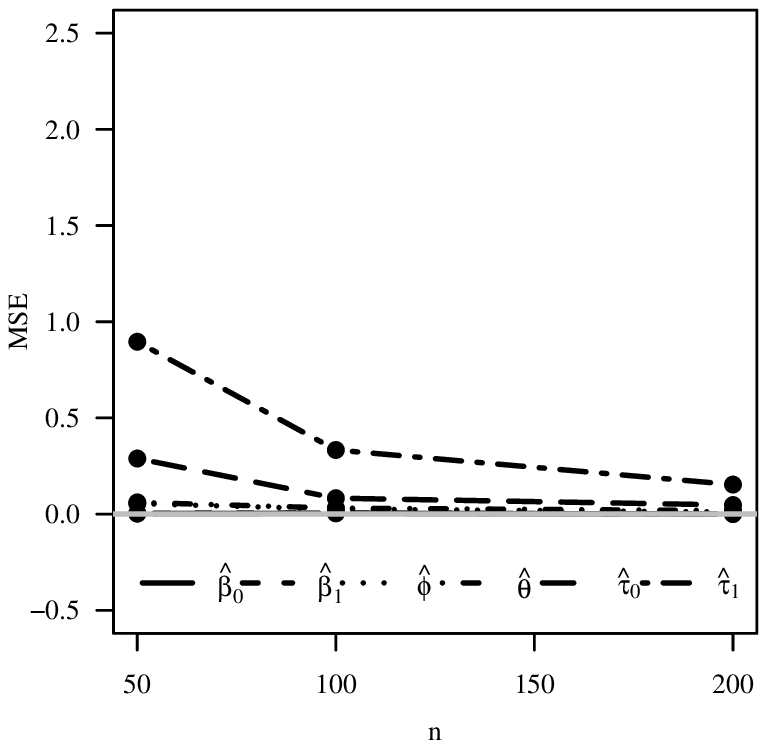}}
	\subfigure[Log-SN, $q = 0.75$]{\includegraphics[scale = 0.65]{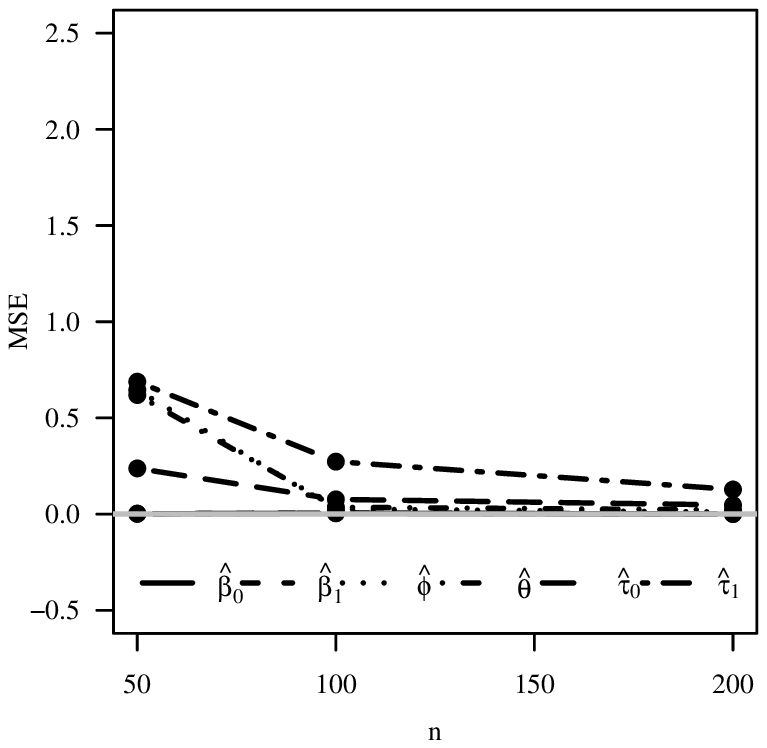}}
	\subfigure[Log-ST, $q = 0.25$]{\includegraphics[scale = 0.65]{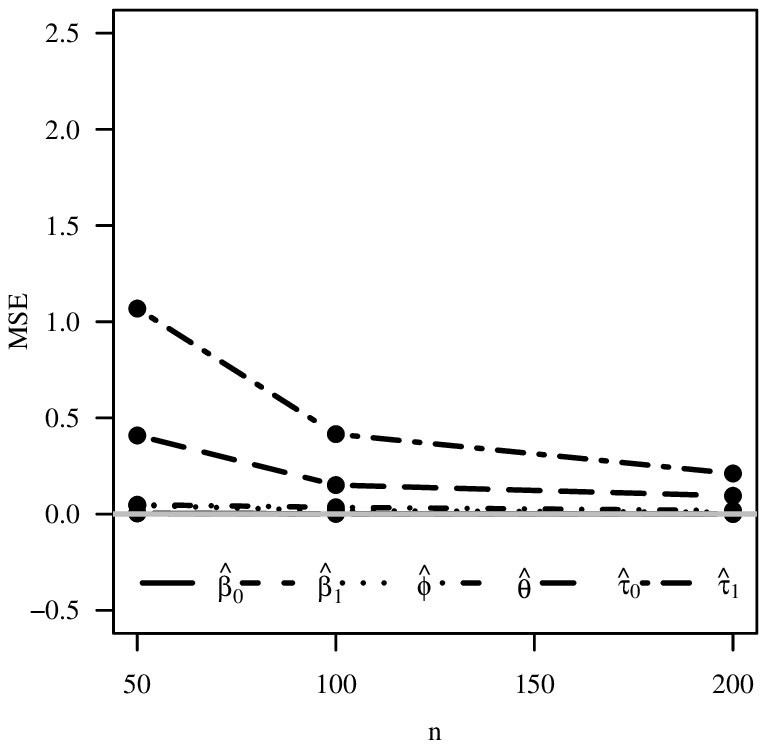}}
	\subfigure[Log-ST, $q = 0.5$]{\includegraphics[scale = 0.65]{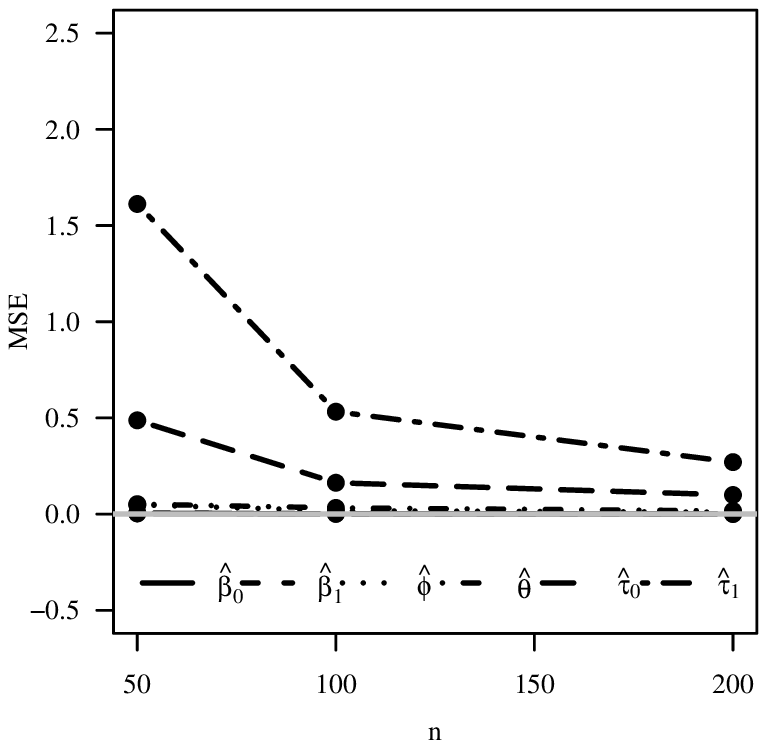}}
	\subfigure[Log-ST, $q = 0.75$]{\includegraphics[scale = 0.65]{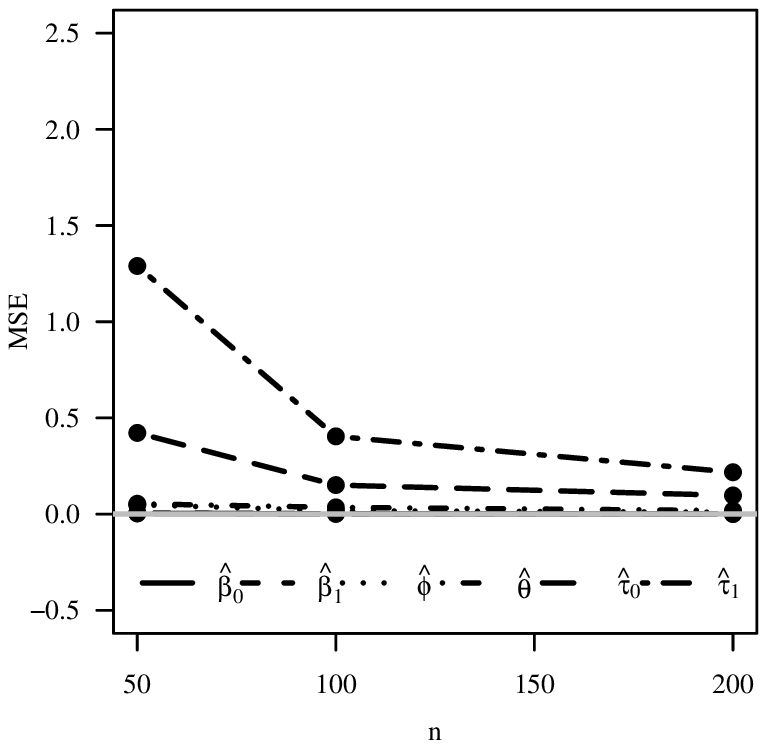}}
	\caption{MSEs of the CML estimators for the model QLS-ARMAX(1,1). (continuation)}
	\label{fig:06}
\end{figure}

In the first study, our goal is to demonstrate the performance of the proposed model and the CML estimation method in retrieving the true parameter values. We do this by assessing the biases and mean squared errors (MSEs) of the CML estimators. The expressions of the empirical bias and MSE of the estimators are given by
\begin{equation}
	\text{Bias}(\widehat{\varphi}) = \frac{1}{R} \sum_{r = 1}^{R} (\widehat{\varphi}^{(r)} - \varphi), \quad \text{MSE}(\widehat{\varphi}) = \frac{1}{R} \sum_{r = 1}^{R} (\widehat{\varphi}^{(r)} - \varphi)^2, \nonumber
\end{equation}
where $R$ is the number of Monte Carlo runs and $\widehat{\varphi}^{(r)}$ is the estimate of the parameter $\varphi$ at the $r$th Monte Carlo run. The steps involved in the simulation study are presented below in the form of an algorithm (see Algorithm \ref{alg:simulation}). We plot the biases and MSEs of the CML estimators and these are presented in Figures \ref{fig:01}-\ref{fig:06}. The plots in these figures show that both biases and MSEs of the estimators approach zero as the sample size increases, as one would expect. Moreover, the model with distribution log-SN($\vartheta = 0.1$) has the lowest MSE of the estimators when compared to other distributions.

In the second simulation study, we assess the adequacies of the theoretical assumptions of the Cox-Snell residuals and randomized quantile residuals. A summary of descriptive statistics (MN = mean, MD = median, SD = standard-deviation, CS = skewness coefficient, CK = ``excess'' kurtosis coefficient) is presented in Tables \ref{tab:gcs-res}-\ref{tab:rq-res}. The results obtained indicate that Cox-Snell residuals follow a standard exponential distribution with the Monte Carlo estimates for the mean and the standard deviation being 1 for both. The results corresponding to the randomized quantile residuals also indicate that the simulated randomized quantile residuals follow a standard normal distribution with the estimated mean and standard deviation as 0 and 1, respectively.

\begin{algorithm}
	\floatname{algorithm}{Algorithm}
	\caption{Steps involved in the simulation study.}\label{alg:simulation}
	\begin{algorithmic}[1]
		\State Choose a QLS-ARMAX($p,q$) model based on some PDF generator $g$ from the log-symmetric family and set the values of the model parameters.
		\State Generate 5\,000 samples based on the chosen model.
		\State {Estimate the model parameters using the CML method for each sample.}
		\State Compute the empirical bias and MSE of the estimators.
		\State For each Monte Carlo run, calculate the GCS and RQ residuals and the respective descriptive statistics:
		mean, median, SD, CS and CK.
		\State Determine the means of the descriptive statistics obtained in Step 5 from the 5,000 runs.
	\end{algorithmic}
\end{algorithm}

	\begin{table}[H]
		\centering
		\caption{Summary statistics for the GCS residuals.}
		\adjustbox{max height=\dimexpr\textheight\relax,
			max width=\dimexpr\textwidth-2.5cm\relax}{
			\begin{tabular}{lccccccccc}
				\toprule
				$n$ & $q$ & & Log-NO & Log-$t$ & Log-PE & Log-HP & Log-SL & Log-SN & Log-ST \\
				\hline
				\multirow{15}{*}{50} & \multirow{5}{*}{0.25} & MN & 0.9991 & 0.9940 & 0.9959 & 0.9919 & 0.9920 & 0.9994 & 0.9907 \\
				& & MD & 0.6997 & 0.7027 & 0.6953 & 0.6964 & 0.6934 & 0.7010 & 0.6971 \\
				& & SD & 0.9833 & 0.9711 & 0.9816 & 0.9719 & 0.9797 & 0.9836 & 0.9708 \\
				& & CS & 1.5226 & 1.4980 & 1.5362 & 1.5028 & 1.5375 & 1.5245 & 1.5061 \\
				& & CK & 2.4895 & 2.4122 & 2.5715 & 2.4168 & 2.5973 & 2.5059 & 2.4421 \\
				\cline{2-10}
				& \multirow{5}{*}{0.5} & MN & 1.0011 & 0.9977 & 1.0005 & 0.9977 & 0.9969 & 1.0021 & 0.9957 \\
				& & MD & 0.6949 & 0.6971 & 0.6943 & 0.6975 & 0.6951 & 0.6991 & 0.6950 \\
				& & SD & 0.9928 & 0.9820 & 0.9871 & 0.9804 & 0.9830 & 0.9896 & 0.9797 \\
				& & CS & 1.5518 & 1.5264 & 1.5342 & 1.5197 & 1.5303 & 1.5299 & 1.5259 \\
				& & CK & 2.6126 & 2.5199 & 2.5535 & 2.4871 & 2.5279 & 2.4962 & 2.5220 \\
				\cline{2-10}
				& \multirow{5}{*}{0.75} & MN & 1.0003 & 1.0048 & 1.0041 & 1.0067 & 1.0019 & 1.0040 & 1.0030 \\
				& & MD & 0.6915 & 0.6963 & 0.6921 & 0.6981 & 0.6945 & 0.6959 & 0.6936 \\
				& & SD & 0.9986 & 0.9948 & 0.9982 & 0.9936 & 0.996 & 0.9990 & 0.9931 \\
				& & CS & 1.5792 & 1.5443 & 1.5688 & 1.5295 & 1.5625 & 1.5689 & 1.5455 \\
				& & CK & 2.7352 & 2.5677 & 2.7090 & 2.5021 & 2.6630 & 2.6998 & 2.5781 \\
				\hline
				\multirow{15}{*}{100} & \multirow{5}{*}{0.25} & MN & 0.9984 & 0.9941 & 0.9944 & 0.9964 & 0.9942 & 0.9985 & 0.992 \\
				& & MD & 0.6967 & 0.6915 & 0.6924 & 0.6949 & 0.6956 & 0.6968 & 0.6933 \\
				& & SD & 0.9882 & 0.9834 & 0.9807 & 0.9793 & 0.9780 & 0.9886 & 0.9748 \\
				& & CS & 1.7086 & 1.6911 & 1.6640 & 1.6567 & 1.6686 & 1.7115 & 1.6539 \\
				& & CK & 3.6534 & 3.5562 & 3.4059 & 3.3872 & 3.4256 & 3.6696 & 3.3767 \\
				\cline{2-10}
				& \multirow{5}{*}{0.5} & MN & 1.0000 & 0.9977 & 0.9991 & 0.9989 & 0.9973 & 1.0001 & 0.9967 \\
				& & MD & 0.6950 & 0.6926 & 0.6944 & 0.6947 & 0.6943 & 0.6950 & 0.6941 \\
				& & SD & 0.9939 & 0.9864 & 0.9847 & 0.9833 & 0.9839 & 0.9944 & 0.9795 \\
				& & CS & 1.7269 & 1.6915 & 1.6676 & 1.662 & 1.6803 & 1.7299 & 1.6560 \\
				& & CK & 3.7369 & 3.5464 & 3.4263 & 3.4079 & 3.4683 & 3.7543 & 3.3764 \\
				\cline{2-10}
				& \multirow{5}{*}{0.75} & MN & 1.0022 & 1.0031 & 1.0054 & 1.0046 & 1.0010 & 1.0023 & 1.0028 \\
				& & MD & 0.6927 & 0.6946 & 0.6969 & 0.6949 & 0.6930 & 0.6926 & 0.6956 \\
				& & SD & 1.0023 & 0.9931 & 0.9934 & 0.9942 & 0.9922 & 1.0028 & 0.9877 \\
				& & CS & 1.7582 & 1.6932 & 1.6813 & 1.6869 & 1.6998 & 1.7623 & 1.6581 \\
				& & CK & 3.9077 & 3.5347 & 3.5085 & 3.5167 & 3.5665 & 3.9331 & 3.3604 \\
				\hline
				\multirow{15}{*}{200} & \multirow{5}{*}{0.25} & MN & 0.9977 & 0.9941 & 0.9958 & 0.9961 & 0.9977 & 0.9977 & 0.9940 \\
				& & MD & 0.6922 & 0.6914 & 0.6917 & 0.6933 & 0.6928 & 0.6922 & 0.6912 \\
				& & SD & 0.9967 & 0.9813 & 0.9827 & 0.9796 & 0.9889 & 0.9968 & 0.9808 \\
				& & CS & 1.8667 & 1.7675 & 1.7594 & 1.7430 & 1.7981 & 1.8681 & 1.7703 \\
				& & CK & 4.7826 & 4.1765 & 4.1439 & 4.0539 & 4.3011 & 4.7930 & 4.2046 \\
				\cline{2-10}
				& \multirow{5}{*}{0.5} & MN & 1.0004 & 0.9973 & 0.9993 & 0.9978 & 0.9984 & 1.0004 & 0.9966 \\
				& & MD & 0.6940 & 0.6939 & 0.6934 & 0.6940 & 0.6937 & 0.6940 & 0.6932 \\
				& & SD & 1.0000 & 0.9831 & 0.9866 & 0.9840 & 0.9902 & 1.0001 & 0.9828 \\
				& & CS & 1.8761 & 1.7688 & 1.7675 & 1.7632 & 1.8040 & 1.8775 & 1.7717 \\
				& & CK & 4.8556 & 4.1837 & 4.1859 & 4.1637 & 4.3490 & 4.8662 & 4.2033 \\
				\cline{2-10}
				& \multirow{5}{*}{0.75} & MN & 1.0044 & 1.0015 & 1.0037 & 1.0004 & 1.0029 & 1.0044 & 1.0004 \\
				& & MD & 0.6956 & 0.6966 & 0.6953 & 0.6941 & 0.6948 & 0.6956 & 0.6956 \\
				& & SD & 1.0069 & 0.9872 & 0.9927 & 0.9879 & 0.9977 & 1.0070 & 0.9868 \\
				& & CS & 1.8957 & 1.7634 & 1.7758 & 1.7687 & 1.8163 & 1.8974 & 1.7682 \\
				& & CK & 4.9945 & 4.1421 & 4.2354 & 4.1946 & 4.3895 & 5.0080 & 4.1700 \\
				\bottomrule
			\end{tabular}
		}
				\label{tab:gcs-res}
	\end{table}

	\begin{table}[H]
		\centering
		\caption{Summary statistics for the RQ residuals.}
		\adjustbox{max height=\dimexpr\textheight\relax,
			max width=\dimexpr\textwidth-2.5cm\relax}{
			\begin{tabular}{lccccccccc}
				\toprule
				$n$ & $q$ & & Log-NO & Log-$t$ & Log-PE & Log-HP & Log-SL & Log-SN & Log-ST \\
				\hline
				\multirow{15}{*}{50} & \multirow{5}{*}{0.25} & MN & -0.0011 & -0.0009 & -0.0043 & -0.0065 & -0.0071 & -0.0008 & -0.0058 \\
				& & MD & 0.0030 & 0.0084 & 0.0015 & 0.0013 & -0.0039 & 0.0046 & 0.0017 \\
				& & SD & 1.0100 & 0.9995 & 1.0088 & 1.0059 & 1.0066 & 1.0100 & 1.0020 \\
				& & CS & -0.0163 & -0.0174 & -0.0196 & -0.0283 & -0.0121 & -0.0155 & -0.0198 \\
				& & CK & -0.3184 & -0.3671 & -0.3095 & -0.3476 & -0.3257 & -0.3218 & -0.3593 \\
				\cline{2-10}
				& \multirow{5}{*}{0.5} & MN & 0.0000 & 0.0012 & 0.0005 & -0.0004 & -0.0004 & 0.0015 & -0.0009 \\
				& & MD & -0.0032 & 0.0016 & -0.0003 & 0.0021 & -0.0021 & 0.0017 & -0.0014 \\
				& & SD & 1.0096 & 1.0017 & 1.0076 & 1.0044 & 1.0028 & 1.0095 & 1.0012 \\
				& & CS & 0.0161 & 0.0004 & -0.0009 & -0.0042 & 0.0052 & 0.0043 & 0.0007 \\
				& & CK & -0.3373 & -0.3482 & -0.3378 & -0.3632 & -0.3446 & -0.3363 & -0.3538 \\
				\cline{2-10}
				& \multirow{5}{*}{0.75} & MN & -0.0013 & 0.0069 & 0.0034 & 0.0073 & 0.0033 & 0.0022 & 0.0055 \\
				& & MD & -0.0074 & 0.0007 & -0.0028 & 0.0033 & -0.0026 & -0.0019 & -0.0027 \\
				& & SD & 1.0091 & 1.0040 & 1.0081 & 1.0070 & 1.005 & 1.0104 & 1.0028 \\
				& & CS & 0.0342 & 0.0182 & 0.0224 & 0.0097 & 0.0245 & 0.0247 & 0.0218 \\
				& & CK & -0.3136 & -0.3383 & -0.3182 & -0.3546 & -0.3205 & -0.3226 & -0.3414 \\
				\hline
				\multirow{15}{*}{100} & \multirow{5}{*}{0.25} & MN & -0.0019 & -0.0045 & -0.0059 & -0.0015 & -0.0033 & -0.0018 & -0.005 \\
				& & MD & 0.0020 & -0.0036 & -0.0017 & 0.0008 & 0.0009 & 0.0020 & -0.0015 \\
				& & SD & 1.0054 & 1.0007 & 1.0040 & 1.0002 & 1.0003 & 1.0056 & 0.9980 \\
				& & CS & -0.0153 & -0.0045 & -0.0134 & -0.0085 & -0.0188 & -0.0155 & -0.0098 \\
				& & CK & -0.1722 & -0.2129 & -0.2390 & -0.2513 & -0.2184 & -0.1679 & -0.2549 \\
				\cline{2-10}
				& \multirow{5}{*}{0.5} & MN & -0.0005 & 0.0002 & -0.0002 & 0.0010 & -0.0002 & -0.0005 & 0.0007 \\
				& & MD & -0.0003 & -0.0024 & 0.0007 & 0.0004 & -0.0011 & -0.0003 & -0.0006 \\
				& & SD & 1.0047 & 0.9991 & 1.0025 & 0.9999 & 0.9993 & 1.0049 & 0.9966 \\
				& & CS & 0.0052 & 0.0026 & -0.0046 & 0.002 & -0.0010 & 0.0051 & -0.0001 \\
				& & CK & -0.1824 & -0.2120 & -0.2427 & -0.2582 & -0.2222 & -0.1787 & -0.2546 \\
				\cline{2-10}
				& \multirow{5}{*}{0.75} & MN & 0.0010 & 0.0051 & 0.0054 & 0.0057 & 0.0026 & 0.0009 & 0.0060 \\
				& & MD & -0.0031 & 0.0002 & 0.0039 & 0.0007 & -0.0023 & -0.0032 & 0.0014 \\
				& & SD & 1.0051 & 1.0005 & 1.0042 & 1.0015 & 1.0008 & 1.0053 & 0.9982 \\
				& & CS & 0.0250 & 0.0089 & 0.0038 & 0.0186 & 0.0111 & 0.0250 & 0.0103 \\
				& & CK & -0.1727 & -0.2118 & -0.2367 & -0.2483 & -0.2113 & -0.1685 & -0.2566 \\
				\hline
				\multirow{15}{*}{200} & \multirow{5}{*}{0.25} & MN & -0.0033 & -0.0035 & -0.0039 & -0.0017 & -0.0007 & -0.0033 & -0.0035 \\
				& & MD & -0.0024 & -0.0030 & -0.0022 & -0.0004 & -0.0014 & -0.0024 & -0.0032 \\
				& & SD & 1.0035 & 0.9969 & 1.0009 & 0.9977 & 0.9990 & 1.0035 & 0.9969 \\
				& & CS & -0.0027 & -0.0010 & -0.0089 & -0.0046 & -0.0002 & -0.0027 & -0.0028 \\
				& & CK & -0.0649 & -0.1794 & -0.1853 & -0.2119 & -0.1369 & -0.0632 & -0.1787 \\
				\cline{2-10}
				& \multirow{5}{*}{0.5} & MN & 0.0001 & 0.0007 & 0.0002 & -0.0003 & 0.0001 & 0.0001 & -0.0002 \\
				& & MD & -0.0003 & 0.0001 & -0.0002 & 0.0003 & -0.0003 & -0.0003 & -0.0008 \\
				& & SD & 1.0024 & 0.9958 & 1.0000 & 0.9982 & 0.9989 & 1.0024 & 0.9959 \\
				& & CS & 0.0066 & 0.0002 & -0.0011 & -0.0003 & 0.0021 & 0.0066 & 0.0001 \\
				& & CK & -0.0697 & -0.1746 & -0.1830 & -0.1899 & -0.1331 & -0.0683 & -0.1740 \\
				\cline{2-10}
				& \multirow{5}{*}{0.75} & MN & 0.0035 & 0.0044 & 0.0040 & 0.0021 & 0.0038 & 0.0035 & 0.0033 \\
				& & MD & 0.0020 & 0.0035 & 0.0023 & 0.0006 & 0.0010 & 0.0020 & 0.0022 \\
				& & SD & 1.0036 & 0.9971 & 1.0010 & 0.9985 & 1.0003 & 1.0036 & 0.9969 \\
				& & CS & 0.0155 & 0.0019 & 0.0067 & 0.0075 & 0.0116 & 0.0155 & 0.0028 \\
				& & CK & -0.0639 & -0.1804 & -0.1835 & -0.1943 & -0.1265 & -0.0622 & -0.1767 \\
				\bottomrule
			\end{tabular}
		}
				\label{tab:rq-res}
	\end{table}

\clearpage

\section{Application to Walmart sales data}\label{sec:4}

\begin{figure}[!ht]
\centering
{\includegraphics[scale = 1]{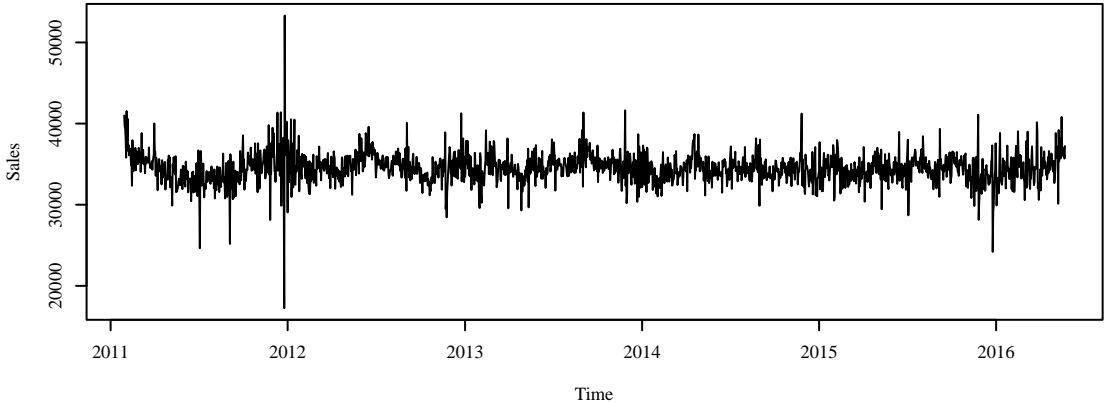}}
\vspace{-14.00cm}
 \caption{Daily sales of 30,491 Walmart products.}
\label{fig:series_sales}
\end{figure}

To illustrate the applicability of the proposed models, we considered the dataset on the competition "M5 Forecasting - Accuracy" \citep{makridakis2020m5}. This dataset corresponds to the daily sales ($Y_{t}$) history of 30,491 Walmart products and is related to 10 stores of the hypermarket chain in three different US states. For the purpose of analysis, the sets of all these time series were grouped to form only one time series composed of the total daily sales. Then, the series was adjusted for trend and multiple seasonality using the MSTL decomposition \citep{hyndman2018forecasting,bandara2021mstl}, and considering three types of seasonality that are typical of daily sales series: annual (cycle of 365.25 observations), monthly (cycle of 30.43 observations) and weekly (cycle of 7 observations). We thus obtained a stationary series resulting from the filtering of the trend and multiple seasonalities; see Figure \ref{fig:series_sales} for a plot of this series. The explanatory variables\footnote{We have studied  twelve explanatory dummy variables referring to important calendar events that do not have a fixed date (Easter, Mother's Day, Father's Day, Superbowl, Thanksgiving, ...). } are $snapca_{1t}$, snapTX ($snaptx_{2t}$), Mother's Day ($mother_{3t}$) and Thanksgiving ($thanks_{4t}$). The dataset prepared for this study is available upon request.

\begin{figure}[h!]
\small
\vspace{-0.5cm}
\centering
\vspace{-0.5cm}
\includegraphics[scale=0.8]{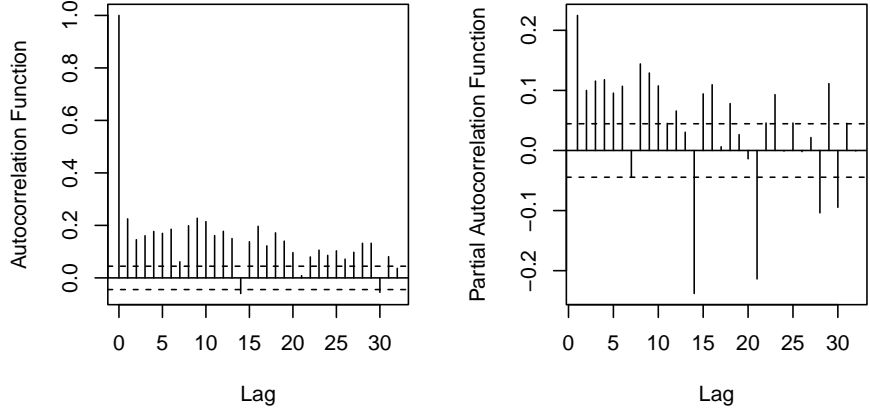} 
\caption{ \small Plots of ACF (left) and PACF (right) of the residuals from the fitted regression model with the Walmart sales data.}
\label{fig_acf_pacf_lm}
\end{figure}

Initially, we can fit a linear regression model and analyze the autocorrelation function (ACF) and partial autocorrelation function (PACF). Considering the daily sales ($Y_{t}$) as the response variable, we can assume the following structure:
\begin{equation} \label{model:lm}
Y_{t} = \beta_0+ \beta_1 snapca_{1t}+ \beta_2 snaptx_{2t} + \beta_3mother_{3t} + \beta_4 thanks_{4t} + \varepsilon_t,
\end{equation}
where $ \varepsilon_t $ is a random error consisting of independent random variables which are identically distributed as normal with zero mean and variance $ \sigma_{\varepsilon} ^ 2$. Figure \ref{fig_acf_pacf_lm} shows the ACF and PACF of the residuals from the least squares fit of \eqref{model:lm}. From this figure, we observe that the residuals are autocorrelated and ARMA models seem to be more appropriate for these data.

In addition to the limitation due to autocorrelated residuals, other important aspects that must be observed are the characteristics in the data. Table~\ref{tab:estdesc} reports descriptive statistics of the observed Walmart sales, including the mean, median, standard deviation (SD), coefficient of variation (CV), skewness (CS), (excess) kurtosis (CK), and minimum and maximum values. From this table, we observe skewness and high
kurtosis in the data. Such observations make the proposed log-symmetric quantile ARMA models good candidates for fitting the data since they account for asymmetry with or without heavy tails.

\begin{table}[!ht]
\centering
\caption{Summary statistics for the Walmart sales data.}
\label{tab:estdesc}
\begin{tabular}{ccccccccccccccccc}
\hline
   Mean   & Median       &  SD     &  CV       &  CS   &  CK    & minimum & maximum & size \\
\hline
34488.87  & 34460.03    &1878.172  & 5.446\%  &  0.165 &10.396 & 17256.78 & 53304.64  & 330 \\
\hline
\end{tabular}
\end{table}

We then analyze the Walmart sales data using the QLS-ARMAX($p,q$) models, expressed as
\begin{eqnarray*}
	\eta_t = \log(Q_t) &=&  \beta_0+ \beta_1 snapca_{1t}+ \beta_2 snaptx_{2t} + \beta_3mother_{3t} + \beta_4 thanks_{4t} \\
	&& + \sum_{i=1}^{p} \phi_i \left[ h(Y_{t-i}) - (\beta_0+ \beta_1 snapca_{1{t-i}}+ \beta_2 snaptx_{2{t-i}} + \beta_3mother_{3{t-i}}  \right. \\
	&& \left. + \beta_4 thanks_{4{t-i}}) \right] + \sum_{j = 1}^{q} \theta_j r_{t-j}, \quad t = \max\{p,q\},\ldots, n, \nonumber \\
	 \gamma_t = \log(\kappa_t) &=&  \tau_0 + \tau_1 mother_{1t}+ \tau_2 thanks_{2t} , \quad t = \max\{p,q\},\ldots, n. \nonumber
\end{eqnarray*}
First, we have to find the best model among the class of QLS-ARMAX($p,q$) models that include the log-normal (log-NO), log-Student-$t$ (log-$t$),
log-power-exponential (log-PE), log-hyperbolic (log-HP), log-slash (log-SL), log-contaminated-normal (log-CN), extended Birnbaum-Saunders (EBS) and extended Birnbaum-Saunders-$t$ (EBS-$t$) distributions. To this end, we fit the models based on a grid of values of $q \in \{0.01,0.02, \dots,0.99\}$. Then, we compute the averages of the corresponding vAkaike (AIC), Bayesian (BIC), corrected Akaike (CAIC) and Hannan-Quinn (HQIC) information criteria values. From Table \ref{tab:criterions}, we observe that the log-$t$ quantile ARMA model provides the best adjustment compared to the other log-symmetric quantile ARMA models based on the values of AIC, BIC, CAIC and HIC.

\begin{table}[!ht]
\centering
\footnotesize
	\caption{Averages of the AIC, BIC, CAIC and HQIC values computed across $q \in \{ 0.01,0.02, \ldots, 0.98, 0.99\}$ for the indicated QLS-ARMAX models with the Walmart sales data.} 		 \label{tab:criterions}

		\begin{tabular}{lrrrrrrrrr}
			\toprule
   Indicator   & log-NO   & log-$t$  & log-PE   & log-HP   & log-SL   & log-CN   & EBS      & EBS-$t$ \\
			\midrule
			AIC & -6219.350 & \textbf{-6705.274} & -6643.247 & -6647.953 & -6682.070 & -6604.540 & -6190.254 & -6704.203 \\
           BIC & -6130.214 & \textbf{-6616.138} & -6554.112 & -6558.818 & -6592.935 & -6515.405 & -6095.547 & -6615.068 \\
          CAIC & -6219.137 & \textbf{-6705.062}  & -6643.035 & -6647.741 & -6681.858 & -6604.328 & -6190.010 & -6703.991 \\
           HQIC & -6218.960 & \textbf{-6704.885} & -6642.858 & -6647.564 & -6681.681 & -6604.151 & -6189.840 & -6703.814\\
			\bottomrule
	\end{tabular}
\end{table}

\begin{figure}[!ht]
	\centering
	\subfigure[$\phi_1$]{\includegraphics[scale = 0.25]{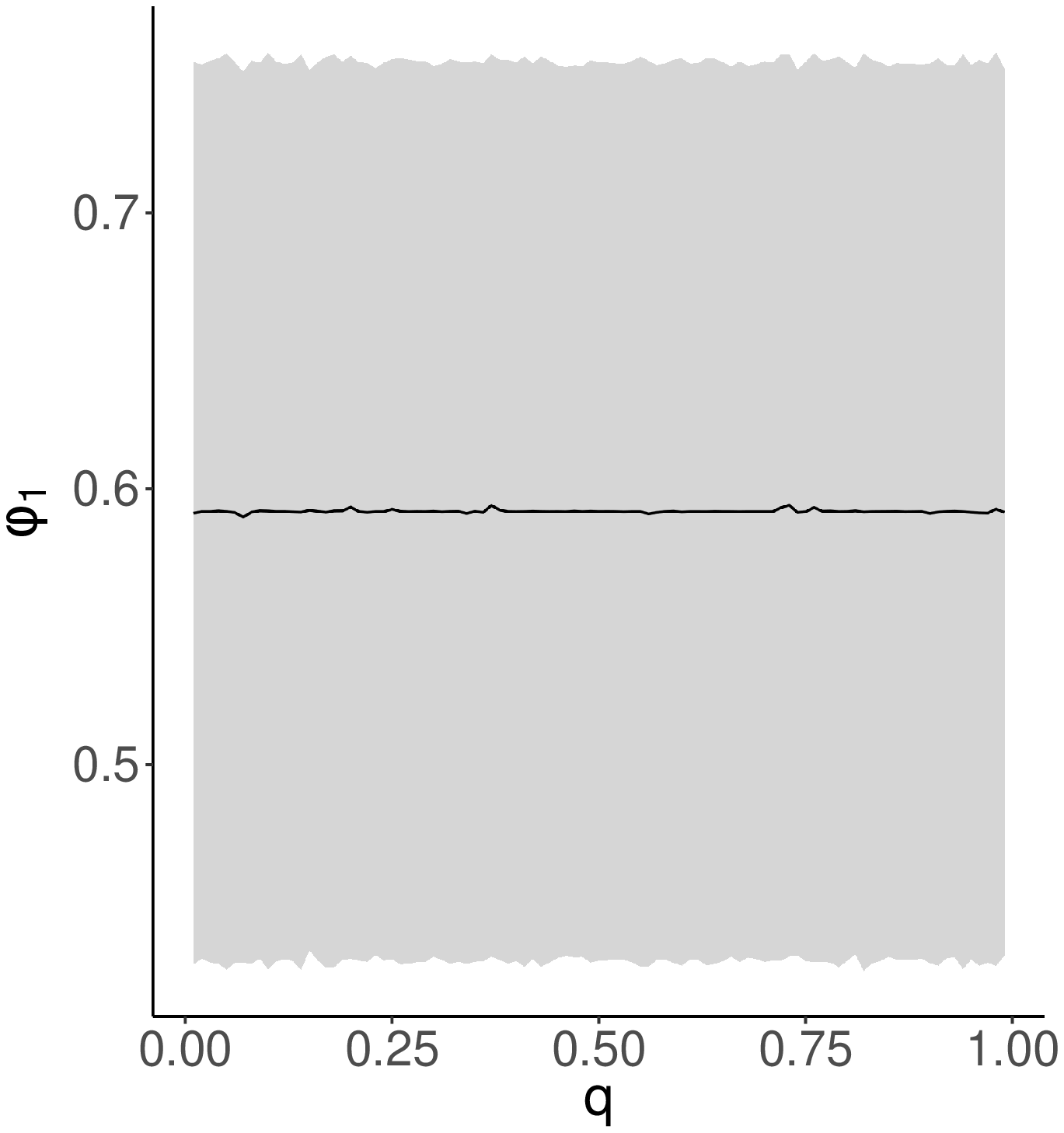}}
	\subfigure[$\theta_1$]{\includegraphics[scale = 0.25]{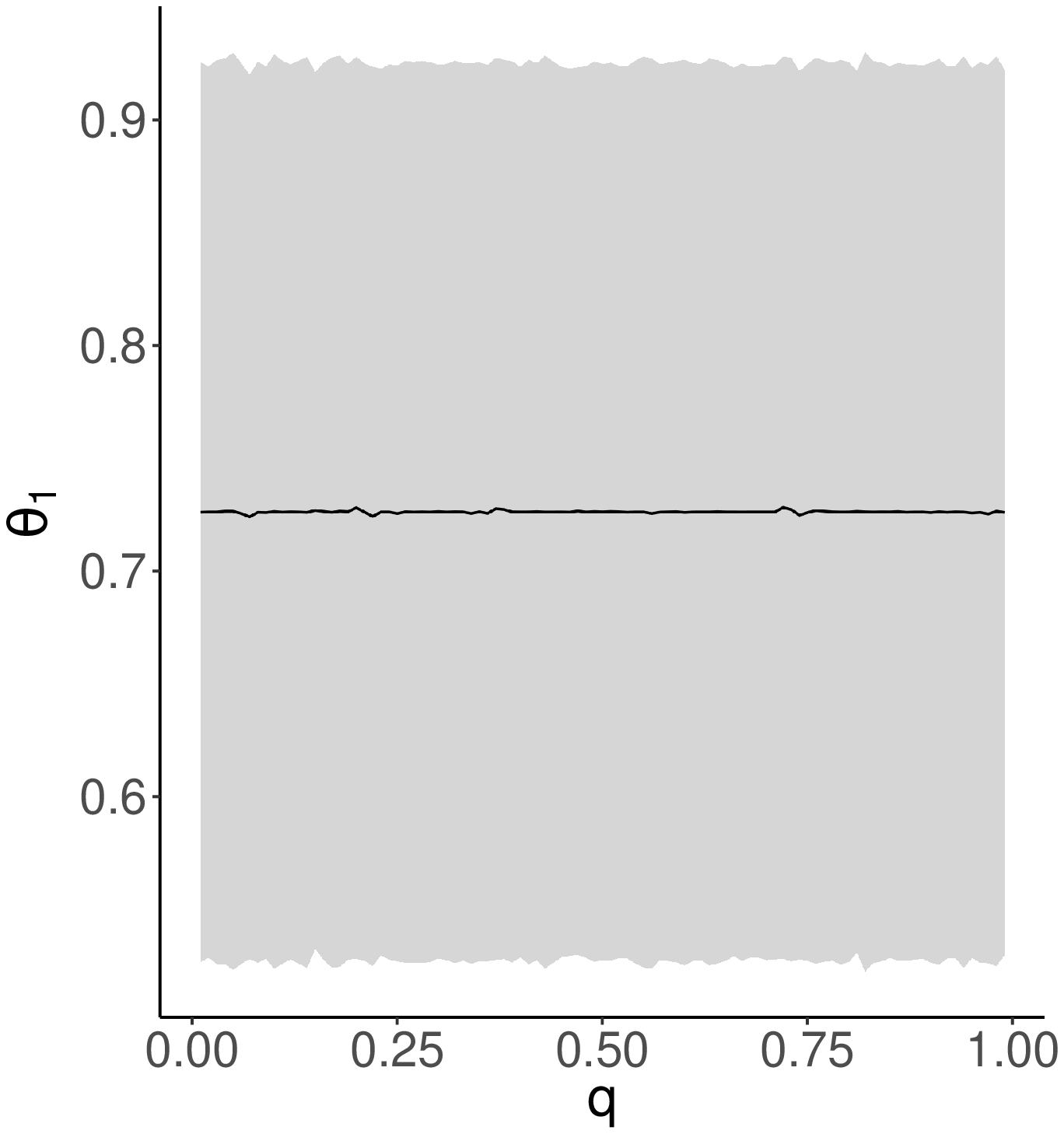}}
	\subfigure[$\beta_0$]{\includegraphics[scale = 0.25]{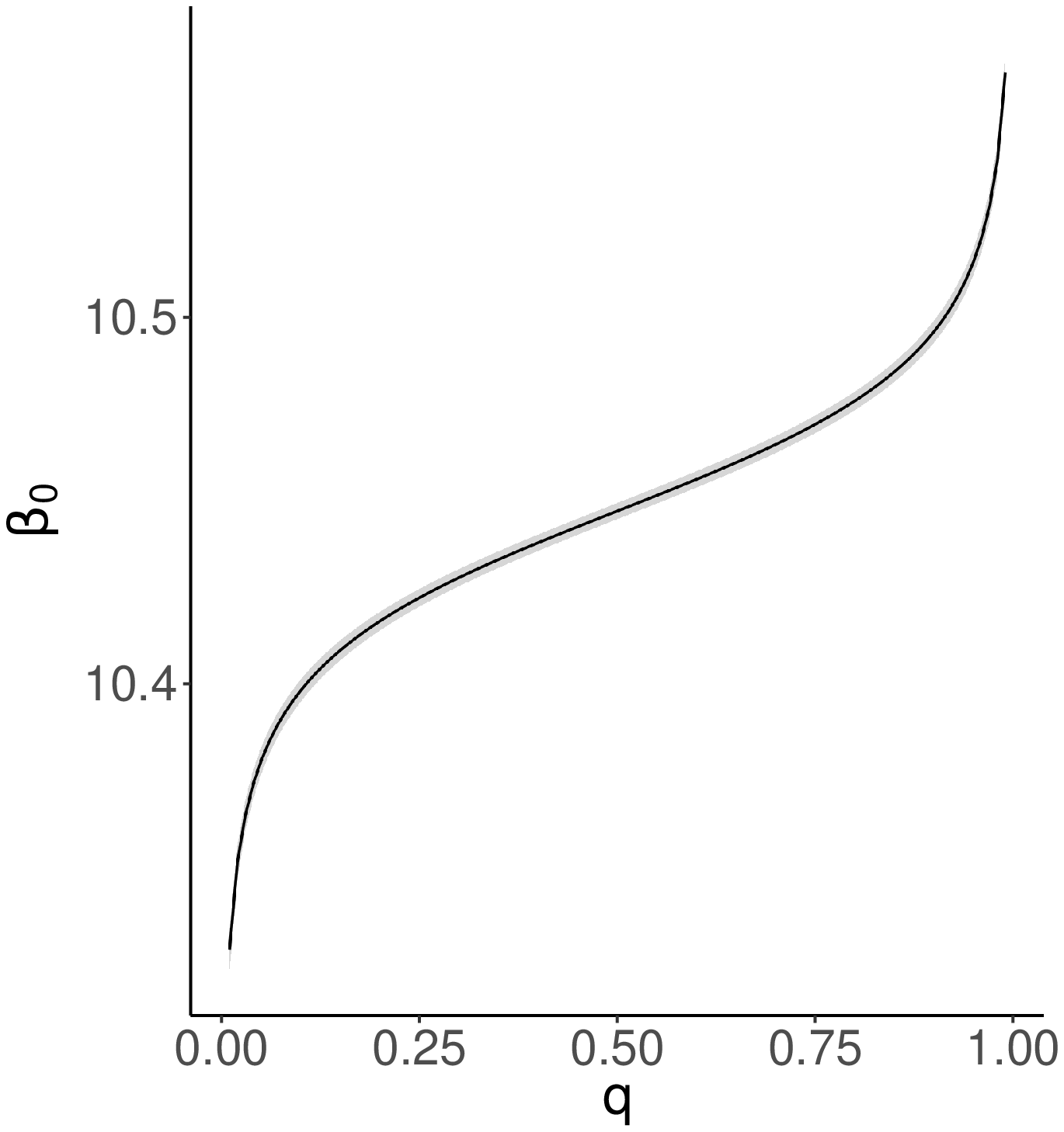}}
	\subfigure[$\beta_1$]{\includegraphics[scale = 0.25]{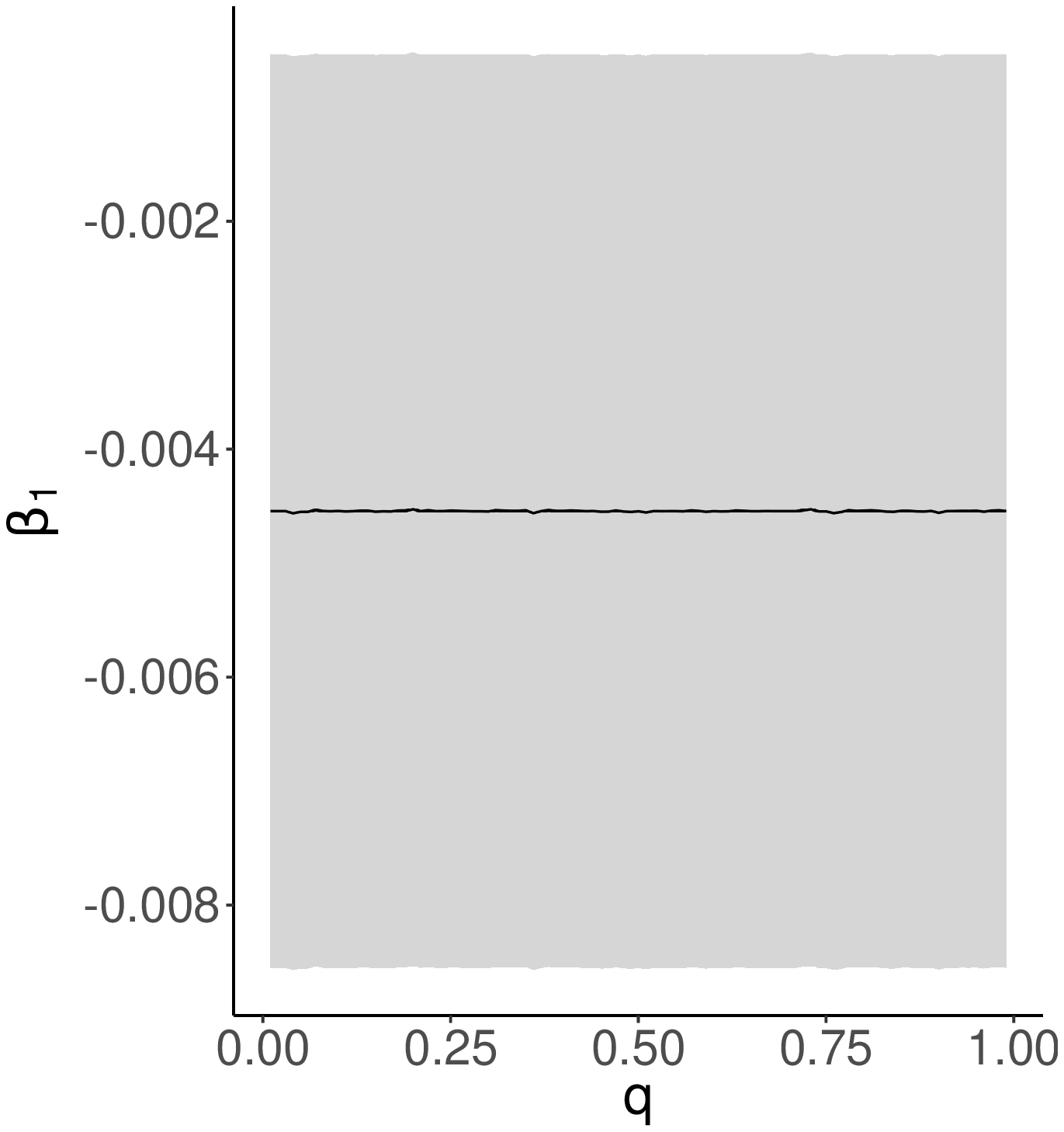}}
	\subfigure[$\beta_2$]{\includegraphics[scale = 0.25]{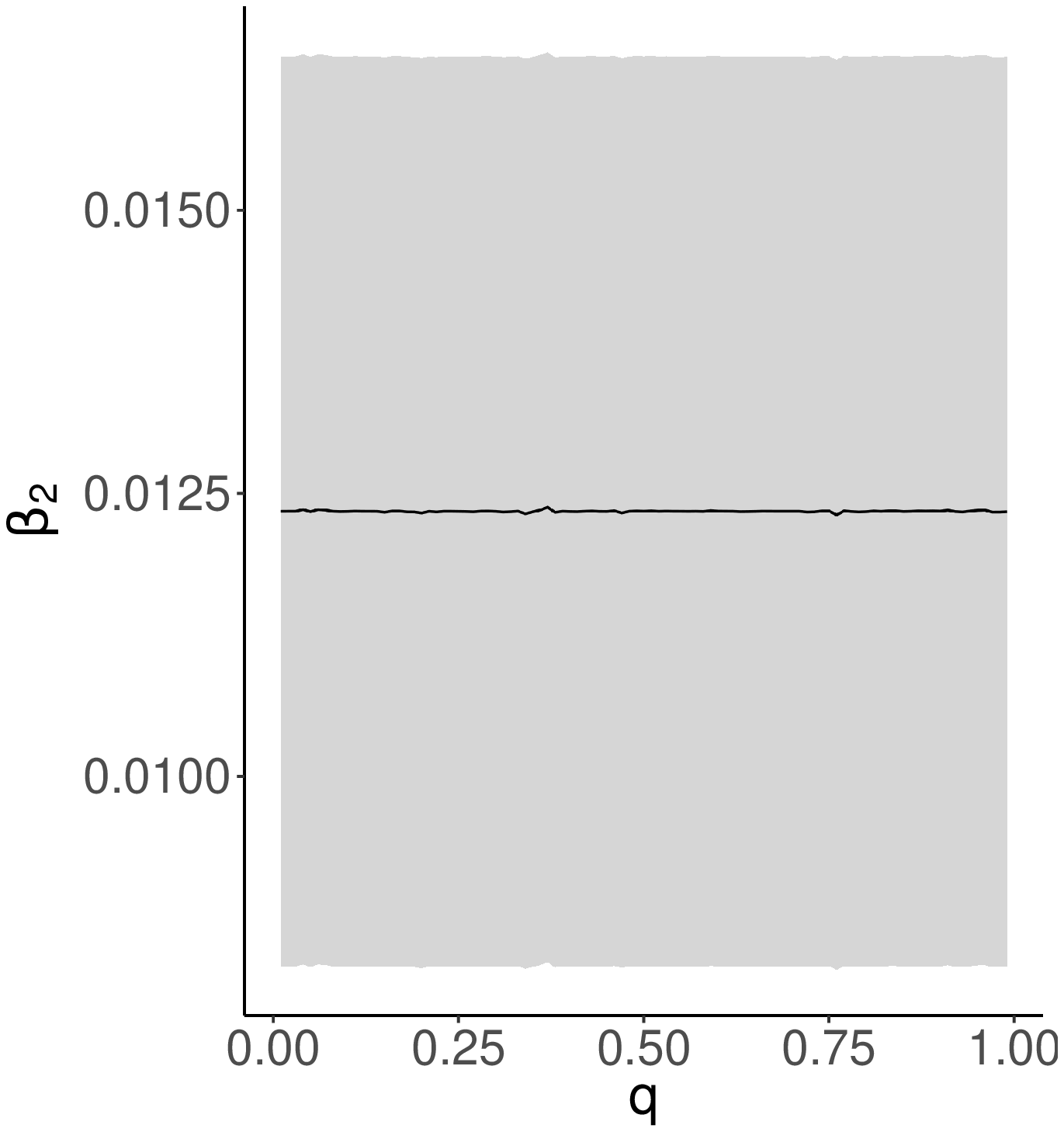}}
	\subfigure[$\beta_3$]{\includegraphics[scale = 0.25]{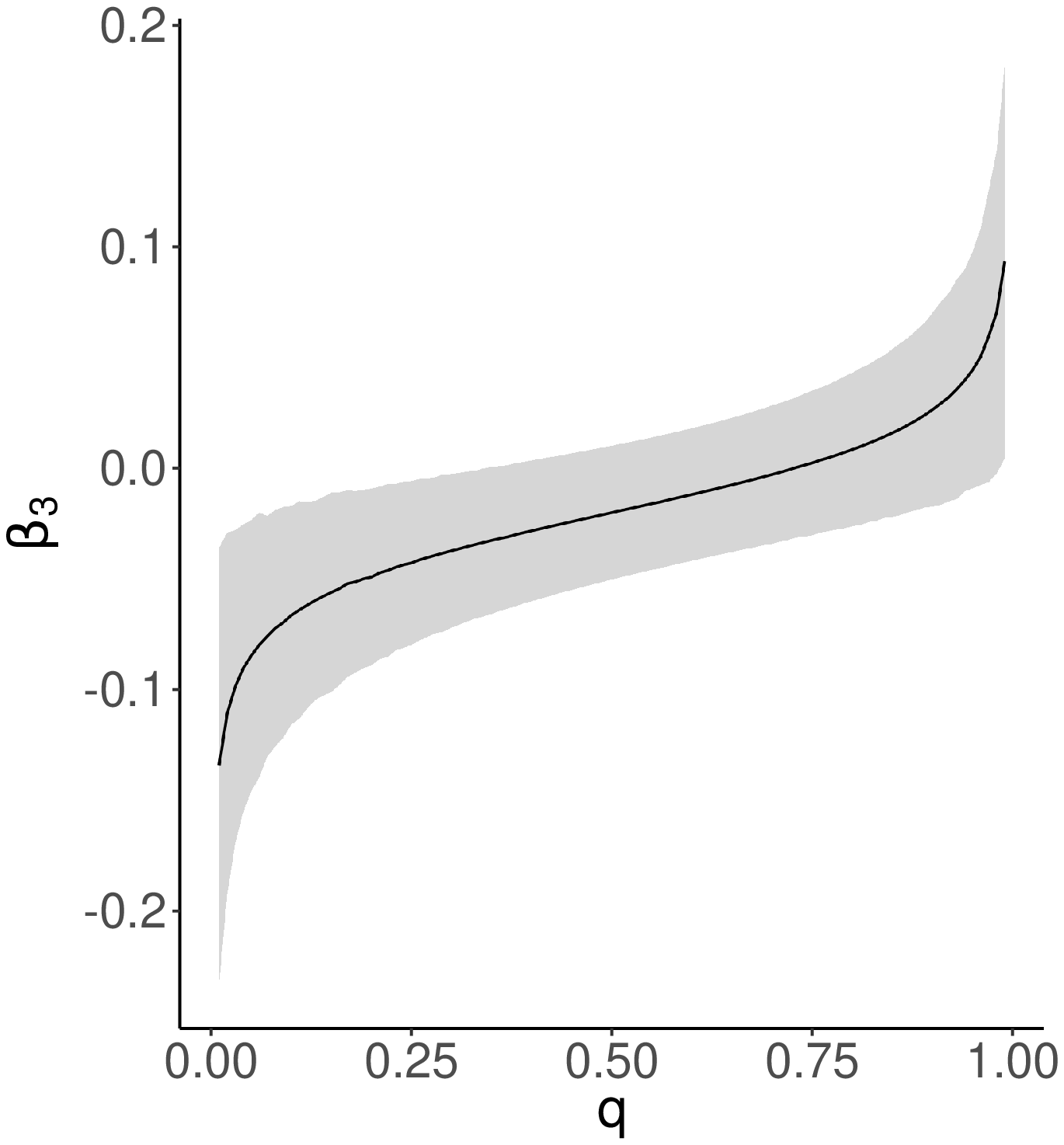}}
	\subfigure[$\beta_4$]{\includegraphics[scale = 0.25]{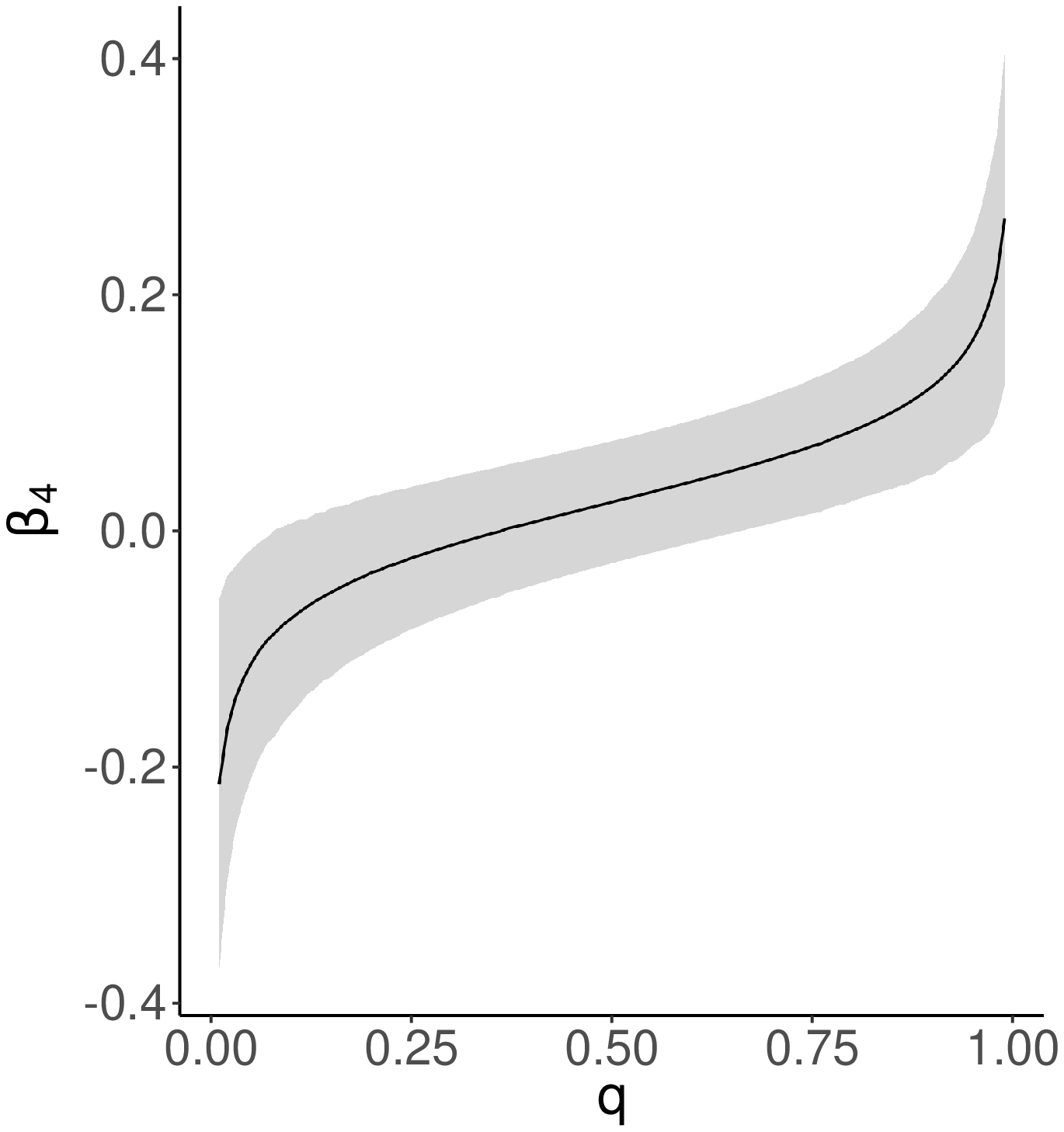}}
	\subfigure[$\tau_0$]{\includegraphics[scale = 0.25]{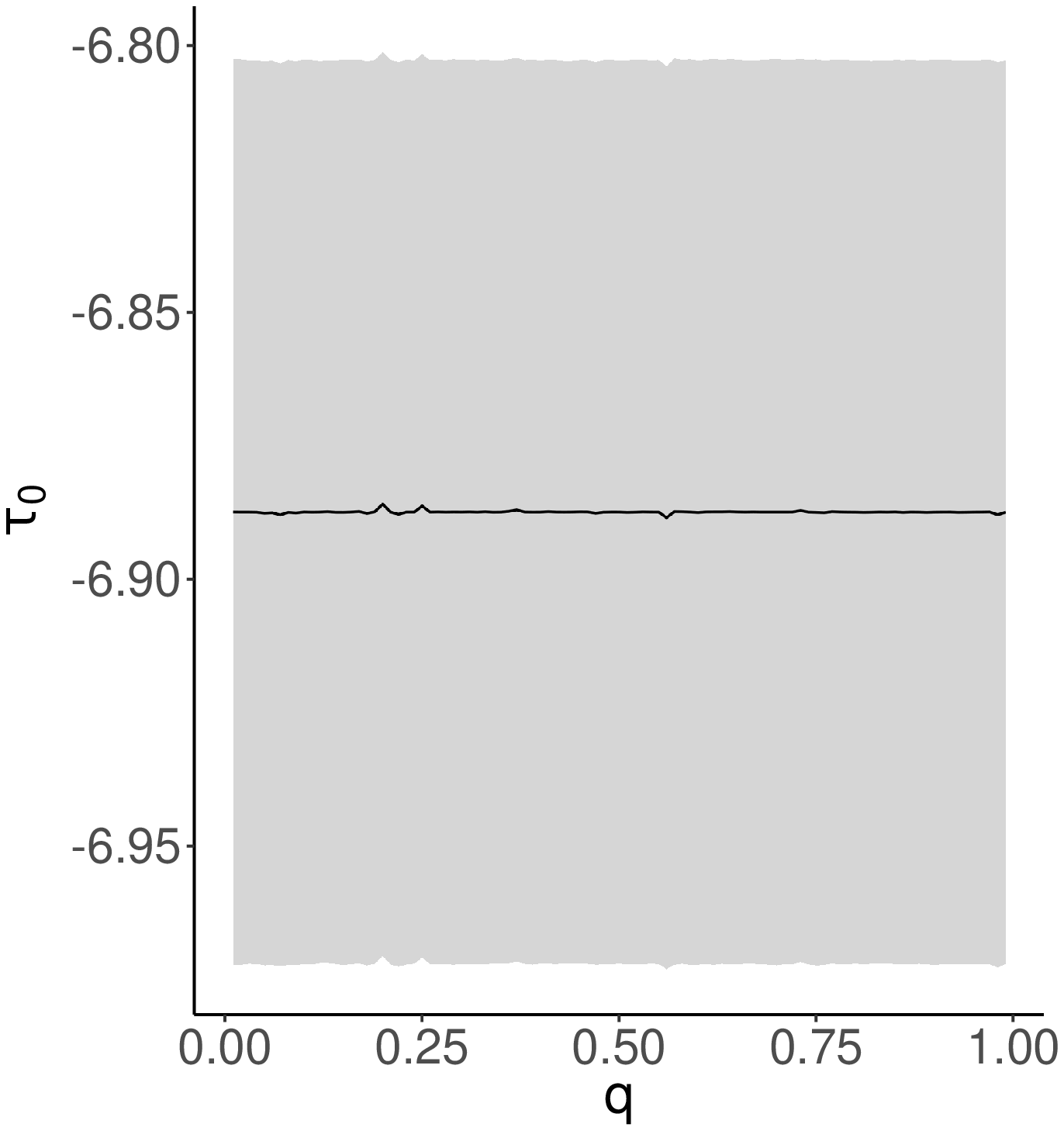}}
	\subfigure[$\tau_1$]{\includegraphics[scale = 0.25]{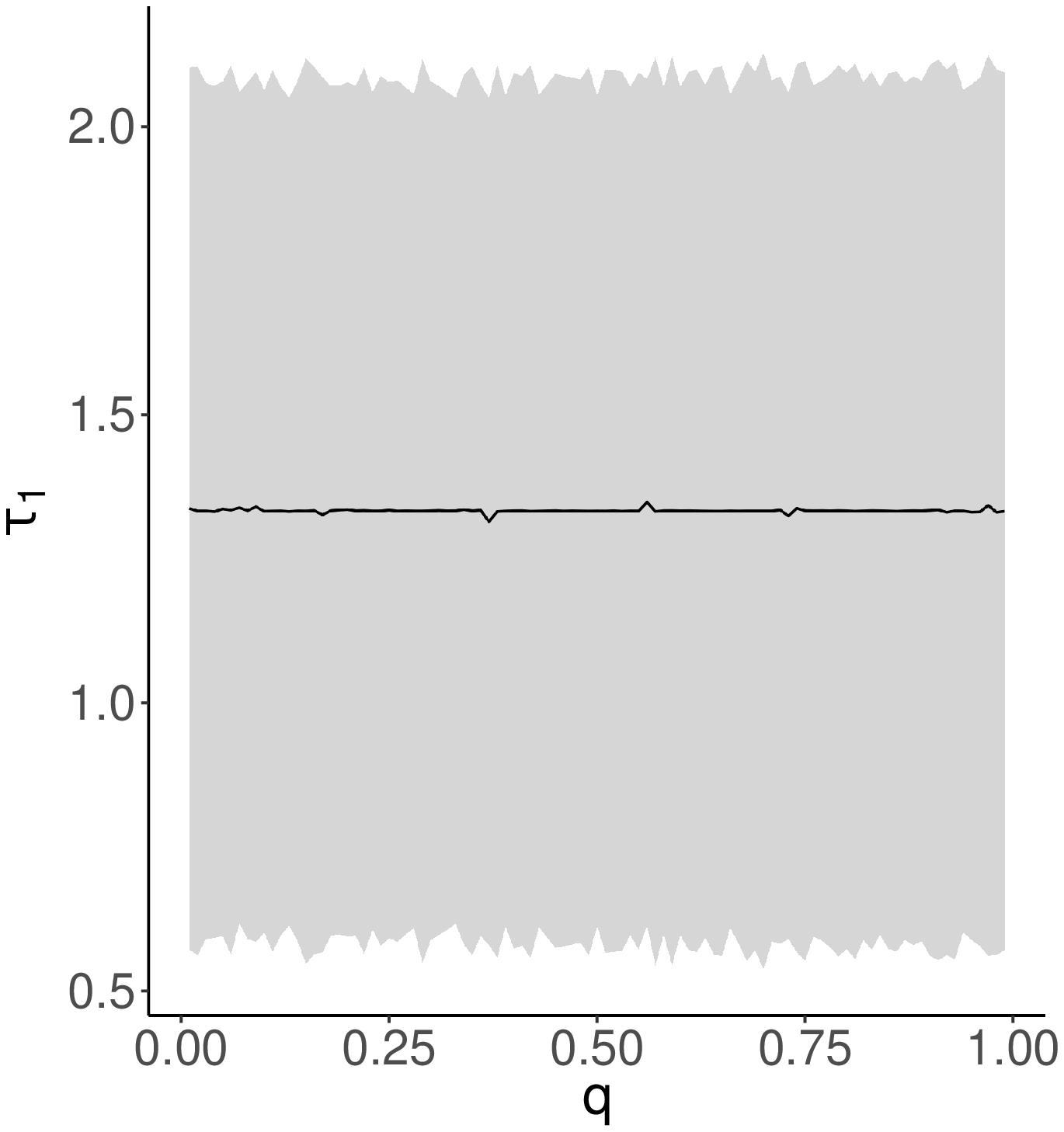}}
	\subfigure[$\tau_2$]{\includegraphics[scale = 0.25]{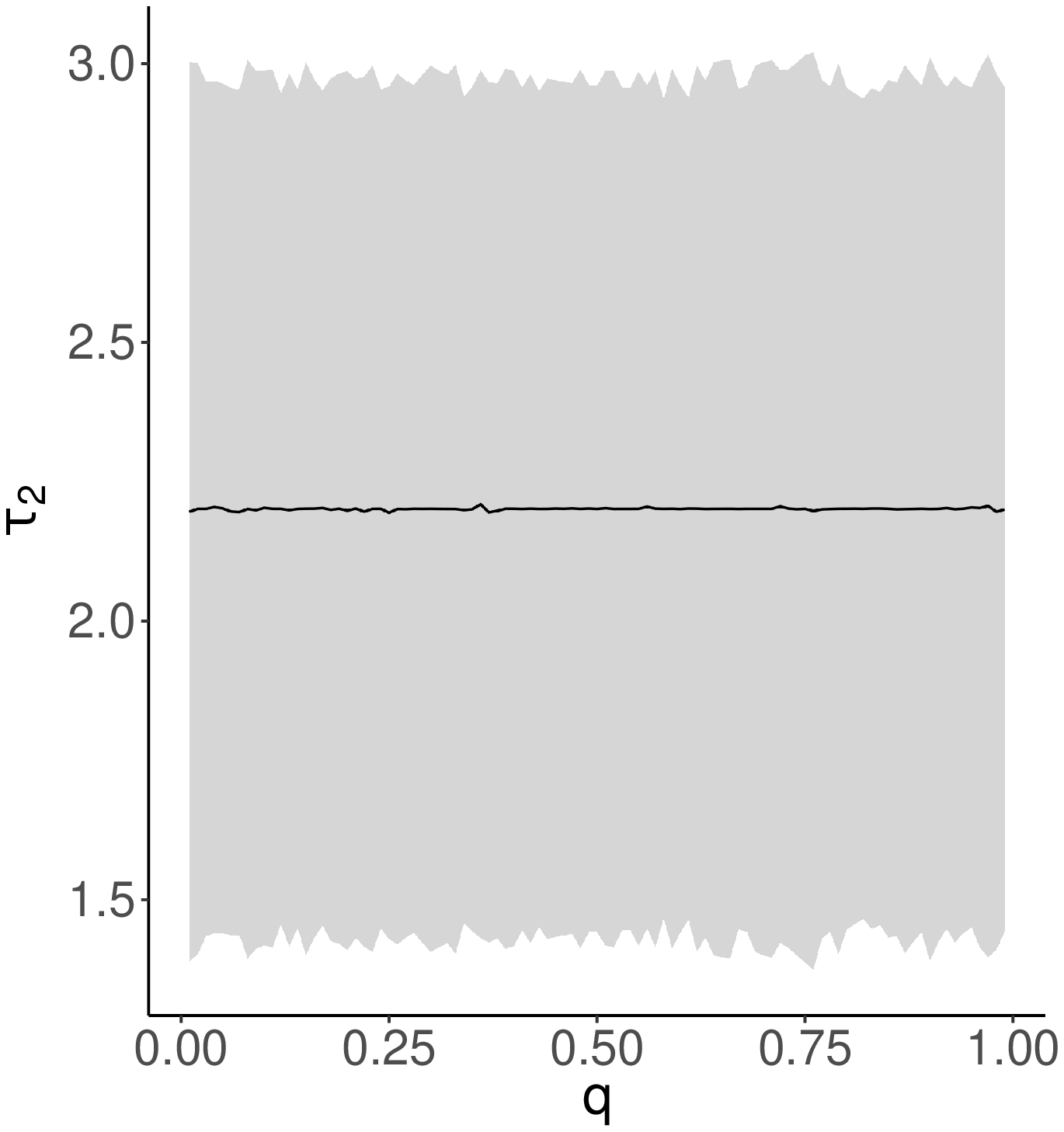}}
	\caption{Estimated parameters in the log-$t$ quantile ARMA model across $q$ for the Walmart sales data.}
	\label{fig:estimates}
\end{figure}

In Figure \ref{fig:estimates}, we plot the estimated parameters of the log-$t$ quantile ARMA model across $q\in\{0.01,\ldots, 0.99\}$. We notice the coefficients $\hat{\beta}_0, \hat{\beta}_1, \hat{\beta}_2$ to increase with an increase in the quantile value. Futhermore, we also note that the other coefficients do not vary according to the quantile fixed in the fitted model.

\begin{figure}[!ht]
	\centering
	\includegraphics[scale = 0.75]{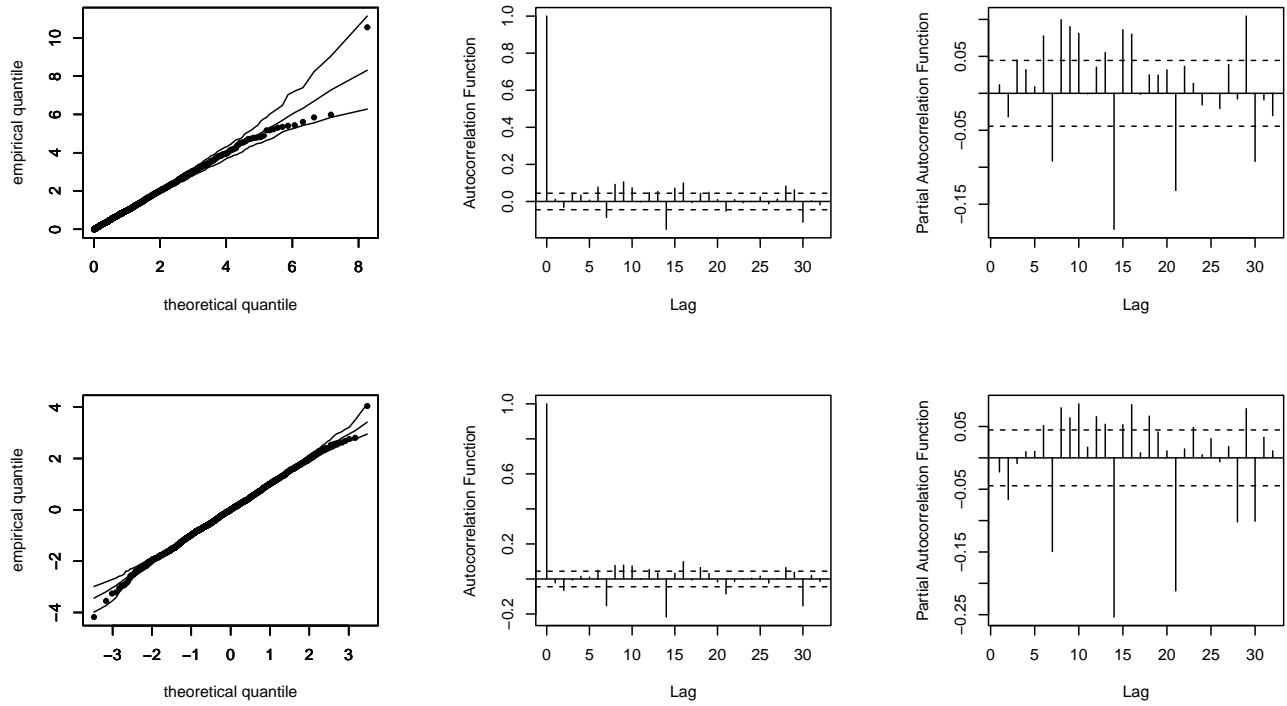} 
	\caption{ \small QQ plot with simulated envelopes (left), ACF (center) and PACF plots (right) of the CS and RQ residuals corresponding to the log-$t$ ARMA model with $q = 0.5$ and $\hat \vartheta = 4$ based on the Walmart sales data.}
	\label{fig:resid}
\end{figure}

In Figure~\ref{fig:resid}, we present the QQ plots with simulated, ACF and PACF of the Cox-Snell residuals (CS) and randomized quantile residuals (RQ) corresponding to the proposed model and considering the indicated quantile as $q = 0.5$.
From Figure~\ref{fig:resid}, we note that the CS and RQ residuals indicate that the postulated model presents a good fit to the Walmart sales data. The ACF and PACF plots of the CS and RQ residuals indicate that the log-$t$-ARMA(1,1) quantile model produces non-autocorrelated residuals.

To assess the quality of the forecasts provided by the fitted log-$t$-ARMA quantile model, we divided the Walmart sales data as follows: the first 1,900 observations were used to fit the model, whereas the last 41 observations were used in assessing the quality of forecasts. For comparison purposes, we also considered the fit of the ARMAX(1,1) model to the Walmart sales data.

In Table~\ref{tab:estimates}, we present the ML estimates of the parameters of the log-$t$-ARMA(1,1) quantile model with $q = 0.5$ and the ARMAX(1,1) model. In Table~\ref{tab:comp}, we present some error measures of the point and interval forecasts. For the point error measures, we consider the following: root of mean squared error (RMSE), mean absolute error (MAE), mean absolute scaled error (MASE) and symmetric mean absolute percentage error (SMAPE). For the interval error measure, we only consider the mean scaled interval score (MSIS); see \cite{makridakis2020m5} for more details about these measures. In particular, for the MSIS metric, we are interested in comparing the 95\% interval obtained with the estimates of the log-$t$-ARMA model fitted for the quantiles $q = 0.025, 0.975$ and the asymptotic prediction interval of 95\% based on the ARMAX model.
The RMSE and MSIS metrics are smaller for the log-$t$ quantile model, indicating that the quantile model is better according to the metrics. On the other hand, for the MAE and MASE metrics, the results are smaller for the ARMAX model, indicates that it presents predictions according to these best metrics. Finally, regarding sMAPE, the value is the same for both models.
\begin{table}[!ht]
	\centering
	\caption{ML estimates and standard errors in parentheses for the log-$t$ ARMA with $q = 0.5$ and $\hat \vartheta = 4$, and ARMAX model for the first 1,900 observations of the Walmart sales data.}
	\adjustbox{max height=\dimexpr\textheight-3.5cm\relax,
			max width=\textwidth}{
	\begin{tabular}{lcccccc}
		\toprule
		& & \multicolumn{5}{c}{Estimate (standard error)} \\
		\cline{3-7}
		Model & Parameter & Intercept & snapca & snapt & mother & thanks \\
		\toprule
		\multirow{4}{*}{log-$t$} & $\hat \beta$ & 10.4448(0.0012) & -0.0052(0.0021) & 0.0143(0.0021) & -0.0268(0.0165) & 0.0227(0.0281) \\
		& $\hat \tau$ & -6.8343(0.0436) & & & 1.2365(0.4200) & 2.2772(0.3859) \\
		\cline{2-7}
		& $\hat \phi$ & 0.9191(0.0459) \\
		& $\hat \theta$ & -0.6618(0.0510) \\
		\toprule
		\multirow{3}{*}{ARMAX} & $\hat \beta$ & 34390.22(193.54) & -151.39(94.71) & 543.87(90.9) & -1251.99(412.52) & 1227.12(413.52) \\
		& $\hat \phi$ & 0.9767(0.0077) \\
		& $\hat \theta$ & -0.8869(0.0152) \\
		\bottomrule
	\end{tabular}}
	\label{tab:estimates}
\end{table}

\begin{figure}[!ht]
	\centering
	\subfigure[log-$t$-ARMA(1,1)]{
		\includegraphics[scale = 0.3]{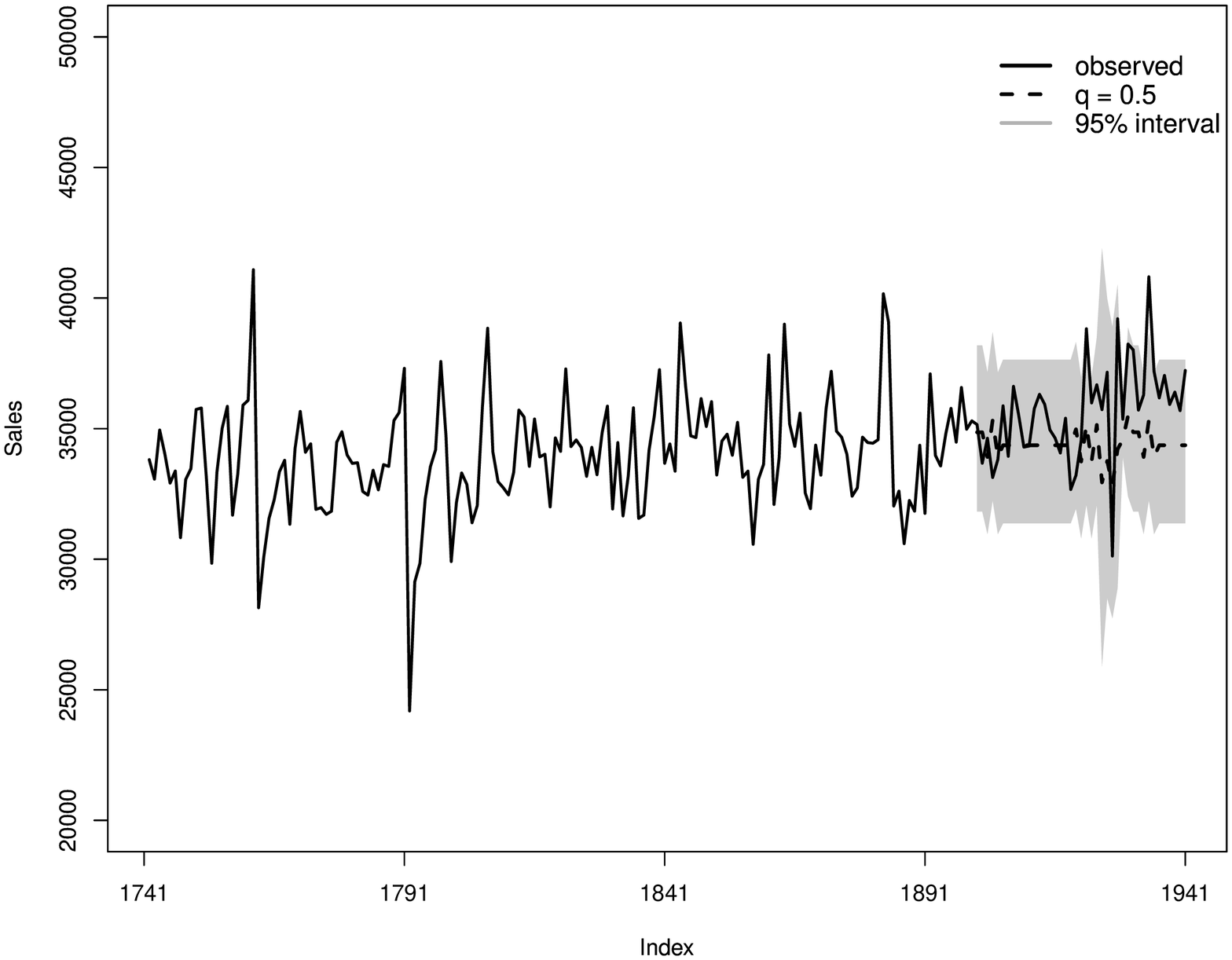}
	}
	\subfigure[ARMAX(1,1)]{
		\includegraphics[scale = 0.3]{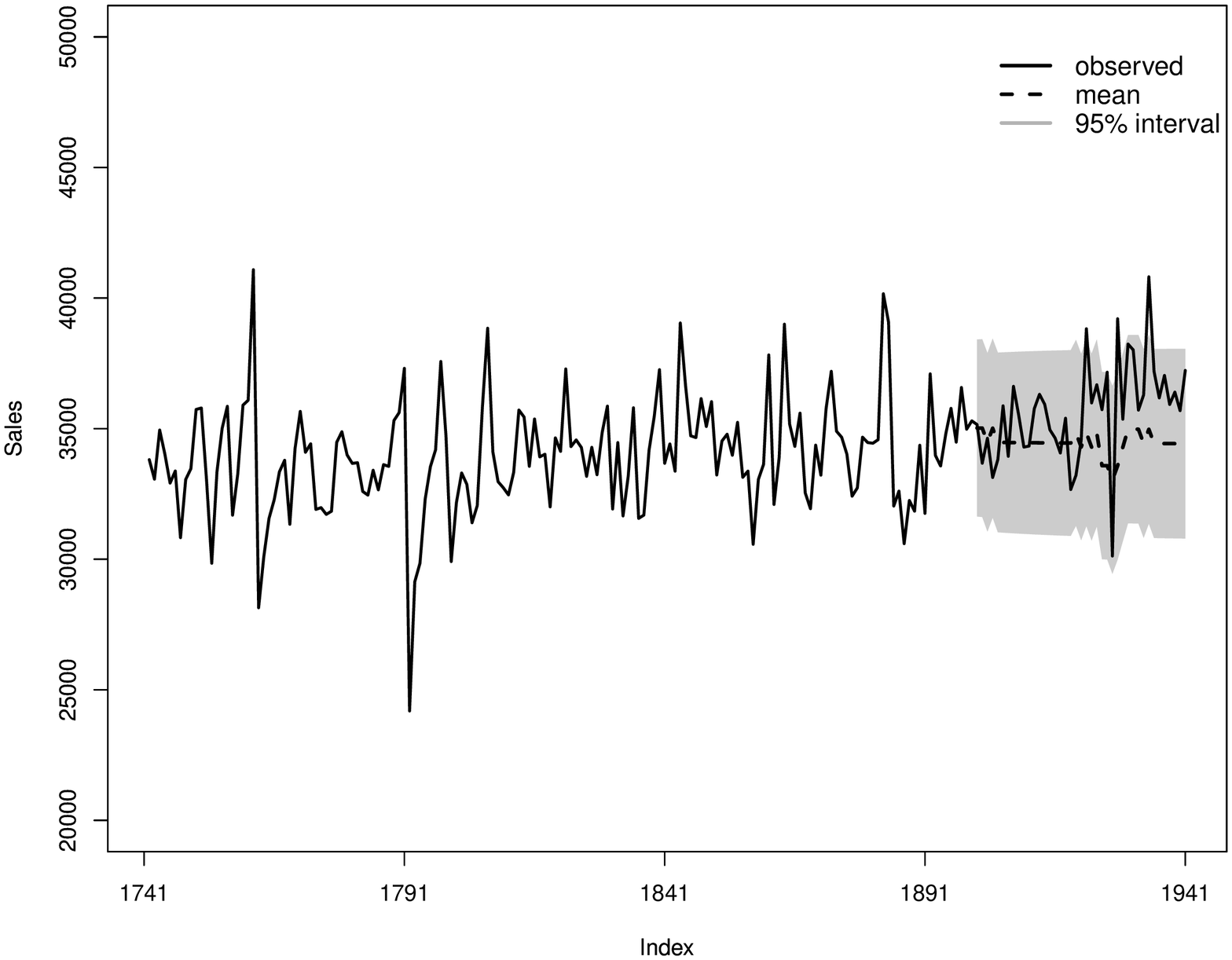}
	}
	\caption{95\% interval prediction for the last 41 observations of the Walmart sales data by the log-$t$-ARMA(1,1) quantile model (left) and the ARMAX(1,1) model (right).}
	\label{fig:forecast}
\end{figure}

In Figure~\ref{fig:forecast}, we present the forecasts for the last 41 observations of the Walmart sales data based on the log-$t$-ARMA and ARMAX models. We can see from the Figure~\ref{fig:forecast} that the prediction intervals contain most of the real values. In addition, we can notice that closer to the prediction horizon, the predicted values for the log-$t$-ARMA quantile and ARMAX models are less accurate.

\begin{table}[!ht]
	\centering
	\caption{Error metrics in the log-$t$-ARMA(1,1) quantile model and the ARMAX(1,1) model based on the last 41 observations of the Walmart sales data.}
	\begin{tabular}{cccccc}
		\toprule
		Model & RMSE & MAE & MASE & SMAPE & MSIS \\
		\hline
		log-$t$ & 2193.11 & 1796.17 & 1.17 & 102.44 & 174.26 \\
		ARMAX & 2210.80 & 1760.71 & 1.15 & 102.44 & 183.63 \\
		\bottomrule
	\end{tabular}
\label{tab:comp}
\end{table}


\section{Concluding remarks}\label{sec:5}

In this paper, we have proposed parametric quantile autoregressive moving average models based on a reparameterized version of the log-symmetric distributions, which is indexed by a quantile parameter. Thus, the proposed autoregressive moving average models is defined in terms of a conditional quantile, allowing different estimates for different quantiles of interest. The proposed model accommodates covariates and can be seen as an extension to the case where there is temporal dependence of the log-symmetric quantile regression models. A Monte Carlo simulation was carried out to evaluate the performance of the proposed models and the estimation method. We have applied the proposed models and some other existing autoregressive moving average models to a real data set correponding to daily sales of 30,491 Walmart products. The results support the suitability of the proposed log-symmetric quantile autoregressive moving average models, both in terms of model fitting and forecasting ability.


\begin{appendices}
	\section{Stationarity conditions}
	\label{sec:apendxa}
	
\begin{Proof}
	Let $\boldsymbol{\Phi} (B) = \sum_{i = 0}^{q} \phi_i B^{i}$ and $\boldsymbol{\Theta} (B) = \sum_{i = 0}^{q} \theta_i B^{i}$ be the moving average polynomial and the autoregressive polynomial, with $\theta_0 = \phi_0 = 1$ e $y_t B^{i} = y_{t-i}$, respectively. The lag operator being $\boldsymbol{\Psi} (B) = \boldsymbol{\Theta}(B) \boldsymbol{\Phi}(B)^{-1} = \sum_{i=0}^{\infty} \psi_i B^{i}$ with $\psi_0 = 1$ e $\boldsymbol{\Phi}(B)$ invertible, the QLS-ARMAX($p,q$) model can be rewritten as
	\begin{eqnarray}
		\boldsymbol{\Phi}(B)(Y_t - h^{-1}(\boldsymbol{x_t^\top \beta}{\color{black} + \varrho_t}))
		&=& \boldsymbol{\Theta}(B) {\color{black}\varepsilon}_t, \nonumber \\
		Y_t &=&
		h^{-1}(\boldsymbol{x_t^\top \beta}{\color{black} + \varrho_t}) + \boldsymbol{\Psi}(B){\color{black}\varepsilon}_t. \nonumber
	\end{eqnarray}
Thus, the marginal expectation of $Y_t$ in the QLS-ARMAX($p,q$) model is given by
\begin{eqnarray}
	\mathrm{E}[Y_t] &=&
	\mathrm{E}[h^{-1}(\boldsymbol{x_t^\top \beta}{\color{black} + \varrho_t}) + \boldsymbol{\Psi}(B){\color{black}\varepsilon}_t], \nonumber \\
	&=&
	{\color{black} \mathrm{E}}[
	h^{-1}(\boldsymbol{x_t^\top \beta}{\color{black} + \varrho_t})
	]
	+
	\boldsymbol{\Psi}(B) \mathrm{E}[{\color{black}\varepsilon}_t], \nonumber \\
	&=&
	{\color{black} \mathrm{E}}
	[
	h^{-1}(\boldsymbol{x_t^\top \beta}{\color{black} + \varrho_t})
	], \nonumber
\end{eqnarray}
where the error ${\color{black}\varepsilon_t = Y_t - Q_t}$, with ${\color{black} \ Q_t=h^{-1}(\boldsymbol{x_t^\top \beta}{\color{black} + \varrho_t})}$, is an MDS sequence with $\mathrm{E}[{\color{black}\varepsilon}_t] = 0$ for all $t$.
\end{Proof}
	

\begin{Proof}
Rewriting $Y_t = Q_t + r_t$, where $\mathrm{E}[r_t] = 0$ and $\mathrm{Cov}[r_t,r_s] = 0$, we have
\begin{eqnarray}\label{eq:vary_t}
	\mathrm{Var}[r_t] &=& \mathrm{E}[\mathrm{Var}[r_t|\mathcal{B}_{t-1}]] + \mathrm{Var}[\mathrm{E}[r_t|\mathcal{B}_{t-1}]] = \mathrm{E}[\mathrm{Var}[r_t|\mathcal{B}_{t-1}]] \nonumber \\
	&=& \mathrm{E}[\mathrm{Var}[Y_t - Q_t|\mathcal{B}_{t-1}]] = \mathrm{E}[\mathrm{Var}[Y_t|\mathcal{B}_{t-1}]].
\end{eqnarray}
Then, by \eqref{eq:vary_t}, the variance of $Y_t$ is given by
\begin{eqnarray}
	\mathrm{Var}[Y_t] &=& \mathrm{Var}[h^{-1}(\boldsymbol{x_t^\top \beta}) + \boldsymbol{\Psi}(B)r_t] = \mathrm{Var}[\boldsymbol{\Psi}(B)r_t] = \mathrm{Var} \left[ \sum_{i = 0}^{\infty} \psi_i r_{t-i} \right] \nonumber \\
	&=& \sum_{i = 0}^{\infty} \psi_i^2 \mathrm{Var} \left[ r_{t-i} \right] = \sum_{i = 0}^{\infty} \psi_i^2 \mathrm{E}[\mathrm{Var}[Y_{t-i}|\mathcal{B}_{t-i-1}]].
\end{eqnarray}
\end{Proof}


\begin{Proof}
From Theorems \ref{teo1} and \ref{teo2}, we have
\begin{eqnarray}\label{eq:covy_t}
	\mathrm{Cov}[Y_t,Y_{t-k}] &=& \mathrm{Cov}\left[ h^{-1}(\boldsymbol{x_t^\top \beta}) + \sum_{i=0}^{\infty} \psi_i r_{t-i}, h^{-1}(\boldsymbol{x_{t-k}^\top \beta}) + \sum_{i=0}^{\infty} \psi_{i+k} r_{t-k-i} \right] \nonumber\\
	&=& \sum_{i=0}^{\infty} \psi_i \psi_{i+k} \mathrm{Cov}\left[ r_{t-i}, r_{t-k-i} \right] = \sum_{i=0}^{\infty} \psi_i \psi_{i+k} \mathrm{Var}\left[ r_{t-i} \right] \nonumber \\
	&=& \sum_{i=0}^{\infty} \psi_i \psi_{i+k} \mathrm{E}[\mathrm{Var}[Y_{t-i}|\mathcal{B}_{t-i-1}]].
\end{eqnarray}
Then, by \eqref{eq:vary_t} and \eqref{eq:covy_t}, we have
\begin{eqnarray}
	\mathrm{Corr}[Y_t,Y_{t-k}] &=& \frac{\mathrm{Cov}[Y_t,Y_{t-k}]}{\sqrt{\mathrm{Var}[Y_t] \mathrm{Var}[Y_{t-k}]}} \nonumber \\
	&=& \frac{\sum_{i=0}^{\infty} \psi_i \psi_{i+k} \mathrm{E}[\mathrm{Var}[Y_{t-i}|\mathcal{B}_{t-i-1}]]}{\sqrt{\sum_{i = 0}^{\infty} \psi_i^2 \mathrm{E}[\mathrm{Var}[Y_{t-i}|\mathcal{B}_{t-i-1}]]} \sqrt{\sum_{i = 0}^{\infty} \psi_{i+k}^2 \mathrm{E}[\mathrm{Var}[Y_{t-k-i}|\mathcal{B}_{t-k-i-1}]]}} \nonumber \\
	&=& \frac{\sum_{i=0}^{\infty} \psi_i \psi_{i+k} \mathrm{E}[\mathrm{Var}[Y_{t-i}|\mathcal{B}_{t-i-1}]]}{\prod_{j \in \left\lbrace 0,k \right\rbrace } \sqrt{ \sum_{i = 0}^{\infty} \psi_{i+j}^2 \mathrm{E}[\mathrm{Var}[Y_{t-j-i}|\mathcal{B}_{t-j-i-1}]]}}. \nonumber
\end{eqnarray}
\end{Proof}

\section{Hessian matrix}
\label{sec:apendxb}

We denoted the Hessian matrix by ${\cal J}(\boldsymbol{\zeta})$, which is given by
\begin{equation}
	{\cal J}(\boldsymbol{\zeta}) =
	\begin{bmatrix}
		{\cal J}_{\bm{\beta \beta}}(\bm \zeta) & {\cal J}_{\bm{\beta \tau}}(\bm \zeta) & {\cal J}_{\bm{\beta \phi}}(\bm \zeta) & {\cal J}_{\bm{\beta \theta}}(\bm \zeta) \\
		{\cal J}_{\bm{\tau \beta}}(\bm \zeta) & {\cal J}_{\bm{\tau \tau}}(\bm \zeta) & {\cal J}_{\bm{\tau \phi}}(\bm \zeta) & {\cal J}_{\bm{\tau \theta}}(\bm \zeta) \\
		{\cal J}_{\bm{\phi \beta}}(\bm \zeta) & {\cal J}_{\bm{\phi \tau}}(\bm \zeta) & {\cal J}_{\bm{\phi \phi}}(\bm \zeta) & {\cal J}_{\bm{\phi \theta}}(\bm \zeta) \\
		{\cal J}_{\bm{\theta \beta}}(\bm \zeta) & {\cal J}_{\bm{\theta \tau}}(\bm \zeta) & {\cal J}_{\bm{\theta \phi}}(\bm \zeta) & {\cal J}_{\bm{\theta \theta}}(\bm \zeta)
	\end{bmatrix} \nonumber
\end{equation}
so that we get the second derivative of $\ell(\boldsymbol{\zeta})_{m,n}$ with respect to $\boldsymbol{\zeta} = (\bm{\beta, \tau, \phi, \theta})^\top$. For the main diagonal elements of Hessian matrix, first, we have $\beta_s$, $s = 0, \ldots, k$,
\begin{eqnarray}
	{\cal J}_{\bm{\beta \beta}}(\bm \zeta) &=&  \frac{\p^2}{\p \beta_s^2} \ell(\boldsymbol{\zeta})_{m,n}, \quad s = 0,1\ldots , k \nonumber \\
	&=&  \sum_{t = m + 1}^{n} \frac{1}{\sqrt{\kappa_t}} \left\lbrace \frac{\p^2 Q_t}{\p \eta_t^2} \frac{\p \eta_t}{\p \beta_s} \left[ v(z_t) \frac{z_t}{Q_t} \frac{\p \eta_t}{\p \beta_s} \right] + \frac{\p Q_t}{\p \eta_t} \left(  \frac{\p}{\p \beta_s} \left[ v(z_t) \frac{z_t}{Q_t} \frac{\p \eta_t}{\p \beta_s} \right] \right) \right\rbrace \nonumber \\
	&=&  \sum_{t = m + 1}^{n} \frac{1}{\sqrt{\kappa_t}} \left\lbrace v(z_t) \frac{z_t}{Q_t} \frac{\p^2 Q_t}{\p \eta_t^2} \left( \frac{\p \eta_t}{\p \beta_s} \right)^2 + \frac{\p Q_t}{\p \eta_t} \left[ \frac{\p v(z_t)}{\p \beta_s} \left( \frac{z_t}{Q_t} \frac{\p \eta_t}{\p \beta_s} \right) + v(z_t) \frac{\p}{\p \beta_s} \left( \frac{z_t}{Q_t} \frac{\p \eta_t}{\p \beta_s} \right) \right] \right\rbrace \nonumber \\
	&=&  \sum_{t = m + 1}^{n} \frac{1}{\sqrt{\kappa_t}} \left\lbrace v(z_t) \frac{z_t}{Q_t} \frac{\p^2 Q_t}{\p \eta_t^2} \left( \frac{\p \eta_t}{\p \beta_s} \right)^2 + \frac{\p Q_t}{\p \eta_t} \left[ \frac{\p v(z_t)}{\p \beta_s} \left( \frac{z_t}{Q_t} \frac{\p \eta_t}{\p \beta_s} \right) + v(z_t) \left[  \left( \frac{\p z_t}{\p \beta_s} \frac{1}{Q_t}\frac{\p \eta_t}{\p \beta_s} \right) +  \right. \right. \right. \nonumber \\
	&& \left. \left. \left.  z_t \frac{\p}{\p \beta_s} \left( \frac{1}{Q_t}\frac{\p \eta_t}{\p \beta_s} \right) \right] \right] \right\rbrace \nonumber
\end{eqnarray}
where $v'(u) = -2 [g''(u)g(u) - g'(u)^2]/g(u)^2,$ $g''(u) = \text{d}^2 g(u)/\text{d}u^2$, with the partial derivatives
\begin{equation}
	\frac{\p v(z_t)}{\p \beta_s} = - 2 \frac{v'(z_t)}{\sqrt{\kappa_t}} \frac{z_t}{Q_t} \frac{\p Q_t}{\p \eta_t} \frac{\p \eta_t}{\p \beta_s} , \qquad \frac{\p z_t}{\p \beta_s} = -\frac{1}{\sqrt{\kappa_t}}\frac{1}{Q_t} \frac{\p Q_t}{\p \eta_t} \frac{\p \eta_t}{\p \beta_s} \nonumber
\end{equation}
and the partial derivatives are given by
\begin{equation}
	\frac{\p}{\p \beta_s} \left( \frac{1}{Q_t}\frac{\p \eta_t}{\p \beta_s} \right) = - \frac{1}{Q_t^2} \frac{\p Q_t}{\p \eta_t} \left(  \frac{\p \eta_t}{\p \beta_s} \right)^2 - \frac{1}{Q_t} \frac{\p^2 \eta_t}{\p \beta_s^2}, \qquad \frac{\p \eta_t}{\p \beta_s} = x_{t,s} - \left( \sum_{i=1}^{p} \phi_i x_{t-i,s} + \sum_{j=1}^{q} \theta_j \frac{\p Q_{t-j}}{\p \eta_{t-j}} \frac{\p \eta_{t-j}}{\p \beta_s} \right), \nonumber
\end{equation}
\begin{equation}
	\frac{\p^2 \eta_t}{\p \beta_s^2} = - \sum_{j=1}^{q} \theta_j \left( \frac{\p^2 Q_{t-j}}{\p \eta_{t-j}^2} \left( \frac{\p \eta_{t-j}}{\p \beta_s} \right)^2 + \frac{\p Q_{t-j}}{\p \eta_{t-j}} \frac{\p^2 \eta_{t-j}}{\p \beta_s^2} \right). \nonumber
\end{equation}
For $\tau_s$, $s = 0, \ldots, l$, we have
\begin{eqnarray}
	{\cal J}_{\bm{\tau \tau}}(\bm \zeta) &=&  \frac{\p^2}{\p \tau_s^2} \ell(\boldsymbol{\zeta})_{m,n}, \quad s = 0,1\ldots , l \nonumber \\
	&=&  \frac{1}{2} \sum_{t = m + 1}^{n} w_{t,s} \left\lbrace  \frac{\p}{\p \tau_s} \left(  \frac{1}{\kappa_t} \frac{\p \kappa_t}{\p \gamma_t} \right) [v(z_t)z_t(z_t - z_p) - 1] + \left( \frac{1}{\kappa_t} \frac{\p \kappa_t}{\p \gamma_t} \right) \frac{\p}{\p \tau_s} [v(z_t)z_t(z_t - z_p) - 1] \right\rbrace, \nonumber
\end{eqnarray}
where, the partial derivatives which given by
\begin{equation}
	\frac{\p}{\p \tau_s} \left( \frac{1}{\kappa_t} \frac{\p \kappa_t}{\p \gamma_t} \right) = - \frac{1}{\kappa_t^2} \left( \frac{\p \kappa_t}{\p \gamma_t} \right)^2 w_{t,s} + \frac{1}{\kappa_t} \frac{\p^2 \kappa_t}{\p \gamma_t^2} w_{t,s} = \frac{1}{\kappa_t} w_{t,s} \left[ \frac{\p^2 \kappa_t}{\p \gamma_t^2} - \frac{1}{\kappa_t} \left( \frac{\p \kappa_t}{\p \gamma_t} \right)^2 \right], \nonumber
\end{equation}
with
\begin{eqnarray}
	\frac{\p}{\p \tau_s} [v(z_t)z_t(z_t - z_p) - 1] &=& \frac{\p v(z_t)}{\p \tau_s} [z_t(z_t - z_p) - 1] + v(z_t) \frac{\p}{\p \tau_s} \left[ z_t(z_t - z_p) - 1 \right] \nonumber \\
	&=& - \frac{v'(z_t)}{\kappa_t} z_t (z_t - z_p) \frac{\p \kappa_t}{\p \gamma_t} w_{t,s} [z_t(z_t - z_p) - 1] + v(z_t) \nonumber \\
	&& \left[ \frac{\p z_t}{\p \tau_s}(z_t - z_p) + z_t \frac{\p}{\p \tau_s} (z_t - z_p) \right] \nonumber \\
	&=& - \frac{v'(z_t)}{\kappa_t} z_t (z_t - z_p) [z_t(z_t - z_p) - 1] \frac{\p \kappa_t}{\p \gamma_t} w_{t,s} + v(z_t) \nonumber \\
	&& \left[ - \frac{1}{2} \frac{1}{\kappa_t} (z_t - z_p)^2 \frac{\p \kappa_t}{\p \gamma_t} w_{t,s} - \frac{1}{2} \frac{z_t}{\kappa_t} (z_t - z_p) \frac{\p \kappa_t}{\p \gamma_t} w_{t,s} \right] \nonumber \\
	&=& - \frac{v'(z_t)}{\kappa_t} z_t (z_t - z_p) [z_t(z_t - z_p) - 1] \frac{\p \kappa_t}{\p \gamma_t} w_{t,s} - \frac{1}{2} \frac{v(z_t)}{\kappa_t}(z_t - z_p) \frac{\p \kappa_t}{\p \gamma_t} w_{t,s} \nonumber \\
	&& [(z_t - z_p) + z_t]. \nonumber
\end{eqnarray}
For $\phi_s$, $s = 0, \ldots, p$, we have the partial derivatives
\begin{eqnarray}
	{\cal J}_{\bm{\phi \phi}}(\bm \zeta) &=&  \frac{\p^2}{\p \phi_s^2} \ell_{m,n}({\boldsymbol{\zeta}}), \quad s = 1,\ldots, p \nonumber \\
	&=&  \sum_{t = m+1}^{n} \frac{1}{\sqrt{\kappa_t}} \left\lbrace \frac{\p v(z_t)}{\p \phi_s} \left[  \frac{z_t}{Q_t} \frac{\p Q_t}{\p \eta_t} \frac{\p \eta_t}{\p \phi_s} \right] + v(z_t) \left[ \frac{\p}{\p \phi_s} \left( \frac{z_t}{Q_t} \frac{\p Q_t}{\p \eta_t} \frac{\p \eta_t}{\p \phi_s} \right)  \right]  \right\rbrace \nonumber \\
	&=&  \sum_{t = m+1}^{n} \frac{1}{\sqrt{\kappa_t}} \left\lbrace - 2 \frac{v'(z_t)}{\sqrt{\kappa_t}} \frac{z_t}{Q_t} \frac{\p Q_t}{\p \eta_t} \frac{\p \eta_t}{\p \phi_s} \left[  \frac{z_t}{Q_t} \frac{\p Q_t}{\p \eta_t} \frac{\p \eta_t}{\p \phi_s} \right] + v(z_t) \left[ \frac{\p}{\p \phi_s} \left( \frac{z_t}{Q_t} \frac{\p Q_t}{\p \eta_t} \frac{\p \eta_t}{\p \phi_s} \right)  \right]  \right\rbrace, \nonumber
\end{eqnarray}
where, the partial derivatives which given by
\begin{eqnarray}
	\frac{\p}{\p \phi_s} \left( \frac{z_t}{Q_t} \frac{\p Q_t}{\p \eta_t} \frac{\p \eta_t}{\p \phi_s} \right) &=& \frac{\p z_t}{\p \phi_s} \left( \frac{1}{Q_t} \frac{\p Q_t}{\p \eta_t} \frac{\p \eta_t}{\p \phi_s} \right) + z_t \left[ \frac{\p}{\p \phi_s} \left(  \frac{1}{Q_t} \frac{\p Q_t}{\p \eta_t} \frac{\p \eta_t}{\p \phi_s} \right) \right] \nonumber \\
	&=& \frac{\p z_t}{\p \phi_s} \left( \frac{1}{Q_t} \frac{\p Q_t}{\p \eta_t} \frac{\p \eta_t}{\p \phi_s} \right) + z_t \left[ \frac{\p Q_t^{-1}}{\p \phi_s} \left( \frac{\p Q_t}{\p \eta_t} \frac{\p \eta_t}{\p \phi_s} \right) + \frac{1}{Q_t} \frac{\p}{\p \phi_s} \left( \frac{\p Q_t}{\p \eta_t} \frac{\p \eta_t}{\p \phi_s} \right)  \right] \nonumber \\
	&=& \frac{\p z_t}{\p \phi_s} \left( \frac{1}{Q_t} \frac{\p Q_t}{\p \eta_t} \frac{\p \eta_t}{\p \phi_s} \right) + z_t \left[ \frac{\p Q_t^{-1}}{\p \phi_s} \left( \frac{\p Q_t}{\p \eta_t} \frac{\p \eta_t}{\p \phi_s} \right) + \frac{1}{Q_t} \left\{ \frac{\p^2 Q_t}{\p \eta_t^2} \left( \frac{\p \eta_t}{\p \phi_s} \right)^2 + \right. \right. \nonumber \\
	&& \left. \left.  \frac{\p Q_t}{\p \eta_t} \frac{\p^2 \eta_t}{\p \phi_s^2} \right\} \right] \nonumber \nonumber
\end{eqnarray}
with the partial derivatives are given by
\begin{equation}
	\frac{\p z_t}{\p \phi_s} = - \frac{1}{\sqrt{\kappa_t}} \frac{1}{Q_t} \frac{\p Q_t}{\p \eta_t} \frac{\p \eta_t}{\p \phi_s}, \qquad \frac{\p Q_t^{-1}}{\p \phi_s} = - \frac{1}{Q_t^2} \frac{\p Q_t}{\p \eta_t} \frac{\p \eta_t}{\p \phi_s} \nonumber
\end{equation}
and
\begin{equation}
	\frac{\p \eta_t}{\p \phi_s} = \left[ h(y_{t-s}) - \boldsymbol{x_{t-s}^\top \beta} \right]  - \sum_{j=1}^{q} \theta_j \frac{\p Q_{t-j}}{\p \eta_{t-j}} \frac{\p \eta_{t-j}}{\p \phi_s}, \qquad  \frac{\p^2 \eta_t}{\p \phi_s^2} = - \sum_{j=1}^{q} \theta_j \left( \frac{\p^2 Q_{t-j}}{\p \eta_{t-j}^2} \frac{\p \eta_{t-j}}{\p \phi_s} + \frac{\p Q_{t-j}}{\p \eta_{t-j}} \frac{\p^2 \eta_{t-j}}{\p \phi_s^2} \right). \nonumber
\end{equation}

\begin{eqnarray}
	{\cal J}_{\bm{\theta \theta}}(\bm \zeta) &=&  \frac{\p^2}{\p \theta_s^2} \ell_{m,n}{(\boldsymbol{\zeta})}, \quad s = 1,\ldots, q \nonumber \\
	&=&  \sum_{t = m + 1}^{n} \frac{1}{\sqrt{\kappa_t}} \left\lbrace \frac{\p v(z_t)}{\p \theta_s} \left[ \frac{z_t}{Q_t} \frac{\p Q_t}{\p \eta_t} \frac{\p \eta_t}{\p \theta_s} \right] + v(z_t) \frac{\p}{\p \phi_s} \left[ \frac{z_t}{Q_t} \frac{\p Q_t}{\p \eta_t} \frac{\p \eta_t}{\p \theta_s} \right] \right\rbrace \nonumber \\
	&=&  \sum_{t = m + 1}^{n} \frac{1}{\sqrt{\kappa_t}} \left\lbrace - 2 \frac{v'(z_t)}{\sqrt{\kappa_t}} \frac{z_t}{Q_t} \frac{\p Q_t}{\p \eta_t} \frac{\p \eta_t}{\p \theta_s} \left[ \frac{z_t}{Q_t} \frac{\p Q_t}{\p \eta_t} \frac{\p \eta_t}{\p \theta_s} \right] + v(z_t) \frac{\p}{\p \phi_s} \left[ \frac{z_t}{Q_t} \frac{\p Q_t}{\p \eta_t} \frac{\p \eta_t}{\p \theta_s} \right] \right\rbrace \nonumber \\
	&=&  \sum_{t = m + 1}^{n} \frac{1}{\sqrt{\kappa_t}} \left\lbrace - 2 \frac{v'(z_t)}{\sqrt{\kappa_t}}  \left( \frac{z_t}{Q_t} \frac{\p Q_t}{\p \eta_t} \frac{\p \eta_t}{\p \theta_s} \right)^2 + v(z_t) \frac{\p}{\p \phi_s} \left( \frac{z_t}{Q_t} \frac{\p Q_t}{\p \eta_t} \frac{\p \eta_t}{\p \theta_s} \right) \right\rbrace \nonumber
\end{eqnarray}
where
\begin{eqnarray}
	\frac{\p}{\p \theta_s} \left( \frac{z_t}{Q_t} \frac{\p Q_t}{\p \eta_t} \frac{\p \eta_t}{\p \theta_s} \right) &=& \frac{\p z_t}{\p \theta_s} \left( \frac{1}{Q_t} \frac{\p Q_t}{\p \eta_t} \frac{\p \eta_t}{\p \theta_s} \right) + z_t \frac{\p}{\p \theta_s} \left\lbrace  \frac{1}{Q_t} \frac{\p Q_t}{\p \eta_t} \frac{\p \eta_t}{\p \phi_s} \right\rbrace \nonumber \\
	&=& - \frac{1}{\sqrt{\kappa_t}} \left(  \frac{1}{Q_t} \frac{\p Q_t}{\p \eta_t} \frac{\p \eta_t}{\p \theta_s} \right)^2 + z_t \left\lbrace \frac{\p Q_t^{-1}}{\p \theta_s} \left[ \frac{\p Q_t}{\p \eta_t} \frac{\p \eta_t}{\p \theta_s} \right] + \frac{1}{Q_t} \frac{\p}{\p \theta_s} \left[ \frac{\p Q_t}{\p \eta_t} \frac{\p \eta_t}{\p \theta_s} \right]  \right\rbrace \nonumber \\
	&=& - \frac{1}{\sqrt{\kappa_t}} \left(  \frac{1}{Q_t} \frac{\p Q_t}{\p \eta_t} \frac{\p \eta_t}{\p \theta_s} \right)^2 + z_t \left\lbrace \frac{\p Q_t^{-1}}{\p \phi_s} \left[ \frac{\p Q_t}{\p \eta_t} \frac{\p \eta_t}{\p \phi_s} \right] + \frac{1}{Q_t} \left[ \frac{\p^2 Q_t}{\p \eta_t^2} \left( \frac{\p \eta_t}{\p \phi_s} \right)^2 + \right. \right. \nonumber \\
	&& \left. \left.  \frac{\p Q_t}{\p \eta_t} \frac{\p^2 \eta_t}{\p \phi_s^2} \right]  \right\rbrace \nonumber
\end{eqnarray}
with
\begin{equation}
	\frac{\p Q_t^{-1}}{\p \theta_s} = - \frac{1}{Q_t^2} \frac{\p Q_t}{\p \eta_t} \frac{\p \eta_t}{\p \theta_s}, \quad \frac{\p \eta_t}{\p \theta_s} = r_{t-s} - \theta_s \frac{\p Q_{t-s}}{\p \eta_{t-s}} \frac{\p \eta_{t-s}}{\p \theta_s}, \nonumber
\end{equation}
\begin{equation}
	\frac{\p^2 \eta_t}{\p \theta_s^2} = -2 \frac{\p Q_{t-s}}{\p \eta_{t-s}} \frac{\p \eta_{t-s}}{\p \theta_s} - \left[ \frac{\p^2 Q_{t-s}}{\p \eta_{t-s}^2} \left( \frac{\p \eta_{t-s}}{\p \theta_s} \right)^2 + \frac{\p Q_{t-s}}{\p \eta_{t-s}} \frac{\p^2 \eta_{t-s}}{\p \theta_s^2}  \right]. \nonumber
\end{equation}
Now, for the elements off-diagonal of the Hessian matrix,
based on the first derivatives in \eqref{eq:beta_score}-\eqref{eq:theta_score},
 we have, for $\beta_s$, $s = 0, \ldots, k$ and $\tau_u$, $u = 0,\ldots, l,$
\begin{equation}
	{\cal J}_{\bm {\beta \tau}}(\bm \zeta) = \sum_{t = m + 1}^{n} \left\lbrace \left(  \frac{\p}{\p \tau_u} \frac{v(z_t)}{\sqrt{\kappa_t}} \right) z_t + \frac{v(z_t)}{\sqrt{\kappa_t}} \left( \frac{\p z_t}{\p \tau_u} \right) \right\rbrace \frac{1}{Q_t} \frac{\p Q_t}{\p \eta_t} \left[ x_{t,s} - \sum_{i=1}^{p} \phi_i x_{t-i,s} - \sum_{j=1}^{q} \theta_j \frac{\p Q_{t-j}}{\p \eta_{t-j}} \right], \nonumber
\end{equation}
where, the partial derivative is given by
\begin{equation}
	\frac{\p}{\p \tau_u} \left( \frac{v(z_t)}{\sqrt{\kappa_t}} \right)  = -\frac{v'(z_t)}{\kappa_t} \frac{z_t}{\sqrt{\kappa_t}}(z_t - z_p) \frac{\p \kappa_t}{\p \gamma_t} w_{t,u} - \frac{1}{2} \frac{v(z_t)}{\kappa_t} \frac{1}{\sqrt{\kappa_t}} \frac{\p \kappa_t}{\p \gamma_t} w_{t,u} \quad \text{and} \quad \frac{\p z_t}{\p \tau_l} = -\frac{1}{2} \frac{1}{\kappa_t} \frac{\p \kappa_t}{\p \gamma_t} w_{t,u}. \nonumber
\end{equation}
For $\beta_s$, $s = 0, \ldots, k$ and $\phi_u$, $u = 0,\ldots, p,$ we have,
\begin{equation}
	{\cal J}_{\bm {\beta \phi}}(\bm \zeta) = \sum_{t = m+1}^{n} \frac{1}{\sqrt{\kappa_t}} \left\{ \frac{\p v(z_t)}{\p \phi_u} \left( \frac{z_t}{Q_t} \frac{\p Q_t}{\p \eta_t} \frac{\p \eta_t}{\p \beta_s} \right) + v(z_t) \frac{\p}{\p \phi_u} \left( \frac{z_t}{Q_t} \frac{\p Q_t}{\p \eta_t} \frac{\p \eta_t}{\p \beta_s} \right) \right\},  \nonumber
\end{equation}
where, the partial derivative with respect to $\phi_u$ is given by
\begin{eqnarray}
	\frac{\p}{\p \phi_u} \left( \frac{z_t}{Q_t} \frac{\p Q_t}{\p \eta_t} \frac{\p \eta_t}{\p \beta_s} \right) &=& \frac{\p z_t}{\p \phi_u} \left( \frac{1}{Q_t} \frac{\p Q_t}{\p \eta_t} \frac{\p \eta_t}{\p \beta_s} \right) + z_t \frac{\p}{\p \phi_u} \left\{  \frac{1}{Q_t} \frac{\p Q_t}{\p \eta_t} \frac{\p \eta_t}{\p \beta_s} \right\} \nonumber \\
	&=& \frac{\p z_t}{\p \phi_u} \left( \frac{1}{Q_t} \frac{\p Q_t}{\p \eta_t} \frac{\p \eta_t}{\p \beta_s} \right) + z_t \left\{ \frac{\p Q^{-1}_t}{\p \phi_l} \left( \frac{\p Q_t}{\p \eta_t} \frac{\p \eta_t}{\p \beta_s} \right) +  \frac{1}{Q_t} \frac{\p }{\p \phi_u} \left( \frac{\p Q_t}{\p \eta_t} \frac{\p \eta_t}{\p \beta_s} \right) \right\} \nonumber \\
	&=& -\frac{1}{\sqrt{\kappa_t}} \left( \frac{1}{Q_t} \frac{\p Q_t}{\p \eta_t} \right)^2 \frac{\p \eta_t}{\p \beta_s} \frac{\p \eta_t}{\p \phi_u} + z_t \left\{ \frac{\p Q^{-1}_t}{\p \phi_u} \left( \frac{\p Q_t}{\p \eta_t} \frac{\p \eta_t}{\p \beta_s} \right) +  \frac{1}{Q_t} \left[ \left( \frac{\p^2 Q_t}{\p \eta_t^2} \frac{\p \eta_t}{\p \phi_u} \frac{\p \eta_t}{\p \beta_s} \right) + \right. \right.  \nonumber \\
	&&  \left. \left.  \left( \frac{\p Q_t}{\p \eta_t} \frac{\p^2 \eta_t}{\p \beta_s \p \phi_u} \right) \right] \right\}. \nonumber
\end{eqnarray}
For the partial derivatives with respect to $\beta_s$, $s = 0,\ldots, k$ and $\theta_u$, $u = 0,\ldots, q$, we have
\begin{equation}
	{\cal J}_{\bm {\beta \theta}}(\bm \zeta) = \sum_{t = m+1}^{n} \frac{1}{\sqrt{\kappa_t}} \left\{ \frac{\p v(z_t)}{\p \theta_u} \left( \frac{z_t}{Q_t} \frac{\p Q_t}{\p \eta_t} \frac{\p \eta_t}{\p \beta_s} \right) + v(z_t) \frac{\p}{\p \theta_u} \left( \frac{z_t}{Q_t} \frac{\p Q_t}{\p \eta_t} \frac{\p \eta_t}{\p \beta_s} \right) \right\},  \nonumber
\end{equation}
where, the partial derivative with respect to $\theta_u$ is given by
\begin{eqnarray}
	\frac{\p}{\p \theta_u} \left( \frac{z_t}{Q_t} \frac{\p Q_t}{\p \eta_t} \frac{\p \eta_t}{\p \beta_s} \right) &=& \frac{\p z_t}{\p \theta_u} \left( \frac{1}{Q_t} \frac{\p Q_t}{\p \eta_t} \frac{\p \eta_t}{\p \beta_s} \right) + z_t \frac{\p}{\p \theta_u} \left\{  \frac{1}{Q_t} \frac{\p Q_t}{\p \eta_t} \frac{\p \eta_t}{\p \beta_s} \right\} \nonumber \\
	&=& \frac{\p z_t}{\p \theta_u} \left( \frac{1}{Q_t} \frac{\p Q_t}{\p \eta_t} \frac{\p \eta_t}{\p \beta_s} \right) + z_t \left\{ \frac{\p Q^{-1}_t}{\p \theta_l} \left( \frac{\p Q_t}{\p \eta_t} \frac{\p \eta_t}{\p \beta_s} \right) +  \frac{1}{Q_t} \frac{\p }{\p \theta_u} \left( \frac{\p Q_t}{\p \eta_t} \frac{\p \eta_t}{\p \beta_s} \right) \right\} \nonumber \\
	&=& -\frac{1}{\sqrt{\kappa_t}} \left( \frac{1}{Q_t} \frac{\p Q_t}{\p \eta_t} \right)^2 \frac{\p \eta_t}{\p \beta_s} \frac{\p \eta_t}{\p \theta_u} + z_t \left\{ \frac{\p Q^{-1}_t}{\p \theta_u} \left( \frac{\p Q_t}{\p \eta_t} \frac{\p \eta_t}{\p \beta_s} \right) +  \frac{1}{Q_t} \left[ \left( \frac{\p^2 Q_t}{\p \eta_t^2} \frac{\p \eta_t}{\p \theta_u} \frac{\p \eta_t}{\p \beta_s} \right) + \right. \right.  \nonumber \\
	&&  \left. \left.  \left( \frac{\p Q_t}{\p \eta_t} \frac{\p^2 \eta_t}{\p \beta_s \p \theta_u} \right) \right] \right\}. \nonumber
\end{eqnarray}
For the partial derivatives with respect to $\tau_s$, $s = 0,\ldots, l$ and $\phi_u$, $u = 0,\ldots, p$, we have
\begin{eqnarray}
	{\cal J}_{\bm {\tau \phi}}(\bm \zeta) &=& \frac{1}{2} \sum_{t = m+1}^{n} \frac{w_{t,s}}{\kappa_t} \frac{\p \kappa_t}{\p \gamma_t} \left\{ \frac{\p v(z_t)}{\p \phi_u} [z_t (z_t - z_p) - 1] + v(z_t) \left[ \frac{\p}{\p \phi_u} (z_t (z_t - z_p)) \right] \right\} \nonumber \\
	&=& \frac{1}{2} \sum_{t = m+1}^{n} \frac{w_{t,s}}{\kappa_t} \frac{\p \kappa_t}{\p \gamma_t} \left\{ \left[ - 2\frac{ v'(z_t)}{\sqrt{\kappa_t}} \frac{z_t}{Q_t} \frac{\p Q_t}{\p \eta_t} \frac{\p \eta_t}{\p \phi_u} \right] [z_t (z_t - z_p) - 1] + v(z_t) \left[ \frac{\p}{\p \phi_u} (z_t (z_t - z_p)) \right] \right\} \nonumber
\end{eqnarray}
where, the partial derivative with respect to $\phi_u$ is given by
\begin{eqnarray}
	\frac{\p}{\p \phi_l} z_t (z_t - z_p) &=& \frac{\p z_t}{\p \phi_l}(z_t - z_p) + z_t \frac{\p z_t}{\p \phi_l}
	= - \frac{1}{\sqrt{\kappa_t}} \frac{1}{Q_t} \frac{\p Q_t}{\p \eta_t} \frac{\p \eta_t}{\p \phi_s} (z_t - z_p) - \frac{1}{\sqrt{\kappa_t}} \frac{1}{Q_t} \frac{\p Q_t}{\p \eta_t} \frac{\p \eta_t}{\p \phi_l} z_t \nonumber \\
	&=& - \frac{1}{\sqrt{\kappa_t}} \frac{1}{Q_t} \frac{\p Q_t}{\p \eta_t} \frac{\p \eta_t}{\p \phi_l} (2 z_t - z_p). \nonumber
\end{eqnarray}
For the partial derivatives with respect to $\tau_s$, $s = 0,\ldots, l$ and $\theta_u$, $u = 0,\ldots, q$, we have
\begin{eqnarray}
	{\cal J}_{\bm {\tau \theta}}(\bm \zeta) &=& \frac{1}{2} \sum_{t = m+1}^{n} \frac{w_{t,s}}{\kappa_t} \frac{\p \kappa_t}{\p \gamma_t} \left\{ \frac{\p v(z_t)}{\p \theta_u} [z_t (z_t - z_p) - 1] + v(z_t) \left[ \frac{\p}{\p \theta_u} (z_t (z_t - z_p)) \right] \right\} \nonumber \\
	&=& \frac{1}{2} \sum_{t = m+1}^{n} \frac{w_{t,s}}{\kappa_t} \frac{\p \kappa_t}{\p \gamma_t} \left\{ \left[ - 2\frac{ v'(z_t)}{\sqrt{\kappa_t}} \frac{z_t}{Q_t} \frac{\p Q_t}{\p \eta_t} \frac{\p \eta_t}{\p \theta_u} \right] [z_t (z_t - z_p) - 1] + v(z_t) \left[ \frac{\p}{\p \theta_l} (z_t (z_t - z_p)) \right] \right\} \nonumber
\end{eqnarray}
where, the partial derivative with respect to $\theta_u$ is given by
\begin{eqnarray}
	\frac{\p}{\p \theta_l} z_t (z_t - z_p) &=& \frac{\p z_t}{\p \theta_l}(z_t - z_p) + z_t \frac{\p z_t}{\p \theta_l}
	= - \frac{1}{\sqrt{\kappa_t}} \frac{1}{Q_t} \frac{\p Q_t}{\p \eta_t} \frac{\p \eta_t}{\p \theta_l} (z_t - z_p) - \frac{1}{\sqrt{\kappa_t}} \frac{1}{Q_t} \frac{\p Q_t}{\p \eta_t} \frac{\p \eta_t}{\p \theta_l} z_t \nonumber \\
	&=& - \frac{1}{\sqrt{\kappa_t}} \frac{1}{Q_t} \frac{\p Q_t}{\p \eta_t} \frac{\p \eta_t}{\p \theta_l} (2 z_t - z_p). \nonumber
\end{eqnarray}
For the partial derivatives with respect to $\phi_s$, $s = 0,\ldots, p$ and $\theta_u$, $u = 0,\ldots, q$, we have
\begin{eqnarray}
	{\cal J}_{\bm {\phi \theta}}(\bm \zeta) &=& \frac{\p}{\p \theta_u} \left\{ \sum_{t = m + 1}^{n} \frac{v(z_t)}{\sqrt{\kappa_t}} \frac{z_t}{Q_t} \frac{\p Q_t}{\p \eta_t} \frac{\p \eta_t}{\p \phi_s} \right\} \nonumber \\
	&=&  \sum_{t = m + 1}^{n} \frac{1}{\sqrt{\kappa_t}} \left\{ \frac{\p v(z_t)}{\p \theta_u} \left( \frac{z_t}{Q_t} \frac{\p Q_t}{\p \eta_t} \frac{\p \eta_t}{\p \phi_s} \right) + v(z_t) \left[ \frac{\p}{\p \theta_u} \left( \frac{z_t}{Q_t} \frac{\p Q_t}{\p \eta_t} \frac{\p \eta_t}{\p \phi_s} \right) \right] \right\} \nonumber \\
	&=& \sum_{t = m + 1}^{n} \frac{1}{\sqrt{\kappa_t}} \left\{ -2 \frac{v'(z_t)}{\sqrt{\kappa_t}} \left( \frac{z_t}{Q_t} \frac{\p Q_t}{\p \eta_t} \right)^2 \frac{\p \eta_t}{\p \phi_s} \frac{\p \eta_t}{\p \theta_u} + v(z_t) \left[ \frac{\p}{\p \theta_u} \left( \frac{z_t}{Q_t} \frac{\p Q_t}{\p \eta_t} \frac{\p \eta_t}{\p \phi_s} \right) \right] \right\} \nonumber
\end{eqnarray}
where, the partial derivative with respect to $\theta_u$ is given by
\begin{eqnarray}
	\frac{\p}{\p \theta_u} \left( \frac{z_t}{Q_t} \frac{\p Q_t}{\p \eta_t} \frac{\p \eta_t}{\p \phi_s} \right) &=& \frac{\p z_t}{\p \theta_u} \left( \frac{1}{Q_t} \frac{\p Q_t}{\p \eta_t} \frac{\p \eta_t}{\p \phi_s} \right) + z_t \frac{\p}{\p \theta_u} \left\{  \frac{1}{Q_t} \frac{\p Q_t}{\p \eta_t} \frac{\p \eta_t}{\p \phi_s} \right\} \nonumber \\
	&=& \frac{\p z_t}{\p \theta_u} \left( \frac{1}{Q_t} \frac{\p Q_t}{\p \eta_t} \frac{\p \eta_t}{\p \phi_s} \right) + z_t \left\{ \frac{\p Q^{-1}_t}{\p \theta_l} \left( \frac{\p Q_t}{\p \eta_t} \frac{\p \eta_t}{\p \phi_s} \right) +  \frac{1}{Q_t} \frac{\p }{\p \theta_u} \left( \frac{\p Q_t}{\p \eta_t} \frac{\p \eta_t}{\p \phi_s} \right) \right\} \nonumber \\
	&=& -\frac{1}{\sqrt{\kappa_t}} \left( \frac{1}{Q_t} \frac{\p Q_t}{\p \eta_t} \right)^2 \frac{\p \eta_t}{\p \phi_s} \frac{\p \eta_t}{\p \theta_u} + z_t \left\{ \frac{\p Q^{-1}_t}{\p \theta_l} \left( \frac{\p Q_t}{\p \eta_t} \frac{\p \eta_t}{\p \phi_s} \right) +  \frac{1}{Q_t} \left[ \left( \frac{\p^2 Q_t}{\p \eta_t^2} \frac{\p \eta_t}{\p \theta_u} \frac{\p \eta_t}{\p \phi_s} \right) + \right. \right.  \nonumber \\
	&&  \left. \left.  \left( \frac{\p Q_t}{\p \eta_t} \frac{\p^2 \eta_t}{\p \phi_s \p \theta_u} \right) \right] \right\}. \nonumber
\end{eqnarray}
\end{appendices}

\end{document}